\documentclass[prx,twocolumn,nofootinbib,citeautoscript,eqsecnum,10pt,longbibliography,notitlepage,superscriptaddress,]{revtex4-1}

\usepackage{graphicx}% Include figure files
\usepackage{dcolumn}% Align table columns on decimal point
\usepackage{bm}% bold math
\usepackage{color}
\usepackage{amsmath}
\usepackage{tabularx,graphicx}
\usepackage{epstopdf}
\usepackage{latexsym}
\usepackage{amssymb}
\usepackage{amsmath}
\usepackage{color, colortbl}
\usepackage{psfrag}
\usepackage{bbm}
\usepackage{bm}
\usepackage{titlesec}
\usepackage{dsfont}
\usepackage{feynmp}
\usepackage{slashed}
\usepackage{multirow}
\usepackage[tight]{subfigure}

\usepackage[papersize={8.5in,11in}]{geometry}

\usepackage{color}
\definecolor{darkblue}{rgb}{0.,0.,0.4}
\definecolor{darkred}{rgb}{0.5,0.,0.}
\definecolor{BlueViolet}{RGB}{138,43,226}
\definecolor{SkyBlue}{RGB}{30,144,255}
\definecolor{DarkGreen}{RGB}{0,100,0}
\usepackage[pdftex,colorlinks=true,linkcolor=darkblue,citecolor=blue,urlcolor=darkred]{hyperref}

\usepackage{physics}

\usepackage{cases}

\usepackage{epsfig}
\usepackage{subfigure}
\usepackage{mathrsfs}
\usepackage{cleveref}
\usepackage{cases}

\usepackage{xcolor}
\usepackage{tabularx}
\usepackage{array}
\usepackage{lmodern}
\usepackage{textcomp}
\usepackage{makecell}
\usepackage{threeparttable}
\usepackage{ulem}
\usepackage{booktabs}

\renewcommand{\epsilon}{\varepsilon}

\allowdisplaybreaks[3]

\newcommand{\red}[1]{\textcolor{red}{#1}}
\newcommand{\blue}[1]{\textcolor{blue}{#1}}

\definecolor{light-gray}{gray}{0.65}

\allowdisplaybreaks[3]

\begin{document}

%\begin{CJK*}{GBK}{song}

%%%%%%%%%%%%%%%%%%%%%%%%%%%%%%%%%%%%%%%%%%%%%%%%%%%%%%%%%%%%%%%%%%%%%%%%%%%%%%%%%%%%%%
%%%%%%%%%%%%%%%%%%%%%%%%%%%%%%%%%%%%%%%%%%%%%%%%%%%%%%%%%%%%%%%%%%%%%%%%%%%%%%%%%%%%%%

\title{Favorable phase transitions induced by spinful electron-electron interactions in two-dimensional
semimetal with a quadratic band crossing point}

\date{\today}

\author{Yi-Sheng Fu}
\affiliation{Department of Physics, Tianjin University, Tianjin 300072, P.R. China}

\author{Jing Wang}
\altaffiliation{Corresponding author: jing$\textunderscore$wang@tju.edu.cn}
\affiliation{Department of Physics, Tianjin University, Tianjin 300072, P.R. China}
\affiliation{Tianjin Key Laboratory of Low Dimensional Materials Physics and
Preparing Technology, Tianjin University, Tianjin 300072, P.R. China}

\begin{abstract}
We study the effects of marginally spinful electron-electron interactions on the low-energy instabilities and
favorable phase transitions in a two-dimensional (2D) spin-$1/2$ semimetal that owns a quadratic band crossing point (QBCP)
parabolically touched by the upper and lower bands. In the framework of a renormalization group procedure,
all sorts of interactions are treated on the equal footing to derive the coupled energy-dependent evolutions of all
interaction couplings that govern the low-energy properties. Deciphering the essential physical
information of such flows, we at first find that the tendencies of interaction parameters fall into three
categories including Limit case, Special case, and General case based on the initial conditions.
In addition, the 2D QBCP system is attracted to several distinct kinds of fixed points (FPs) in the
interaction-parameter space, namely $\mathrm{FP}_1^{+}$/$\mathrm{FP}_2^{-}$, $\mathrm{FP}_1^{\pm}$/
$\mathrm{FP}_2^{\pm}$/$\mathrm{FP}_3^{\pm}$, and $\mathrm{FP}_1^{\pm}/\mathrm{FP}_3^{\pm}/\mathrm{FP}_{41,42,43}^{\pm}$
with the subscripts characterizing the features of FPs for the Limit, Special, and General cases, respectively.
Furthermore, as approaching these FPs, we demonstrate that the spinful fermion-fermion
interactions can induce a number of favorable instabilities accompanied by certain phase transitions.
Specifically, the quantum anomalous Hall (QAH), quantum spin Hall (QSH), and nematic (Nem.) site(bond) states are
dominant for $\mathrm{FP}_{1}^{\pm}$, $\mathrm{FP}_{2}^{\pm}$, and $\mathrm{FP}_{3}^{\pm}$,
respectively. Rather, QSH becomes anisotropic nearby $\mathrm{FP}_{41,42,43}^{\pm}$ with
one component leading and the others subleading. Besides, Nem.site(bond), chiral superconductivity,
and nematic-spin-nematic (NSN.) site(bond) are subleading candidates around these FPs.
Our findings provide valuable insights for further research into the 2D QBCP and similar systems.

\end{abstract}


\maketitle

%%%%%%%%%%%%%%%%%%%%%%%%%%%%%%%%%%%%%%%%%%%%%%%%%%%%%%%%%%%%%%%%%%%%%%%%%%%%%%%%%%%%%%%
%%%%%%%%%%%%%%%%%%%%%%%%%%%%%%%%%%%%%%%%%%%%%%%%%%%%%%%%%%%%%%%%%%%%%%%%%%%%%%%%%%%%%%%

%\tableofcontents{}

\section{Introduction}

The study of semimetal materials is one of the
hottest research fields in contemporary condensed matter physics~\cite{Lee2005Nature,
Neto2009RMP,Fu2007PRL,Roy2009PRB,Moore2010Nature,Hasan2010RMP,Qi2011RMP,
Sheng2012book,Bernevig2013book,Herbut2018Science}.
Last two decades have witnessed a phenomenally rapid
development on these materials~\cite{Lee2005Nature,
Neto2009RMP,Fu2007PRL,Roy2009PRB,Moore2010Nature,Hasan2010RMP,Qi2011RMP,
Sheng2012book,Bernevig2013book,Herbut2018Science,Armitage2018RMP,Roy2018RRX}, which typically include the Dirac
semimetals~\cite{Wang2012PRB,Young2012PRL,Steinberg2014PRL,Liu2014NM,Liu2014Science,
Xiong2015Science} and Weyl semimetals~\cite{Neto2009RMP,Burkov2011PRL,Yang2011PRB,
Wan2011PRB,Huang2015PRX,Xu2015Science,Xu2015NP,Lv2015NP,Weng2015PRX}.
Such materials are equipped with well-known discrete Dirac points, around which gapless quasiparticles
are excited with linear energy dispersions along two or three directions~\cite{Lee2005Nature,
Neto2009RMP,Fu2007PRL,Roy2009PRB,Moore2010Nature,Hasan2010RMP,
Qi2011RMP,Sheng2012book,Bernevig2013book,Korshunov2014PRB,Hung2016PRB,
Nandkishore2013PRB,Potirniche2014PRB,Nandkishore2017PRB,Sarma2016PRB,
Herbut2018Science}. Recently, there has been a gradual shift of interest from linear-dispersion toward
quadratic-dispersion semimetal-like materials~\cite{Chong2008PRB,Fradkin2008PRB,
Fradkin2009PRL,Cvetkovic2012PRB,Vafek2014PRB,Herbut2012PRB,Mandal2019CMP,Zhu2016PRL,
Vafek2010PRB,Yang2010PRB,Wang2017PRB,Wang2020-arxiv,Roy2020-arxiv,Janssen2020PRB,Shah2011.00249,
Nandkishore2017PRB,Luttinger1956PR,Murakami2004PRB,Janssen2015PRB,Boettcher2016PRB,Janssen2017PRB,Boettcher2017PRB,
Mandal2018PRB,Lin2018PRB,Savary2014PRX,Savary2017PRB,Vojta1810.07695,Lai2014arXiv,Goswami2017PRB,
Szabo2018arXiv,Foster2019PRB,Wang1911.09654,Wang2303.10163}. In particular, significant attention has been focused on
the two-dimensional (2D) electronic system with the upper and lower bands parabolically touching at
certain quadratic band crossing point (QBCP)~\cite{Chong2008PRB,Fradkin2008PRB,
Fradkin2009PRL,Cvetkovic2012PRB,Vafek2014PRB,Herbut2012PRB,Mandal2019CMP,Zhu2016PRL,
Vafek2010PRB,Yang2010PRB,Wang2017PRB,Wang2020-arxiv,Roy2020-arxiv,Janssen2020PRB,Shah2011.00249}.
These QBCPs can be established by distinct kinds of models consisting of
the kagom\'{e} lattice~\cite{Huse2003PRB,Fradkin2009PRL,Janssen2020PRB},
checkerboard lattice~\cite{Fradkin2008PRB,Vafek2014PRB},
collinear spin density wave state~\cite{Chern2012PRL}, and  Lieb lattice~\cite{Tsai2015NJP}.
Besides, recent studies using the large-scale density-matrix renormalization group have demonstrated a series of essential properties of
the QBCP systems~\cite{Schollwock2005RMP,Stoudenmire2012ARCMP,Hubig2015PRB,Sarma2016PRB,
He2017PRX,Chatterjee2022PRB,Nishimoto2010PRL,Zeng2018npj,Zhu2016PRL}.

In marked contrast to their 2D Dirac/Weyl counterparts, in which the density of state (DOS)
vanishes at Dirac points,  the 2D QBCP materials possess a finite DOS exactly at their reduced Fermi surfaces
(i.e., QBCP)~\cite{Fradkin2009PRL,Vafek2014PRB}.
This unique feature together with the gapless quasiparticles (QPs) from discrete QBCPs
plays an essential role in activating a plethora of critical behavior
in the low-energy regime~\cite{Fradkin2009PRL,
Vafek2010PRB,Yang2010PRB,Vafek2014PRB,Venderbos2016PRB,Wu2016PRL,
Zhu2016PRL,Wang2017PRB}. It is of remarkable significance to highlight that the
2D QBCP systems are unstable under the electron-electron interactions, giving rise to
the possibility of weak-coupling interaction-driven phase transitions~\cite{Fradkin2009PRL,
Venderbos2016PRB,Wu2016PRL,Vafek2014PRB,
Vafek2010PRB,Yang2010PRB,Wang2020-arxiv}.
As delivered recently~\cite{Fradkin2009PRL,Vafek2014PRB,Vafek2010PRB,Yang2010PRB},
one can expect the development of the quantum anomalous Hall (QAH) with breaking time-reversal symmetry
and quantum spin Hall (QSH) protected by time-reversal symmetry states in
the presence of electron-electron repulsions in the checkerboard
lattice~\cite{Fradkin2009PRL,Vafek2014PRB} or two-valley bilayer graphene
with QBCPs~\cite{Vafek2010PRB,Yang2010PRB}.

However, the spinful effects can play a critical role in modifying the low-energy behavior as well.
The authors in Refs.~\cite{Vafek2010PRB,Cvetkovic2012PRB} carefully investigated the spinful effects on the honeycomb lattice
and showed several interesting results. In addition, an investigation of such effects on the instability was
examined for a checkerboard lattice~\cite{Fradkin2009PRL,Vafek2014PRB}, which principally
considered the contributions from spin up and down are equivalent and hence employed the
two-component spinor to describe the low-energy excitations. Motivated by these works and given
the important role of spinful effects, we will explicitly take into account the spinful ingredients
by adopting a four-component spinor to establish the low-energy theory as presented in Sec.~\ref{Sec_model}.
This expansion can involve more kinds of electron-electron interactions and enable us to capture all
the potential instabilities induced by the spinful degrees of freedom. In consequence,
this implies that it is inadequate to capture all the potential
instabilities generated by the electron-electron interactions without
considering the spinful effects. Therefore, it is possible that other phase transitions from 2D QBCP
semimetals to certain fascinating states, aside from QAH and QSH states, may occur once the spinful contributions are taken
into account. Unambiguously clarifying this issue would be particularly helpful to improve our
understandings on the low-energy properties of 2D QBCP and analogous materials.

To this goal, we explicitly consider all the marginal
spinful electron-electron interactions in this work. This involves
sixteen different types, which are distinguished by the coupling vertexes (matrices) shown in
Sec.~\ref{Sec_model}, compared to only four spinless sorts of interactions in earlier studies~\cite{Vafek2014PRB,Wang2017PRB}.
In order to unbiasedly treat all these kinds of interactions and their intimate interplay,
it is suitable to adopt the momentum-shell renormalization-group (RG)
approach~\cite{Shankar1994RMP,Wilson1975RMP,Polchinski9210046}, which
is a powerful tool to unravel the energy-dependent hierarchical
properties in the presence of various types of physical ingredients.
Performing the RG analysis yields a set of coupled energy-dependent evolutions
of all fermion-fermion interaction parameters, from which the several interesting results
are obtained in the low-energy regime.

At first, we realize the electron-electron interactions are closely coupled to exhibit
various energy-dependent tendencies, which are broken down into three categories including the Limit case,
Special case, and General case as designated in Sec.~\ref{Sec_Fixed_Points}.
With variations of the initial values of interaction parameters and sign of structure parameter,
the 2D QBCP systems flow towards several distinct sorts of fixed points in the low-energy regime.
In the Limit and Special cases, the system can be driven to the fixed
points $\mathrm{FP}_1^{+}$/$\mathrm{FP}_2^{-}$ and $\mathrm{FP}_1^{\pm}$/
$\mathrm{FP}_2^{\pm}$/$\mathrm{FP}_3^{\pm}$, respectively.
The General case, in addition to $\mathrm{FP}_1^{\pm}$ and $\mathrm{FP}_3^{\pm}$, also harbors
the $\mathrm{FP}_{41,42,43}^{\pm}$ (all these fixed points will be designated
and explained in Sec.~\ref{Sec_Fixed_Points}).

Additionally, accessing the fixed points is always accompanied by certain instabilities that result in breaking some symmetries~\cite{Cvetkovic2012PRB,Vafek2014PRB,Wang2017PRB, Maiti2010PRB, Altland2006Book,Vojta2003RPP,Halboth2000RPL,Halboth2000RPB, Eberlein2014PRB,Chubukov2012ARCMP,Nandkishore2012NP,Chubukov2016PRX,Roy2018RRX,Wang2020NPB}.
This motives us to examine and carefully select the favorable phase
transitions from the candidate states shown in Table~\ref{table:phase} nearby these distinct kinds of fixed points.
Detailed analysis reveals that the spinful fermion-fermion
interactions can induce a number of leading and subleading instabilities as collected
in Table~\ref{table-phase-summary}. Notably, the QAH, QSH, and Nem.site(bond) states are
dominant nearby $\mathrm{FP}_{1}^{\pm}$, $\mathrm{FP}_{2}^{\pm}$, and $\mathrm{FP}_{3}^{\pm}$,
respectively. Instead, around $\mathrm{FP}_{41,42,43}^{\pm}$, QSH becomes anisotropic,
with one component being the leading instability and the others
being subleading. Besides, Nem.site(bond), Chiral SC-I,
and NSN.site(bond) are subleading candidates for these fixed points.
%%%\textbf{It is worth noting that several possible subleading phases may also compete
%%% with the leading ones and become dominant under different conditions.}
It is worth highlighting that the spinful fermion-fermion interactions, compared to the spinless
case~\cite{Vafek2014PRB,Wang2017PRB}, generate more fixed points and induce more
favorable phase transitions, and henceforth play an essential role in reshaping the
low-energy properties of 2D QBCP materials.

The rest of this paper is organized as follows. In Sec.~\ref{Sec_model},
we introduce the microscopic model and construct the effective theory and then
carry out the RG analysis in Sec.~\ref{Sec_RGEqs} to derive the coupled RG
equations of all spinful interaction parameters. In Sec.~\ref{Sec_Fixed_Points},
we present the tendencies of interaction parameters and
all potential sorts of fixed points in the interaction-parameter
space that dictate the low-energy behavior of 2D QBCP materials.
Sec.~\ref{Sec_instab_PT} is followed to address the leading and subleading
instabilities around all these fixed points that are induced by the spinful fermion-fermion interactions.
At last, we exhibit a brief summary of the basic results in Sec.~\ref{Sec_summary}.

\section{Microscopic model and effective action}\label{Sec_model}

The microscopic noninteracting model for a 2D QBCP semimetal with spin one-half electrons on a checkerboard lattice in the
low-energy regime can be expressed by the following Hamiltonian~\cite{Fradkin2008PRB,Fradkin2009PRL,Vafek2014PRB},
\begin{eqnarray}
\mathrm{H}_0 = \sum_{\mathbf{k}<|\Lambda|}\Psi_{\mathbf{k}}^{\dagger}\mathcal{H}_{0}\Psi_{\mathbf{k}},\label{Eq_H_0}
\end{eqnarray}
where $\Lambda$ serves as the momentum cutoff that is associated with the lattice constant and
the Hamiltonian density takes the form of
\begin{eqnarray}
\mathcal{H}_{0}(\mathbf{k}) = t_{I}\mathbf{k}^{2}\Sigma_{00}+2t_{x}k_{x}k_{y}\Sigma_{10}
+t_{z}(k_{x}^{2}-k_{y}^{2})\Sigma_{30},\label{Eq_H_density}
\end{eqnarray}
with $t_I$, $t_x$ and $t_z$ being the microscopic structure parameters.
Hereby, $\Psi_{\mathbf{k}}$ characterizes the low-energy quasi-particles excitations coming from two energy bands,
which is a four-component spinor and designated as $\Psi_{\mathbf{k}}^{T} \equiv (c_{A\uparrow}, c_{A\downarrow}, c_{B\uparrow}, c_{B\downarrow})$ with $A$ and $B$ denoting two sublattices in the checkerboard lattice~\cite{Fradkin2008PRB,Fradkin2009PRL}.
In addition, the $4 \times 4$ matrix is introduced by $\Sigma_{\mu\nu} \equiv \tau_\mu \otimes \sigma_\nu $, where $\tau_\mu$ and $\sigma_\nu$ are Pauli matrices and identity matrix, which act on the lattice space and spin space, respectively.

After diagonalizing the free Hamiltonian density~(\ref{Eq_H_density}), we are left with the parabolical
energy eigenvalues~\cite{Fradkin2008PRB,Fradkin2009PRL,Vafek2014PRB}
\begin{eqnarray}
E(\mathbf{k})\!=\!\frac{\mathbf{k}^2}{\sqrt{2}m}\!\left[\!\lambda\pm\!
\sqrt{\cos^2{\eta}\cos^2{\theta_k}+\sin^2{\eta}\sin^2{\theta_k}}\right],
\end{eqnarray}
where the related coefficients are defined as
\begin{eqnarray}
m &\equiv& \frac{1}{\sqrt{2(t_x^2+t_z^2)}},
\lambda \equiv \frac{t_I}{\sqrt{t_x^2+t_z^2}},
\cos{\eta} \equiv \frac{t_z}{\sqrt{t_x^2+t_z^2}},\nonumber\\
\sin{\eta} &\equiv& \frac{t_x}{\sqrt{t_x^2+t_z^2}},
\cos{\theta_k} \equiv \frac{k_x}{\sqrt{k_x^2+k_y^2}},
\sin{\theta_k} \equiv \frac{k_y}{\sqrt{k_x^2+k_y^2}}.\nonumber
\end{eqnarray}
with $\theta_k$ specifying the direction of momentum. There exist one upward and one downward dispersing band at $|t_I|<\mathrm{min}(|t_x|,|t_z|)$, which touch parabolically at $\mathbf{k}=0$ and are invariant under $C_{4v}$ point
group and time-reversal symmetry~\cite{Fradkin2008PRB,Fradkin2009PRL,Vafek2014PRB}.

Without loss of generality, we will consider in the remainder a particle-hole and rotational symmetric
QBCP semimetal, which requires $t_I=0$ and $t_x=t_z\equiv t$. To proceed, the interacting part which includes all the
marginal short-range electron-electron interactions can be introduced as
follows~\cite{Fradkin2009PRL,Vafek2014PRB,Vafek2010PRB,Yang2010PRB,Wang2020-arxiv}
\begin{widetext}
\begin{eqnarray}
S_{\mathrm{int}} &=& \sum_{ \mu, \nu = 0 }^{3}
\frac{2\pi}{m}\lambda_{\mu\nu}\int_{-\infty}^{\infty}\frac{d\omega_{1}d\omega_{2}d\omega_{3}}{(2\pi)^3}
\int^{\Lambda}\frac{d^2\mathbf{k}_{1}d^2\mathbf{k}_{2}d^2\mathbf{k}_{3}}{(2\pi)^6}\Psi^{\dagger}(\omega_1,\mathbf{k}_1)
\Sigma_{\mu\nu}\Psi(\omega_2,\mathbf{k}_2)\nonumber\\
&&\times\Psi^{\dagger}(\omega_3,\mathbf{k}_3)\Sigma_{\mu\nu}
\Psi(\omega_1+\omega_2-\omega_3,\mathbf{k}_1+\mathbf{k}_2-\mathbf{k}_3),\label{Eq_S_int}
\end{eqnarray}
\end{widetext}
where the $\lambda_{\mu\nu}$ with $\mu, \nu = 0,1,2,3 $, which are positive and represent the repulsive interactions between electrons,
are adopted to measure the coupling strengths that are related to
the interactions distinguished by the matrices $\Sigma_{\mu\nu}$.
Given that the fermionic couplings are marginal at the tree level due to the unique features of the 2D QBCP semimetals
and become relevant at the one-loop level, it is worth highlighting that the fermion-fermion interactions are much more important than the
other interactions and play an essential role in determining the low-energy properties of 2D QBCP materials.
%It is worth highlighting that
%although one can utilize the Fierz identity to reduce the number of interactions~\cite{Herbut-Roy2009PRB,Sarma2016PRB,Boettcher2016PRB},
%we within this work will unbiasedly consider all 16 types of interactions and the explanations are briefly
%discussed in Appendix~\ref{Appendix_FI_discussion}.
Accordingly, we obtain our effective action by taking into account both the free part~(\ref{Eq_H_0}) and
the interacting part~(\ref{Eq_S_int}) as follows
\begin{eqnarray}
S_{\mathrm{eff}}
&=& \int_{-\infty}^{\infty}\frac{d\omega}{2\pi}\int^{\Lambda}\frac{d^2\mathbf{k}}{(2\pi)^2}\Psi^{\dagger}(\omega,\mathbf{k})
\left\{ -i\omega \Sigma_{00}+ t[2k_{x}k_{y}\Sigma_{10}\right.\nonumber\\
&&\left.+(k_{x}^{2}-k_{y}^{2})\Sigma_{30}] \right\}\Psi(\omega,\mathbf{k})+S_{\mathrm{int}}.\label{Eq_S_eff}
\end{eqnarray}
The free electron propagator can be extracted from the noninteracting terms and written as
\begin{eqnarray}
G_{0}(i\omega,\mathbf{k})=\frac{1} { -i\omega+ t[2k_{x}k_{y}\Sigma_{10}+(k_{x}^{2}-k_{y}^{2})\Sigma_{30}]}.
\end{eqnarray}
With these in hand, we are in a suitable position to make the RG analysis.

\section{Renormalization group analysis}\label{Sec_RGEqs}

To proceed, we within this section perform the RG analysis to construct the
coupled energy-dependent flows of all spinful electron-electron couplings, which
contain the low-energy behaviors of 2D QBCP materials. Following the spirit of RG framework~\cite{Shankar1994RMP,Wilson1975RMP,Polchinski9210046},
we separate the fermionic fields into the fast and slow modes within the momentum shell $b\Lambda<k<\Lambda$
and $0<k<b\Lambda$, respectively. Hereby, we utilize $\Lambda$ to characterize
the energy scale and a variable parameter $b$ with $b=e^{-l}<1$
to serve as a running energy scale~\cite{Vafek2014PRB,Altland2006Book,Cvetkovic2012PRB,Wang2011PRB,Wang2013PRB,
Wang2017PRB,Huh2008PRB,Kim2008PRB,Maiti2010PRB,She2010PRB,She2015PRB,Cvetkovic2012PRB,
Vafek2014PRB,Roy2016PRB}. On the basis of these, the noninteracting parts of the effective
field action~(\ref{Eq_S_eff}) consequently can be selected as a free fixed point.
Keeping such fixed point invariant under RG transformations gives rise to
the RG rescaling transformations of fields and momenta as
follows~\cite{Vafek2014PRB,Wang2011PRB,Wang2013PRB,Wang2017PRB,Huh2008PRB,Wang2020-arxiv},
\begin{eqnarray}
k_x&\longrightarrow&k'_xe^{-l},\label{Eq_rescale_k_x}\\
k_y&\longrightarrow&k'_ye^{-l},\\
\omega&\longrightarrow&\omega'e^{-2l},\\
\psi(i\omega,\mathbf{k})&\longrightarrow&
\psi'(i\omega',\mathbf{k}')e^{\frac{1}{2}\int dl(6-\eta_f)}.\label{Eq_rescale_psi}
\end{eqnarray}
Here, the parameter $\eta_f$ is so-called anomalous dimension of fermionic spinor which
is equivalent to zero owing to the marginal fermion-fermion interactions for 2D QBCP systems~\cite{Vafek2014PRB}.

In order to include the higher-level contributions,
we endeavor to carry out the analytical calculations of one-loop electron-electron corrections to
interaction parameters as depicted in Fig.~\ref{Fig_Feynman Diagram}
of Appendix~\ref{Appendix-one-loop-corrections}.
For convenience, the cutoff $\Lambda_0$ which is linked to
the lattice constant can be adopted to measure the momenta and
energy with rescaling $k\rightarrow k/\Lambda_0$
and $\omega\rightarrow\omega/\Lambda_0$~\cite{
Vafek2014PRB,Wang2011PRB,Huh2008PRB,She2010PRB}.
Subsequently, we obtain the one-loop corrections in
Appendix~\ref{Appendix-one-loop-corrections} after
paralleling the tedious but straightforward
evaluations~\cite{Vafek2014PRB,Wang2017PRB,Wang2018JPCM,Wang2019JPCM,Wang2020-arxiv}.
At current stage, we are able to derive the
coupled RG flow equations by combining
the RG scalings~(\ref{Eq_rescale_k_x})-(\ref{Eq_rescale_psi}) and
the one-loop corrections in Appendix~\ref{Appendix-one-loop-corrections}~\cite{Wang2011PRB,Wang2013PRB,
Wang2017PRB,Huh2008PRB,Kim2008PRB,Maiti2010PRB,She2010PRB,She2015PRB,Cvetkovic2012PRB,
Vafek2014PRB,Roy2016PRB}, which take the form of
\begin{widetext}
\begin{eqnarray}
\frac{d \lambda_{00}}{dl}
&=&-\frac{|t|}{t}\big(\lambda_{00} \lambda_{10} +\lambda_{01} \lambda_{11} +\lambda_{02} \lambda_{12} +\lambda_{03} \lambda_{13} +\lambda_{00} \lambda_{30} +\lambda_{01} \lambda_{31} +\lambda_{02} \lambda_{32} +\lambda_{03} \lambda_{33} \big),\label{Eq_RG-1}\\
\frac{d \lambda_{01}}{dl}
&=&-\frac{|t|}{t}  (\lambda_{00} \lambda_{11} -2\lambda_{02} \lambda_{03} +\lambda_{01} \lambda_{10} -2\lambda_{12} \lambda_{13} +\lambda_{00} \lambda_{31} +\lambda_{01} \lambda_{30} +\lambda_{12} \lambda_{23} +\lambda_{13} \lambda_{22} -2\lambda_{22} \lambda_{23}\nonumber\\&& +\lambda_{22} \lambda_{33} +\lambda_{23} \lambda_{32} -2\lambda_{32} \lambda_{33}),\\
\frac{d \lambda_{02}}{dl}
&=&-\frac{|t|}{t} (\lambda_{00} \lambda_{12} -2\lambda_{01} \lambda_{03} +\lambda_{02} \lambda_{10} -2\lambda_{11} \lambda_{13} +\lambda_{00} \lambda_{32} +\lambda_{02} \lambda_{30} +\lambda_{11} \lambda_{23} +\lambda_{13} \lambda_{21} -2\lambda_{21} \lambda_{23} \nonumber\\
&& +\lambda_{21} \lambda_{33}+\lambda_{23} \lambda_{31} -2\lambda_{31} \lambda_{33} ),\\
\frac{d \lambda_{03}}{dl}
&=&-\frac{|t|}{t}(\lambda_{00} \lambda_{13} -2\lambda_{01} \lambda_{02} +\lambda_{03} \lambda_{10} -2\lambda_{11} \lambda_{12} +\lambda_{00} \lambda_{33} +\lambda_{03} \lambda_{30} +\lambda_{11} \lambda_{22} +\lambda_{12} \lambda_{21}
\nonumber\\
&& -2\lambda_{21} \lambda_{22} +\lambda_{21} \lambda_{32}+\lambda_{22} \lambda_{31} -2\lambda_{31} \lambda_{32} ),\\
\frac{d \lambda_{10}}{dl}
&=&-\frac{|t|}{2t}({\lambda_{00} }^2 -2\lambda_{00} \lambda_{10} +2\lambda_{00} \lambda_{20} +{\lambda_{01} }^2 -2\lambda_{01} \lambda_{10} +2\lambda_{01} \lambda_{21} +{\lambda_{02} }^2 -2\lambda_{02} \lambda_{10} +2\lambda_{02} \lambda_{22} +{\lambda_{03} }^2\nonumber\\&& -2\lambda_{03} \lambda_{10} +2\lambda_{03} \lambda_{23} +7{\lambda_{10} }^2 -2\lambda_{10} \lambda_{11} -2\lambda_{10} \lambda_{12} -2\lambda_{10} \lambda_{13} +2\lambda_{10} \lambda_{20} +2\lambda_{10} \lambda_{21} +2\lambda_{10} \lambda_{22}\nonumber\\&& +2\lambda_{10} \lambda_{23} +2\lambda_{10} \lambda_{30} +2\lambda_{10} \lambda_{31} +2\lambda_{10} \lambda_{32} +2\lambda_{10} \lambda_{33} +{\lambda_{11} }^2 +{\lambda_{12} }^2 +{\lambda_{13} }^2 +{\lambda_{20} }^2 -4\lambda_{20} \lambda_{30}\nonumber\\&& +{\lambda_{21} }^2 -4\lambda_{21} \lambda_{31} +{\lambda_{22} }^2 -4\lambda_{22} \lambda_{32} +{\lambda_{23} }^2 -4\lambda_{23} \lambda_{33} +{\lambda_{30} }^2 +{\lambda_{31} }^2 +{\lambda_{32} }^2 +{\lambda_{33} }^2 ),\\
\frac{d \lambda_{11}}{dl}
&=&-\frac{|t|}{t}(\lambda_{00} \lambda_{01} -\lambda_{00} \lambda_{11} -\lambda_{01} \lambda_{11} +\lambda_{02} \lambda_{11} +\lambda_{03} \lambda_{11} -2\lambda_{02} \lambda_{13} -2\lambda_{03} \lambda_{12} +\lambda_{00} \lambda_{21} +\lambda_{01} \lambda_{20} +\lambda_{11} \lambda_{12}\nonumber\\&& +\lambda_{11} \lambda_{13} +\lambda_{11} \lambda_{20} +\lambda_{11} \lambda_{21} -\lambda_{11} \lambda_{22} -\lambda_{11} \lambda_{23} +\lambda_{11} \lambda_{30} +\lambda_{20} \lambda_{21} +\lambda_{11} \lambda_{31} -\lambda_{11} \lambda_{32} -\lambda_{11} \lambda_{33}\nonumber\\&& +\lambda_{12} \lambda_{33} +\lambda_{13} \lambda_{32} -2\lambda_{20} \lambda_{31} -2\lambda_{21} \lambda_{30} +\lambda_{30} \lambda_{31} +3{\lambda_{11} }^2 ) ,\\
\frac{d \lambda_{12}}{dl}
&=&-\frac{|t|}{t}(\lambda_{00} \lambda_{02} -\lambda_{00} \lambda_{12} +\lambda_{01} \lambda_{12} -2\lambda_{01} \lambda_{13} -\lambda_{02} \lambda_{12} -2\lambda_{03} \lambda_{11} +\lambda_{03} \lambda_{12} +\lambda_{00} \lambda_{22} +\lambda_{02} \lambda_{20} +\lambda_{11} \lambda_{12}\nonumber\\&& +\lambda_{12} \lambda_{13} +\lambda_{12} \lambda_{20} -\lambda_{12} \lambda_{21} +\lambda_{12} \lambda_{22} -\lambda_{12} \lambda_{23} +\lambda_{12} \lambda_{30} +\lambda_{20} \lambda_{22} -\lambda_{12} \lambda_{31} +\lambda_{11} \lambda_{33} +\lambda_{12} \lambda_{32}\nonumber\\&& +\lambda_{13} \lambda_{31} -\lambda_{12} \lambda_{33} -2\lambda_{20} \lambda_{32} -2\lambda_{22} \lambda_{30} +\lambda_{30} \lambda_{32} +3{\lambda^2_{12} } ) ,\\
\frac{d \lambda_{13}}{dl}
&=&-\frac{|t|}{t}(\lambda_{00} \lambda_{03} -\lambda_{00} \lambda_{13} -2\lambda_{01} \lambda_{12} -2\lambda_{02} \lambda_{11} +\lambda_{01} \lambda_{13} +\lambda_{02} \lambda_{13} -\lambda_{03} \lambda_{13} +\lambda_{00} \lambda_{23} +\lambda_{03} \lambda_{20} +\lambda_{11} \lambda_{13}\nonumber\\&& +\lambda_{12} \lambda_{13} +\lambda_{13} \lambda_{20} -\lambda_{13} \lambda_{21} -\lambda_{13} \lambda_{22} +\lambda_{13} \lambda_{23} +\lambda_{11} \lambda_{32} +\lambda_{12} \lambda_{31} +\lambda_{13} \lambda_{30} +\lambda_{20} \lambda_{23} -\lambda_{13} \lambda_{31}\nonumber\\&& -\lambda_{13} \lambda_{32} +\lambda_{13} \lambda_{33} -2\lambda_{20} \lambda_{33} -2\lambda_{23} \lambda_{30} +\lambda_{30} \lambda_{33} +3{\lambda^2_{13} } ) ,\\
\frac{d \lambda_{20}}{dl}
&=&-\frac{|t|}{t}(\lambda_{00} \lambda_{10} +\lambda_{01} \lambda_{11} +\lambda_{02} \lambda_{12} +\lambda_{03} \lambda_{13} -2\lambda_{00} \lambda_{20} -2\lambda_{01} \lambda_{20} -2\lambda_{02} \lambda_{20} -2\lambda_{03} \lambda_{20} +\lambda_{00} \lambda_{30}\nonumber\\&& +2\lambda_{10} \lambda_{20} +2\lambda_{11} \lambda_{20} +\lambda_{01} \lambda_{31} +2\lambda_{12} \lambda_{20} +2\lambda_{13} \lambda_{20} +\lambda_{02} \lambda_{32} +\lambda_{03} \lambda_{33} -2\lambda_{10} \lambda_{30} -2\lambda_{20} \lambda_{21}-2\lambda_{11} \lambda_{31}\nonumber\\
&&  -2\lambda_{20} \lambda_{22} -2\lambda_{20} \lambda_{23} -2\lambda_{12} \lambda_{32} -2\lambda_{13} \lambda_{33} +2\lambda_{20} \lambda_{30} +2\lambda_{20} \lambda_{31} +2\lambda_{20} \lambda_{32} +2\lambda_{20} \lambda_{33} +6{\lambda^2_{20} } ),\\
\frac{d \lambda_{21}}{dl}
&=&-\frac{|t|}{t}(\lambda_{00} \lambda_{11} +\lambda_{01} \lambda_{10} -2\lambda_{00} \lambda_{21} -2\lambda_{01} \lambda_{21} +2\lambda_{02} \lambda_{21} +2\lambda_{03} \lambda_{21} -2\lambda_{02} \lambda_{23} -2\lambda_{03} \lambda_{22} +\lambda_{00} \lambda_{31}\nonumber\\&& +\lambda_{01} \lambda_{30} +2\lambda_{10} \lambda_{21} +2\lambda_{11} \lambda_{21} -2\lambda_{12} \lambda_{21} -2\lambda_{13} \lambda_{21} +\lambda_{12} \lambda_{23} +\lambda_{13} \lambda_{22} -2\lambda_{10} \lambda_{31} -2\lambda_{11} \lambda_{30}-2\lambda_{20} \lambda_{21}\nonumber\\
&&  +2\lambda_{21} \lambda_{22} +2\lambda_{21} \lambda_{23} +2\lambda_{21} \lambda_{30} +2\lambda_{21} \lambda_{31} -2\lambda_{21} \lambda_{32} -2\lambda_{21} \lambda_{33} +\lambda_{22} \lambda_{33} +\lambda_{23} \lambda_{32}+6{\lambda^2_{21} } ) ,\\
\frac{d \lambda_{22}}{dl}
&=&-\frac{|t|}{t}(\lambda_{00} \lambda_{12} +\lambda_{02} \lambda_{10} -2\lambda_{00} \lambda_{22} +2\lambda_{01} \lambda_{22} -2\lambda_{01} \lambda_{23} -2\lambda_{02} \lambda_{22} -2\lambda_{03} \lambda_{21} +2\lambda_{03} \lambda_{22} +\lambda_{00} \lambda_{32}\nonumber\\&& +\lambda_{02} \lambda_{30} +2\lambda_{10} \lambda_{22} -2\lambda_{11} \lambda_{22} +\lambda_{11} \lambda_{23} +2\lambda_{12} \lambda_{22} +\lambda_{13} \lambda_{21} -2\lambda_{13} \lambda_{22} -2\lambda_{10} \lambda_{32} -2\lambda_{12} \lambda_{30}-2\lambda_{20} \lambda_{22}\nonumber\\
&&  +2\lambda_{21} \lambda_{22} +2\lambda_{22} \lambda_{23} +2\lambda_{22} \lambda_{30} -2\lambda_{22} \lambda_{31} +\lambda_{21} \lambda_{33} +2\lambda_{22} \lambda_{32} +\lambda_{23} \lambda_{31} -2\lambda_{22} \lambda_{33} +6{\lambda^2_{22} } ) ,\\
\frac{d \lambda_{23}}{dl}
&=&-\frac{|t|}{t}(\lambda_{00} \lambda_{13} +\lambda_{03} \lambda_{10} -2\lambda_{00} \lambda_{23} -2\lambda_{01} \lambda_{22} -2\lambda_{02} \lambda_{21} +2\lambda_{01} \lambda_{23} +2\lambda_{02} \lambda_{23} -2\lambda_{03} \lambda_{23} +\lambda_{00} \lambda_{33}\nonumber\\&& +\lambda_{03} \lambda_{30} +2\lambda_{10} \lambda_{23} +\lambda_{11} \lambda_{22} +\lambda_{12} \lambda_{21} -2\lambda_{11} \lambda_{23} -2\lambda_{12} \lambda_{23} +2\lambda_{13} \lambda_{23} -2\lambda_{10} \lambda_{33} -2\lambda_{13} \lambda_{30}-2\lambda_{20} \lambda_{23}\nonumber\\
&&  +2\lambda_{21} \lambda_{23} +2\lambda_{22} \lambda_{23} +\lambda_{21} \lambda_{32} +\lambda_{22} \lambda_{31} +2\lambda_{23} \lambda_{30} -2\lambda_{23} \lambda_{31} -2\lambda_{23} \lambda_{32} +2\lambda_{23} \lambda_{33}+6{\lambda^2_{23} } ),\\
\frac{d \lambda_{30}}{dl}
&=&-\frac{|t|}{2t}({\lambda_{00} }^2 +2\lambda_{00} \lambda_{20} -2\lambda_{00} \lambda_{30} +{\lambda_{01} }^2 +2\lambda_{01} \lambda_{21} -2\lambda_{01} \lambda_{30} +{\lambda_{02} }^2 +2\lambda_{02} \lambda_{22} -2\lambda_{02} \lambda_{30} +{\lambda_{03} }^2\nonumber\\&& +2\lambda_{03} \lambda_{23} -2\lambda_{03} \lambda_{30} +{\lambda_{10} }^2 -4\lambda_{10} \lambda_{20} +2\lambda_{10} \lambda_{30} +{\lambda_{11} }^2 -4\lambda_{11} \lambda_{21} +2\lambda_{11} \lambda_{30} +{\lambda_{12} }^2 -4\lambda_{12} \lambda_{22}\nonumber\\&& +2\lambda_{12} \lambda_{30} +{\lambda_{13} }^2 -4\lambda_{13} \lambda_{23} +2\lambda_{13} \lambda_{30} +{\lambda_{20} }^2 +2\lambda_{20} \lambda_{30} +{\lambda_{21} }^2 +2\lambda_{21} \lambda_{30} +{\lambda_{22} }^2 +2\lambda_{22} \lambda_{30}\nonumber\\&& +{\lambda_{23} }^2 +2\lambda_{23} \lambda_{30} +7{\lambda_{30} }^2 -2\lambda_{30} \lambda_{31} -2\lambda_{30} \lambda_{32} -2\lambda_{30} \lambda_{33} +{\lambda_{31} }^2 +{\lambda_{32} }^2 +{\lambda_{33} }^2 ),\\
\frac{d \lambda_{31}}{dl}
&=&-\frac{|t|}{t}(\lambda_{00} \lambda_{01} +\lambda_{00} \lambda_{21} +\lambda_{01} \lambda_{20} +\lambda_{10} \lambda_{11} -\lambda_{00} \lambda_{31} -2\lambda_{10} \lambda_{21} -2\lambda_{11} \lambda_{20} -\lambda_{01} \lambda_{31} +\lambda_{02} \lambda_{31} +\lambda_{03} \lambda_{31}\nonumber\\&& -2\lambda_{02} \lambda_{33} -2\lambda_{03} \lambda_{32} +\lambda_{10} \lambda_{31} +\lambda_{20} \lambda_{21} +\lambda_{11} \lambda_{31} -\lambda_{12} \lambda_{31} -\lambda_{13} \lambda_{31} +\lambda_{12} \lambda_{33} +\lambda_{13} \lambda_{32} +\lambda_{20} \lambda_{31}\nonumber\\&& +\lambda_{21} \lambda_{31} -\lambda_{22} \lambda_{31} -\lambda_{23} \lambda_{31} +\lambda_{31} \lambda_{32} +\lambda_{31} \lambda_{33} +3{\lambda^2_{31} } ) ,\\
\frac{d \lambda_{32}}{dl}
&=&-\frac{|t|}{t}(\lambda_{00} \lambda_{02} +\lambda_{00} \lambda_{22} +\lambda_{02} \lambda_{20} +\lambda_{10} \lambda_{12} -\lambda_{00} \lambda_{32} -2\lambda_{10} \lambda_{22} -2\lambda_{12} \lambda_{20} +\lambda_{01} \lambda_{32} -2\lambda_{01} \lambda_{33} -\lambda_{02} \lambda_{32}\nonumber\\&& -2\lambda_{03} \lambda_{31} +\lambda_{03} \lambda_{32} +\lambda_{10} \lambda_{32} +\lambda_{20} \lambda_{22} -\lambda_{11} \lambda_{32} +\lambda_{11} \lambda_{33} +\lambda_{12} \lambda_{32} +\lambda_{13} \lambda_{31} -\lambda_{13} \lambda_{32} +\lambda_{20} \lambda_{32}\nonumber\\&& -\lambda_{21} \lambda_{32} +\lambda_{22} \lambda_{32} -\lambda_{23} \lambda_{32} +\lambda_{31} \lambda_{32} +\lambda_{32} \lambda_{33} +3{\lambda^2_{32} } ) ,\\
\frac{d \lambda_{33}}{dl}
&=&-\frac{|t|}{t}(\lambda_{00} \lambda_{03} +\lambda_{00} \lambda_{23} +\lambda_{03} \lambda_{20} +\lambda_{10} \lambda_{13} -\lambda_{00} \lambda_{33} -2\lambda_{01} \lambda_{32} -2\lambda_{02} \lambda_{31} -2\lambda_{10} \lambda_{23} -2\lambda_{13} \lambda_{20} +\lambda_{01} \lambda_{33}\nonumber\\&& +\lambda_{02} \lambda_{33} -\lambda_{03} \lambda_{33} +\lambda_{10} \lambda_{33} +\lambda_{11} \lambda_{32} +\lambda_{12} \lambda_{31} +\lambda_{20} \lambda_{23} -\lambda_{11} \lambda_{33} -\lambda_{12} \lambda_{33} +\lambda_{13} \lambda_{33} +\lambda_{20} \lambda_{33}\nonumber\\&& -\lambda_{21} \lambda_{33} -\lambda_{22} \lambda_{33} +\lambda_{23} \lambda_{33} +\lambda_{31} \lambda_{33} +\lambda_{32} \lambda_{33} +3{\lambda^2_{33} } ).\label{Eq_RG-2}
\end{eqnarray}

\end{widetext}

\begin{figure}[htbp]
\hspace{-0.35cm}
\includegraphics[scale =0.31]{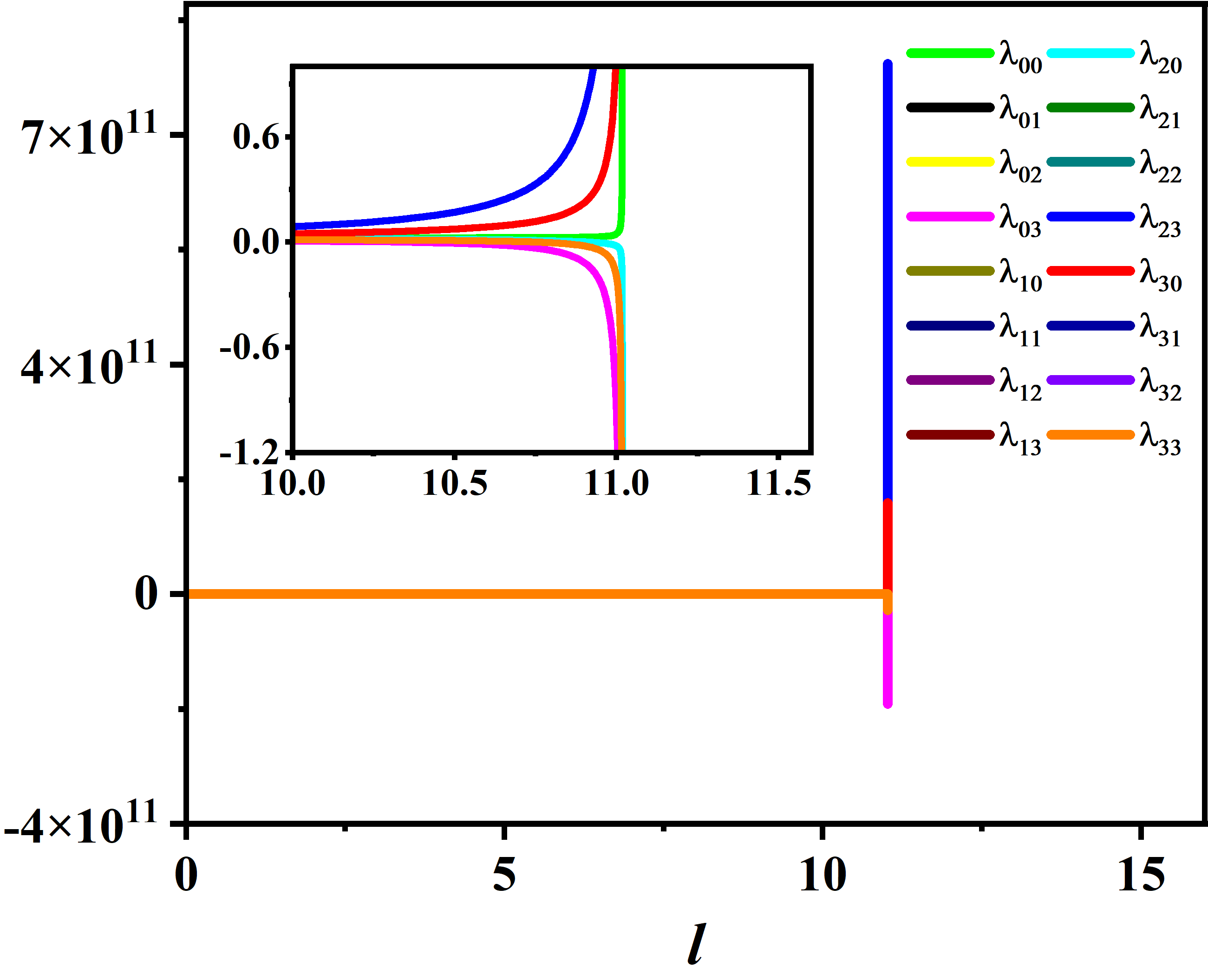}\\ \vspace{-0.1cm}
\caption{(Color online) Energy-dependent flows of all 16 interaction parameters with
a representative initial value $\lambda_{ij}(0)=10^{-2}$ %and a positive $t$
(the basic results are insusceptible to the concrete initial values). Inset: the enlarged regime
around the divergence.}\label{fig1_Limit_case_flows}
\end{figure}

\begin{figure}[htbp]
\centering
\subfigure[]{\includegraphics[width=2.6in]{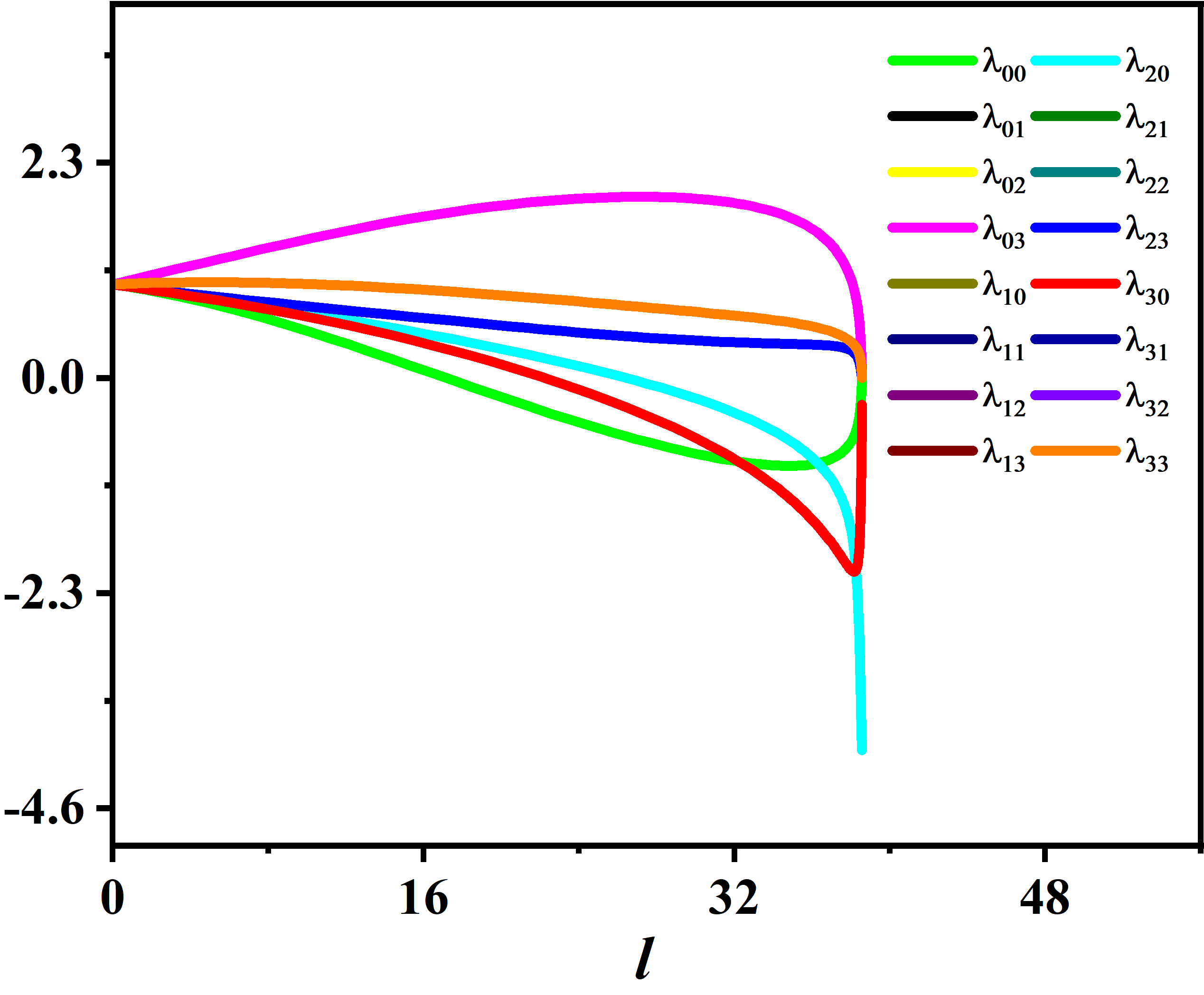}}\\ \vspace{0.15cm}
\subfigure[]{\includegraphics[width=2.62in]{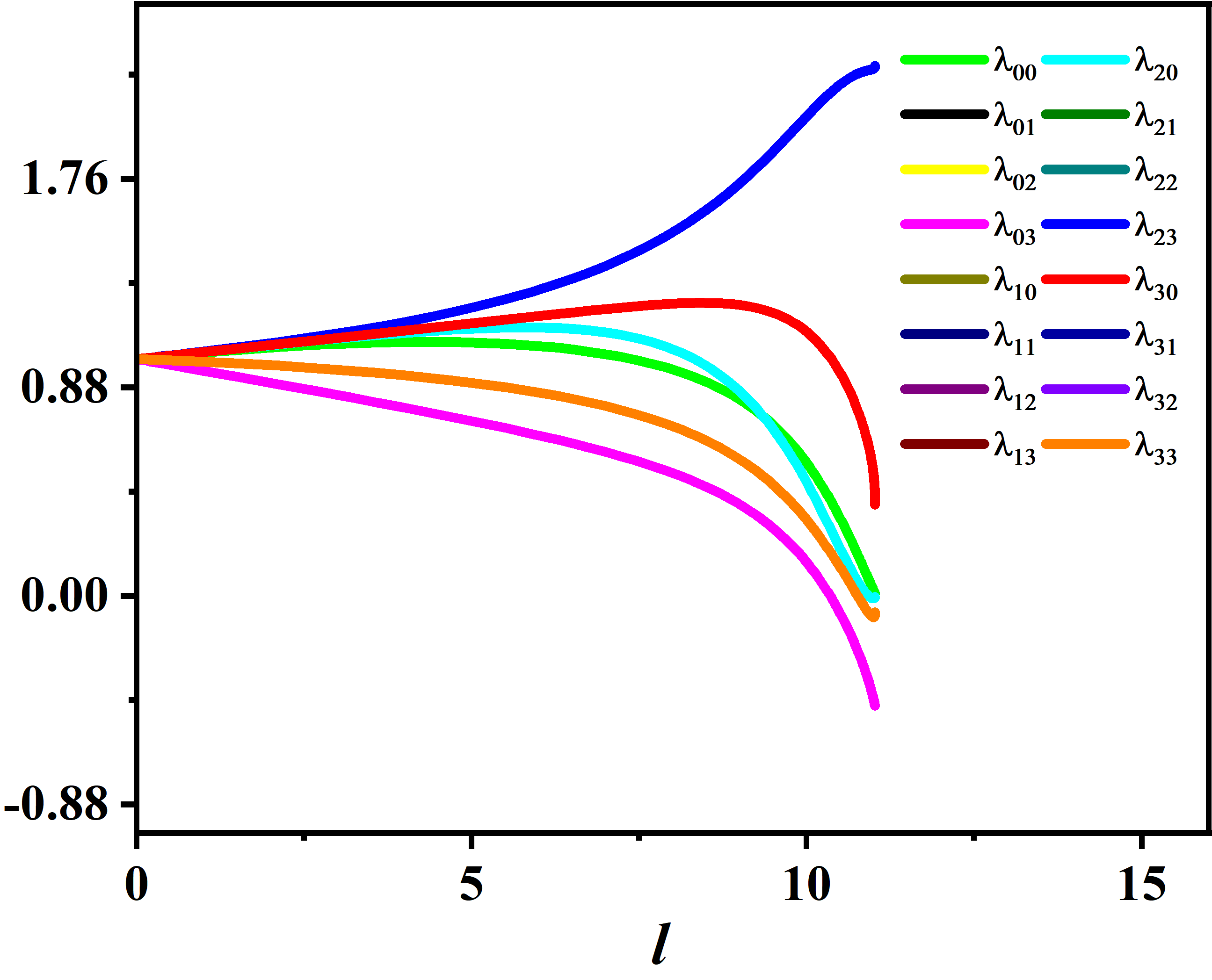}}\\ \vspace{-0.1cm}
\caption{(Color online) Energy-dependent flows of all 16 rescaled interaction parameters
and fixed points in the Limit case with a representative initial value $\lambda_{ij}(0)=10^{-2}$ for (a) $t>0$, and (b) $t<0$,
respectively (the basic results are insusceptible to the concrete initial values).}
\label{fig2_Limit_case_FPs}
\end{figure}

\begin{figure}[htbp]
\centering
%%\subfigure[\label{fig:FP_123_a}]{\includegraphics[scale = 0.27]{fig2a_FP1+.png}}
\subfigure[]{\includegraphics[width=2.6in]{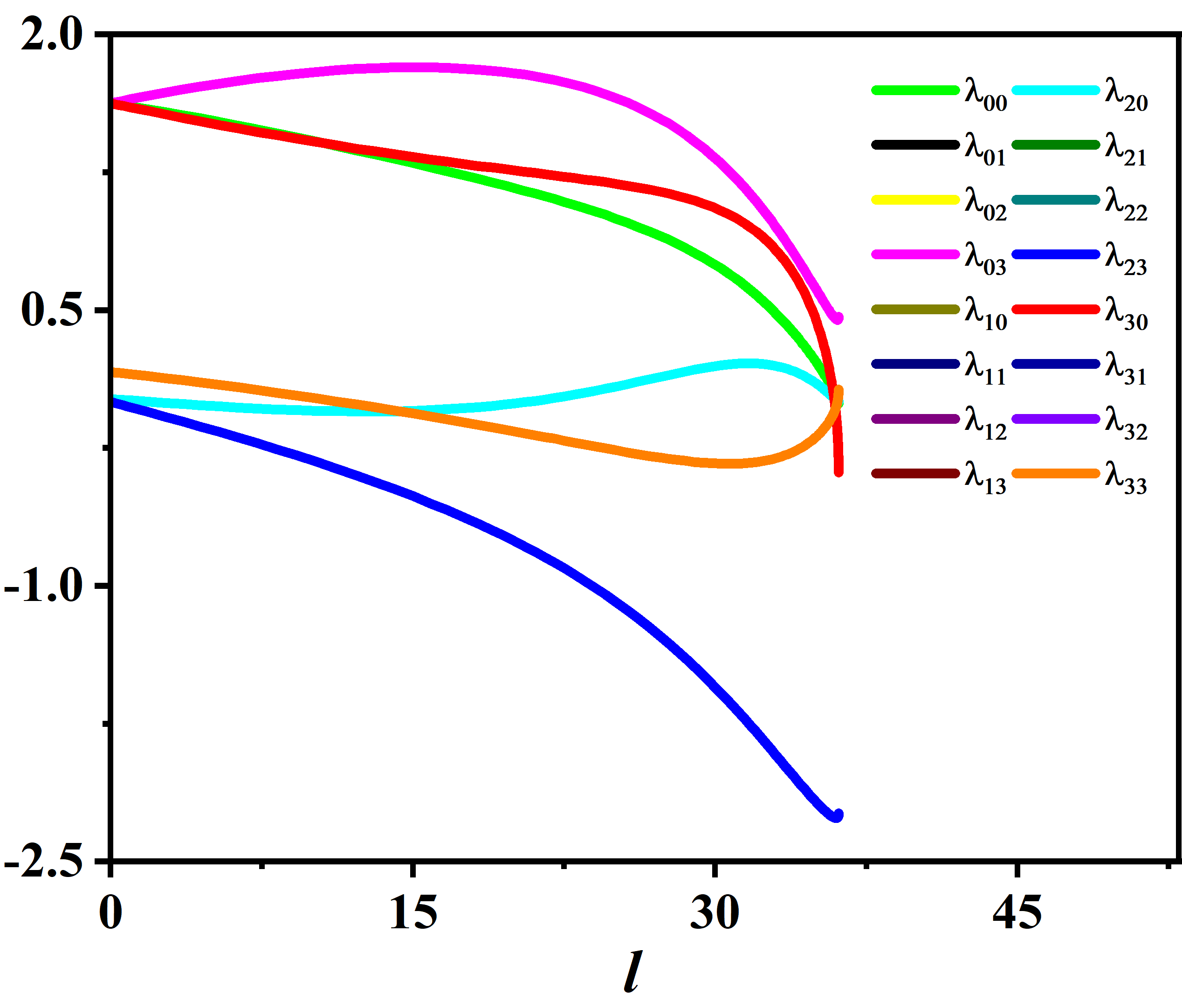}}\vspace{0.15cm}
\subfigure[]{\includegraphics[width=2.6in]{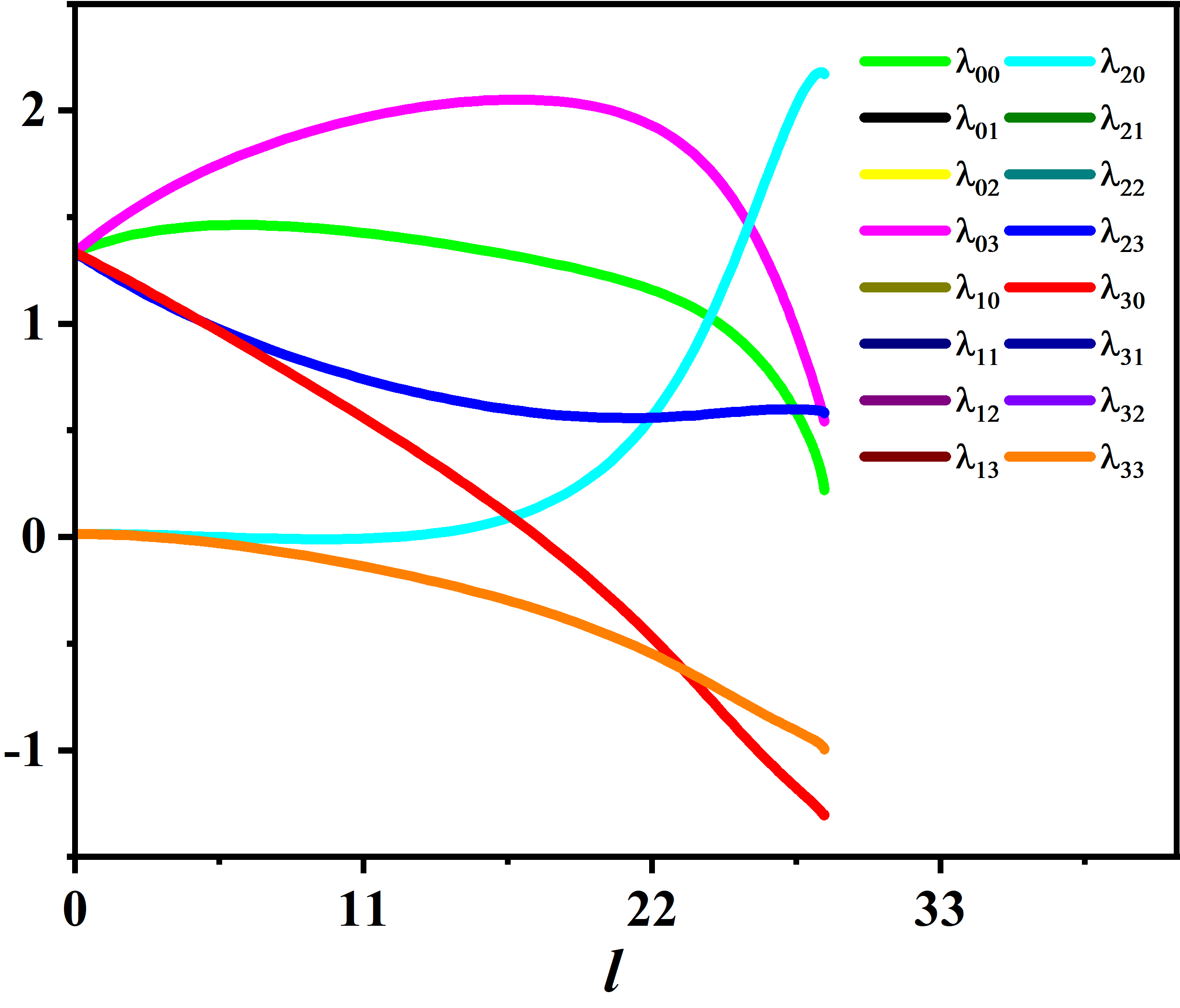}}\\
\vspace{0.05cm}
\caption{(Color online) Energy-dependent flows of electron-electron interaction parameters
in the Special case at $t>0$ and fixed points towards:
(a) $\mathrm{FP}_2^+$  and
(b) $\mathrm{FP}_3^+$ with $(\lambda_{00},\lambda_{01}, \lambda_{10},\lambda_{11},\lambda_{20},\lambda_{21})
=(10^{-2},10^{-2},10^{-2},10^{-3},10^{-4},10^{-7})$ and $(10^{-2},10^{-2},10^{-2},10^{-4},10^{-4},10^{-2})$,
respectively (the basic results for fixed points are insusceptible
to the concrete initial values).}\label{fig3_FP_123}
\end{figure}

These RG equations are closely coupled and ferociously
compete with each other, which give rise to the energy-dependent
interaction parameters and govern the physical behavior in the
low-energy regime~\cite{Shankar1994RMP,Altland2006Book}. In order to unveil the underlying physical information
of 2D QBCP system, we are going to investigate the potential fixed points of such
interaction parameters in the following Sec.~\ref{Sec_Fixed_Points}, and defer the study of accompanying instabilities and phase
transitions induced by fermionic interactions to Sec.~\ref{Sec_instab_PT}, respectively.

\section{Potential fixed points}\label{Sec_Fixed_Points}

\begin{figure}[htbp]
    \hspace{-0.8cm}
    \subfigure[]{\includegraphics[scale = 0.135]{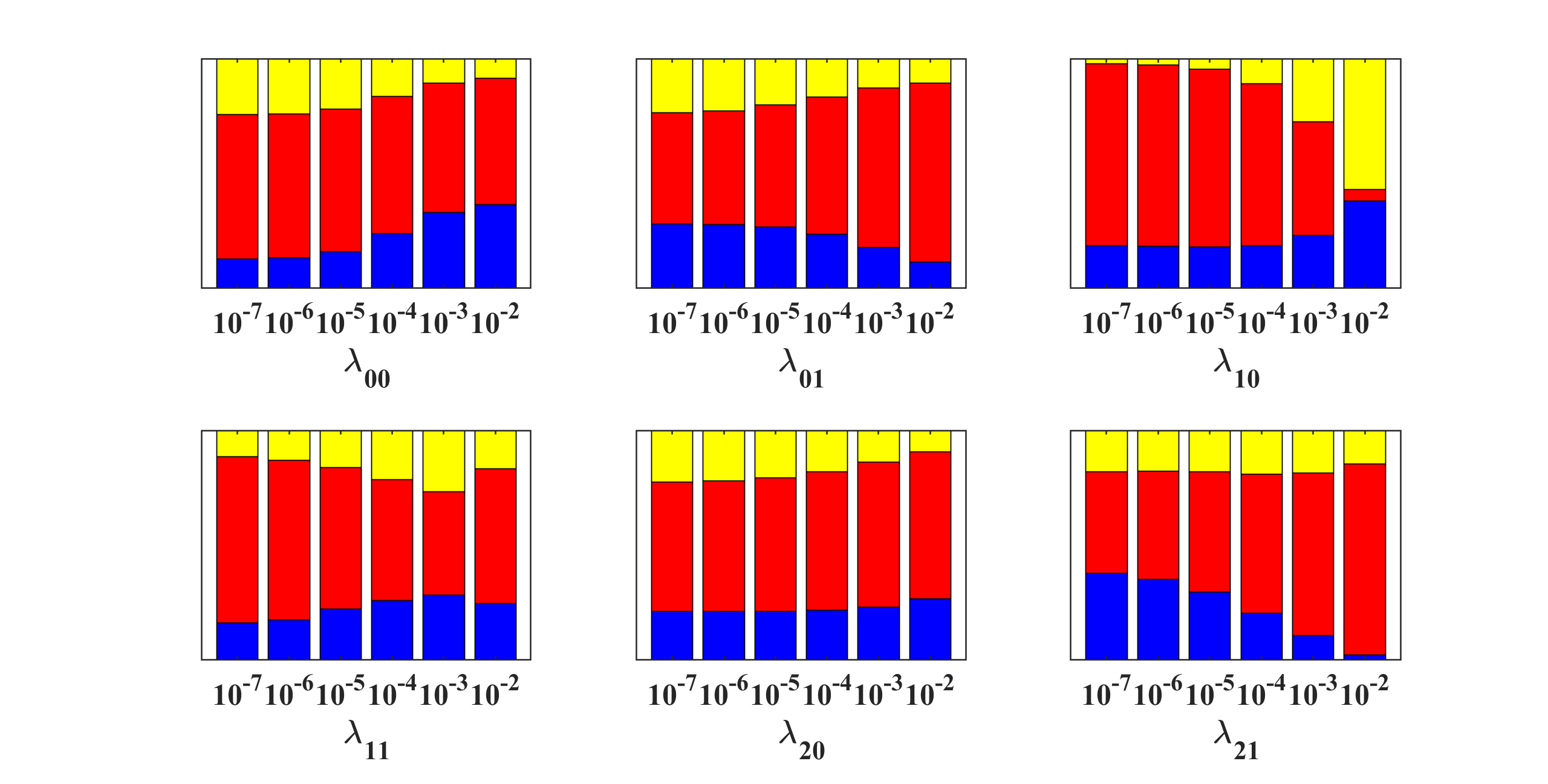}}\vspace{-0.15cm}\\
    \hspace{-0.8cm}
    \subfigure[]{\includegraphics[scale = 0.135]{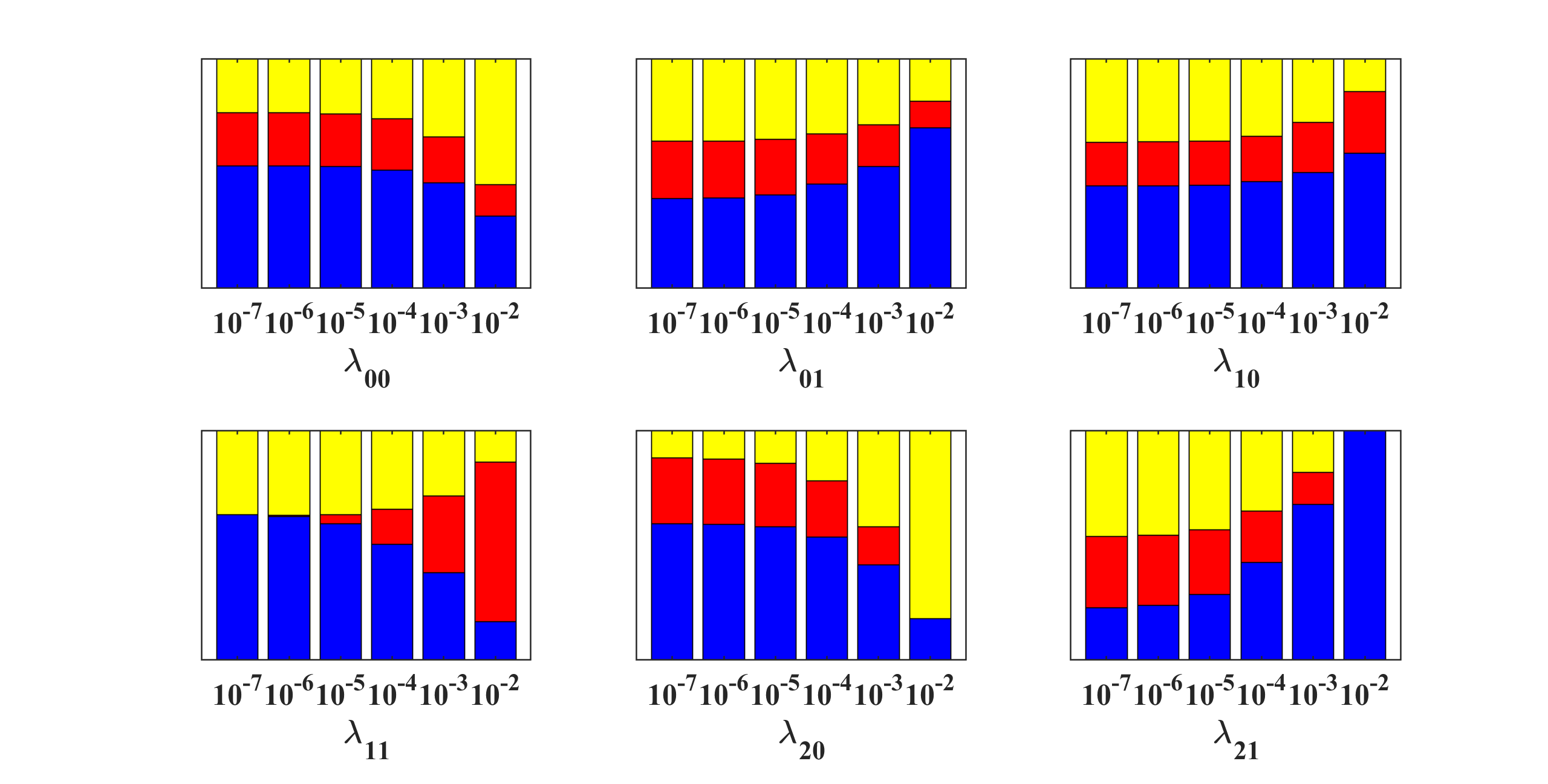}}\\
    \vspace{-0.1cm}
\caption{(Color online) Choosing $(\lambda_{00},\lambda_{01},\lambda_{10},\lambda_{11},\lambda_{20},\lambda_{21})$
from six classes for an example (the basic results are similar for other choices) and showing the competition among different fixed points
by tuning the magnitude of certain parameter (horizontal axis): (a) $t>0$ and (b) $t<0$,
where the vertical axis characterizes the possibility for fixed points, and
yellow, blue, and red correspond to the  $\mathrm{FP}_1^\pm$, $\mathrm{FP}_2^\pm$,
and $\mathrm{FP}_3^\pm$, respectively.}
    \label{fig4_single-para-influence-Specical-case}
\end{figure}

As aforementioned, the low-energy fate of 2D QBCP system is dictated by the
coupled RG equations~(\ref{Eq_RG-1})-(\ref{Eq_RG-2}), which
capture the interplay among all electron-electron interactions.
In this section, we examine the behavior of the interaction parameters as the energy scales
decrease, aiming to reveal their tendencies and identify potential fixed
points at the lowest-energy regime. After carrying out the numerical analysis of RG equations~(\ref{Eq_RG-1})-(\ref{Eq_RG-2}),
we figure out that the energy-dependent interaction parameters exhibit a series of
interesting evolutions and are attracted to distinct kinds of fixed points that are
of close dependence upon the initial conditions.
As the symmetries of the free Hamiltonian do not impose strict constraints on the independence of fermion-fermion interactions,
the initial values of fermion-fermion interactions can be taken independently.
To simplify our analysis, we cluster the starting conditions into three distinct cases: (i) Limit case
in which all 16 interaction parameters are assigned the same value at the beginning, (ii) Special case for which
only parts of interaction parameters share certain initial value,  and (iii) General case where all 16 interaction
parameters are independent and hence randomly take their own starting values.
Hereby, it is necessary to highlight that such three distinct cases are denominated only based on the
initial conditions of fermion-fermion interactions.  As the parameter $t$ in our model~(\ref{Eq_H_0}) is an energy-independent
constant to the one-loop level, the QBCP semimetal owns both the particle-hole and rotational symmetry
for all these three cases unless certain instabilities are induced at the lowest-energy limit. In the following, we are going to
consider these three cases one by one.

\subsection{Limit case}~\label{Subsection_FP_Limit}

For the sake of simplicity, we consider the Limit case at first. In this scenario,
we assume that all interaction couplings have the same value at the start. Based on
our numerical analysis of the RG evolutions, we have identified the basic tendencies
of the interaction parameters as shown in Fig.~\ref{fig1_Limit_case_flows}.

Learning from Fig.~\ref{fig1_Limit_case_flows}, we notice that several interaction parameters
flow towards divergence at the low-energy regime owing to the intimate competition among them.
In order to seek the potential fixed points, we are suggested to rescale the parameters
by an unsign-changed parameter~\cite{Vafek2010PRB,Vafek2014PRB,Chubukov2016PRX}. On the basis of this spirit, we bring out
$\lambda_+\equiv(\sum_{ij}\lambda_{ij}^2/16)^{1/2}$ and then measure all
interactions with $\lambda_+$, namely designating the transformation
$\lambda_{ij}/(\sum_{ij}\lambda_{ij}^2/16)^{1/2}\longrightarrow\lambda_{ij}$.
For convenience, we from now on regard $\lambda_{ij}$ as the rescaled interaction parameters (unless
stated otherwise). In addition to the interaction parameters, the structure parameter $t$ in our model~(\ref{Eq_S_eff})
also appears in the coupled RG flows and can alter the RG equations based on its sign.

\begin{figure*}[htbp]
    \centering
\subfigure[]{\includegraphics[width = 0.26\linewidth]{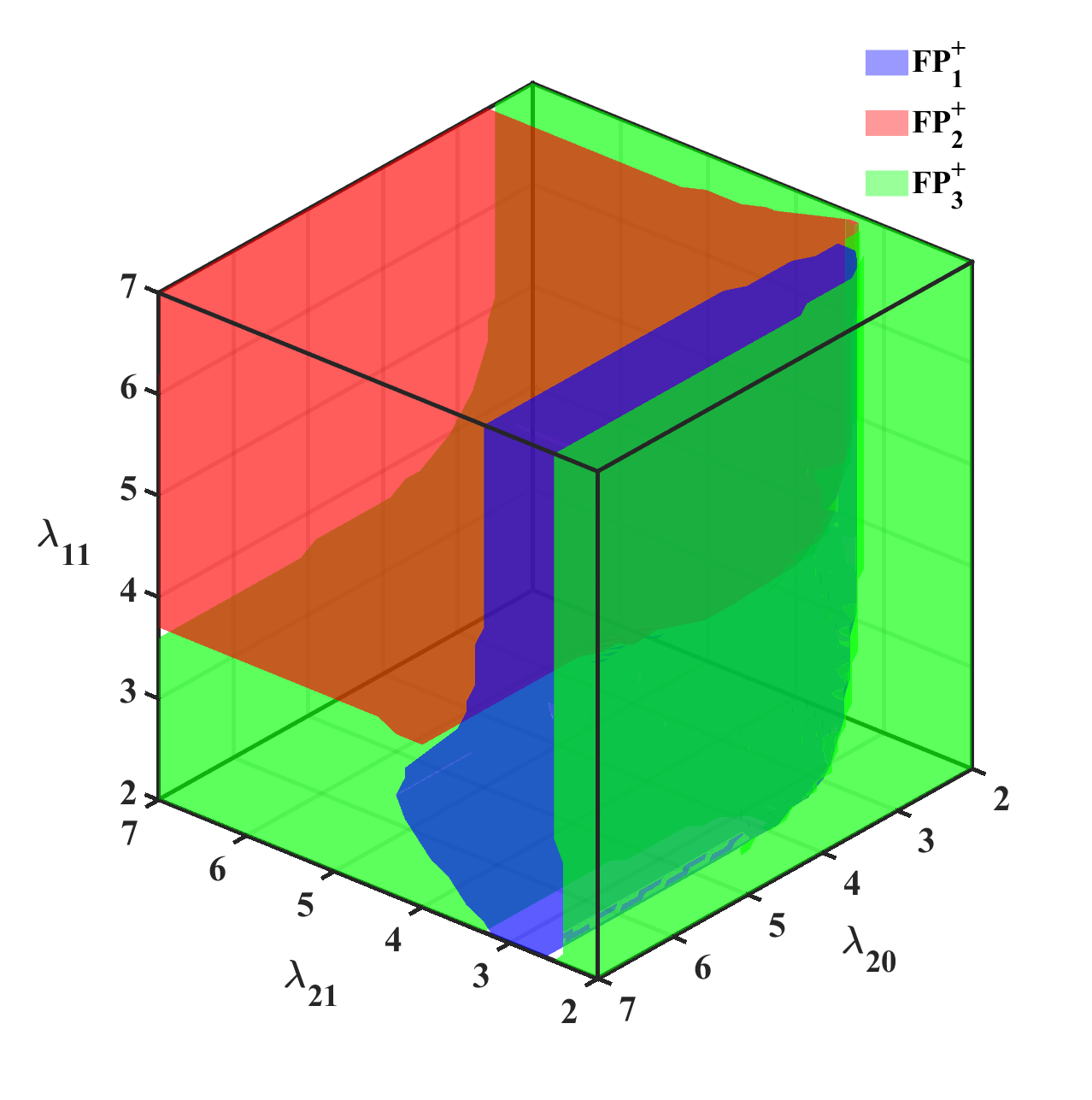}}\hspace{-0.6cm}
\subfigure[]{\includegraphics[width = 0.26\linewidth]{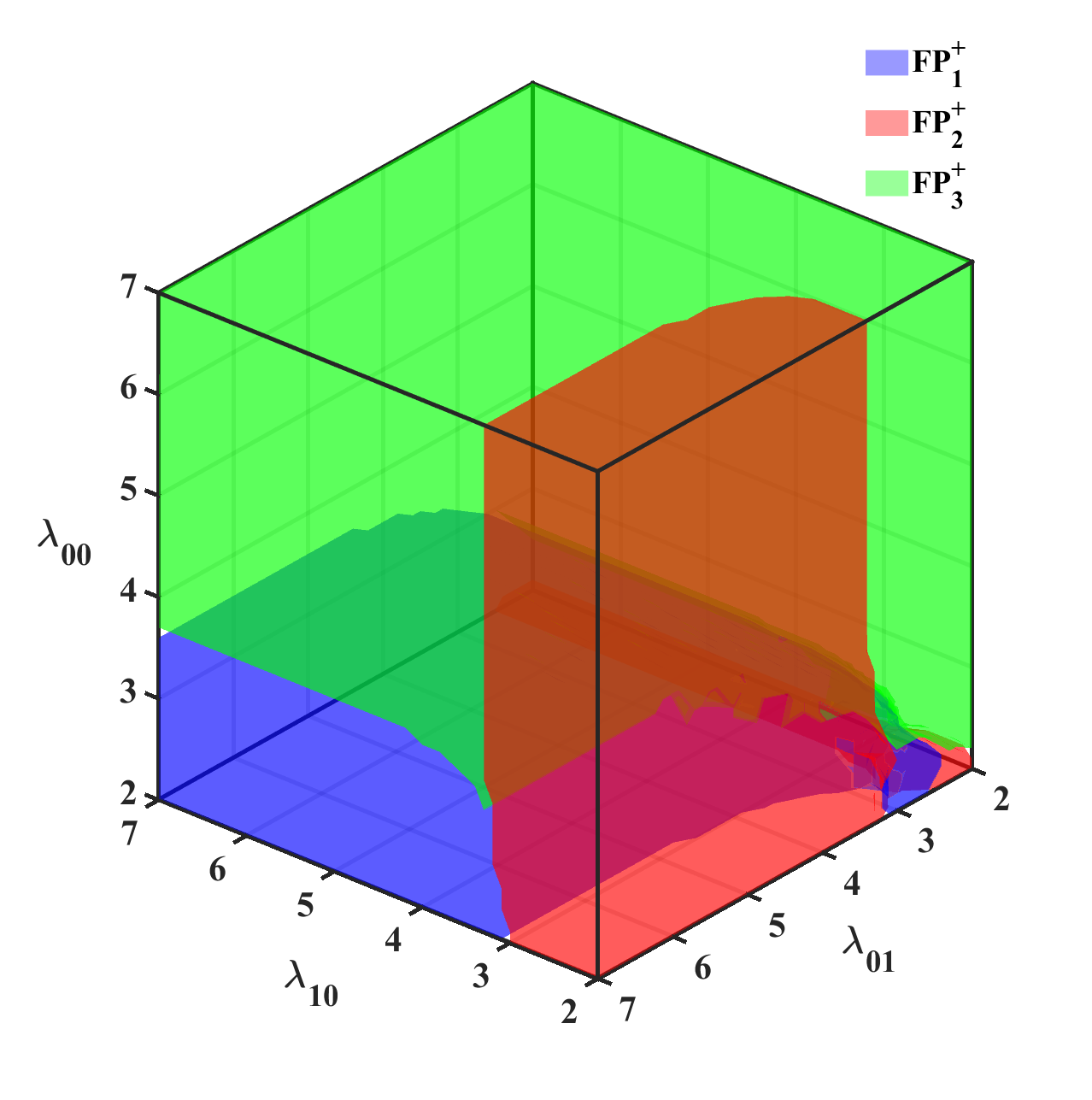}}\hspace{-0.6cm}
\subfigure[]{\includegraphics[width = 0.26\linewidth]{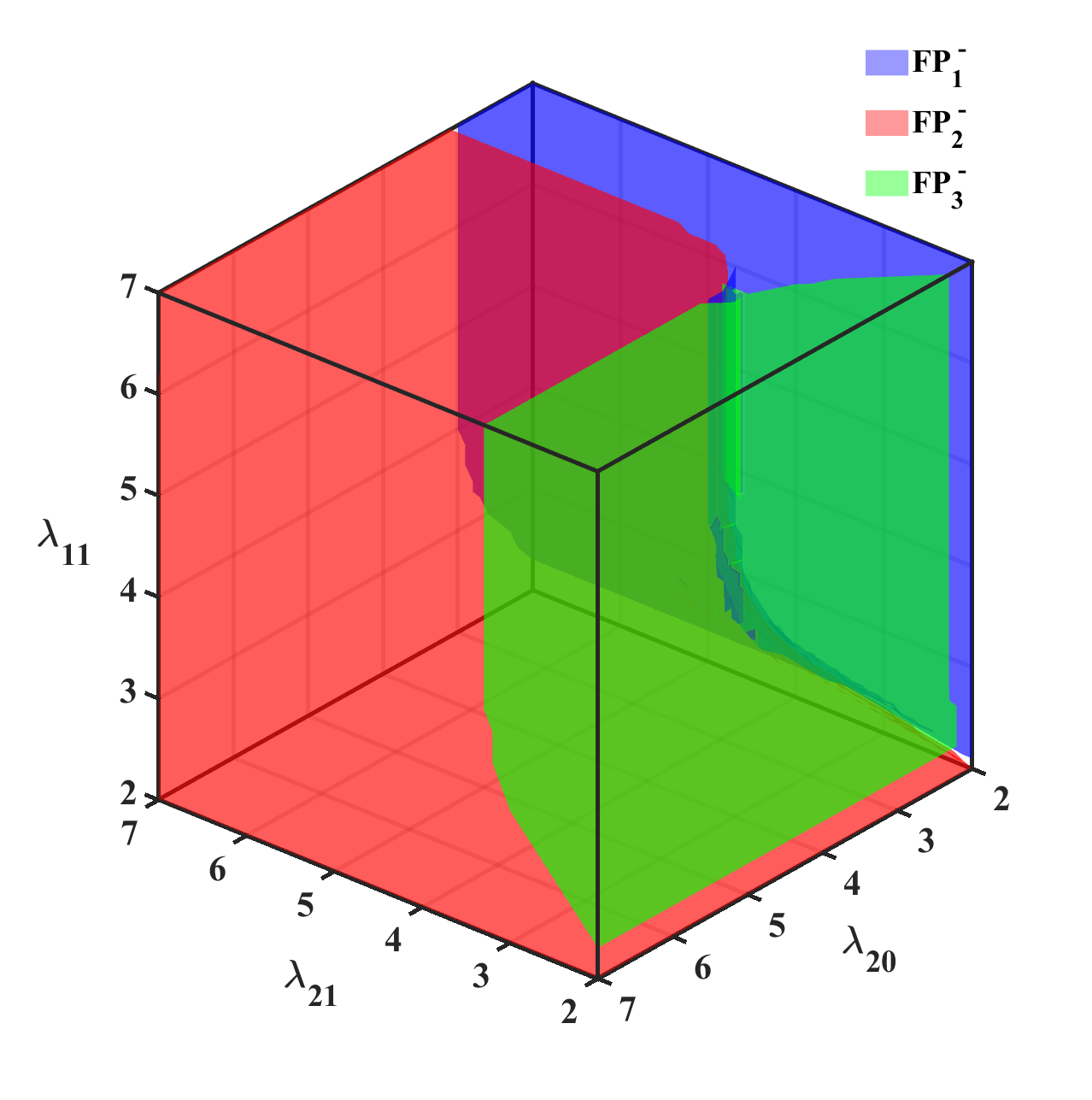}}\hspace{-0.6cm}
\subfigure[]{\includegraphics[width = 0.26\linewidth]{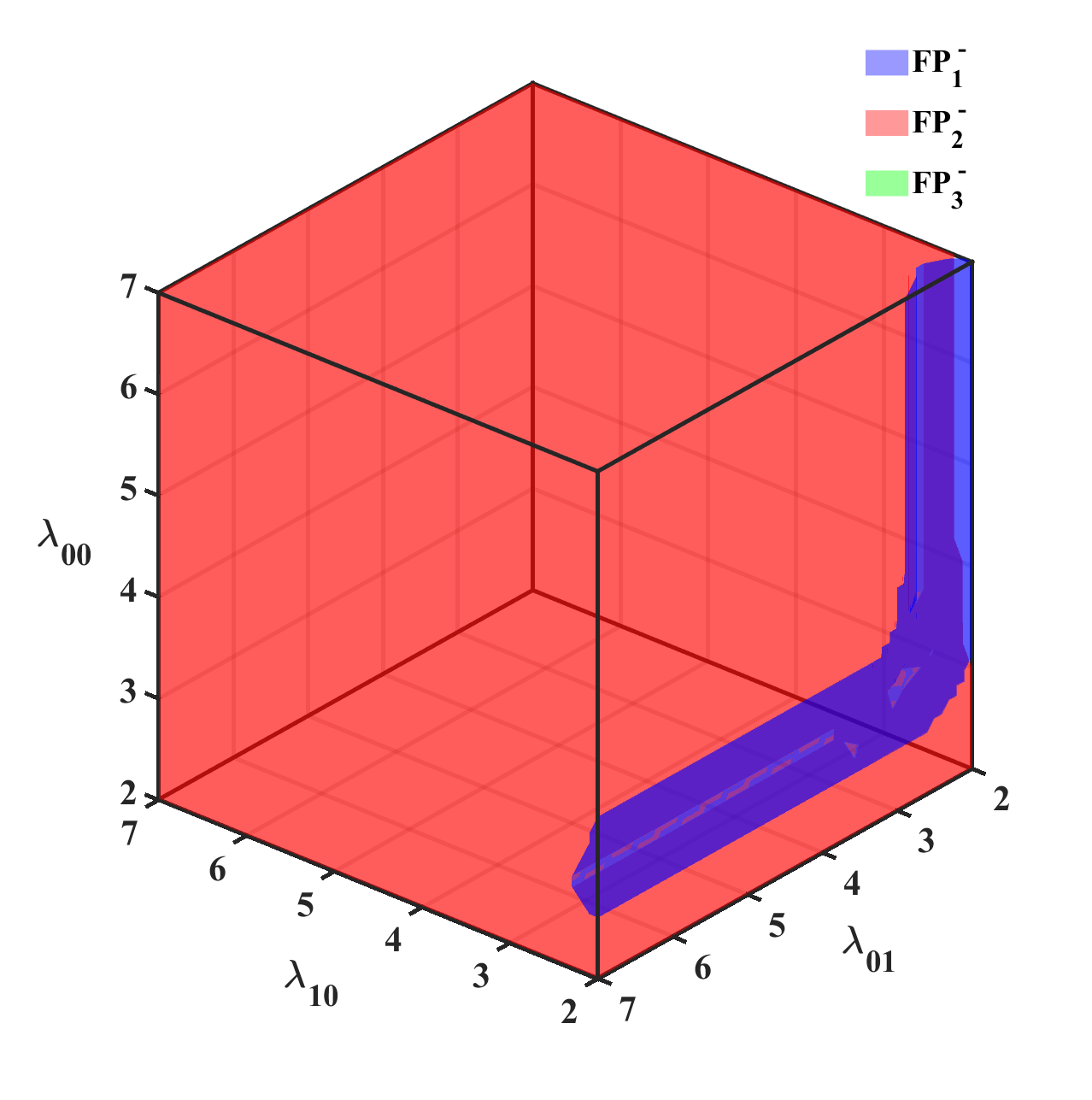}}\\
\vspace{0.05cm}
\caption{(Color online) Choosing $(\lambda_{00},\lambda_{01},\lambda_{10},\lambda_{11},\lambda_{20},\lambda_{21})$
from six classes for an example instance (the basic results are similar for other choices)
and showing the competition among different fixed points by tuning the magnitudes of three parameters:
(a) $\lambda_{11}$, $\lambda_{21}$, and $\lambda_{20}$ with $t>0$ and the initial parameters $(\mathrm{Class}-i)=(10^{-3},10^{-3},10^{-3},10^{-x},10^{-y},10^{-z})$,
(b) $\lambda_{00}$, $\lambda_{10}$, and $\lambda_{01}$ with $t>0$ and the initial parameters $(\mathrm{Class}-i)=(10^{-x},10^{-y},10^{-z},10^{-3},10^{-3},10^{-3})$,
(c) $\lambda_{11}$, $\lambda_{21}$, and $\lambda_{20}$ with $t<0$ and the initial parameters $(\mathrm{Class}-i)=(10^{-3},10^{-3},10^{-3},10^{-x},10^{-y},10^{-z})$,
and (d) $\lambda_{00}$, $\lambda_{10}$, and $\lambda_{01}$ with $t<0$ and the initial parameters $(\mathrm{Class}-i)=(10^{-x},10^{-y},10^{-z},10^{-3},10^{-3},10^{-3})$,
where $x,y,z$ serve as the magnitudes of the related parameters, as well as
blue, red, and green correspond to the $\mathrm{FP}_1^\pm$, $\mathrm{FP}_2^\pm$,
and $\mathrm{FP}_3^\pm$, respectively.}
\label{fig5_three-para-influence-Specical-case}
\end{figure*}

Under this circumstance, we perform the numerical analysis and present the primary results for evolutions of
rescaled parameters in Fig.~\ref{fig2_Limit_case_FPs} for both a positive and negative starting value
of parameter $t$ (For completeness, we have varied the initial parameters from $10^{-2}$ to $10^{-7}$ and found that
the qualitative behavior of the parameters are similar). One can find that the basic evolutions of
parameters in the Limit case are insusceptible to the initial interaction values, but instead
heavily hinge upon the sign of parameter $t$. In other words, there exist two distinct kinds
of fixed points in the Limit case, which are distinguished by the sign of $t$.
For the sake of simplicity, we hereafter introduce the notation $\mathrm{FP_N^\pm}$ to denominate and
distinguish the potential fixed points, where the subscript $N$ is an integer to denote the order of
the fixed points, and $\pm$ capture the sign of $t$. In this sense, we can refer to the fixed
point in Fig.~\ref{fig2_Limit_case_FPs}(a) as $\mathrm{FP_1^+}$
and Fig.~\ref{fig2_Limit_case_FPs}(b) as $\mathrm{FP_2^-}$, respectively.
Specifically, they appropriately take the form of
\begin{eqnarray}
\mathrm{FP}_1^+\!\approx\!\!
    \begin{pmatrix}
         0.0000 & 0.0000 & 0.0000 & 0.0000 \\
        -0.2876 & 0.0000 & 0.0000 & 0.0000 \\
        -3.9792 & 0.0000 & 0.0000 & 0.0000 \\
        -0.2876 & 0.0000 & 0.0000 & 0.0000
    \end{pmatrix}
\end{eqnarray}
and
\begin{eqnarray}
\mathrm{FP}_2^-\!\approx\!\!
    \begin{pmatrix}
         0.0078 & -0.4624 & -0.4624 & -0.4624\\
         0.3848 & -0.0680 & -0.0680 & -0.0680\\
        -0.0037 &  2.2386 &  2.2386 &  2.2386\\
         0.3848 & -0.0680 & -0.0680 & -0.0680
    \end{pmatrix}
\end{eqnarray}
where the $(i,j)$ element corresponds to the interaction parameter $\lambda_{ij}$.

\subsection{Special case}\label{Subsection_FP_special}

Next, we move to study the Special case. Upon closer inspection of Fig.~\ref{fig1_Limit_case_flows},
one can notice that several interaction couplings are overlapped
and all interaction parameters cluster into six new classes, namely
Class-1 ($\lambda_{00}$), Class-2 ($\lambda_{01}$,  $\lambda_{02}$,  $\lambda_{03}$),
Class-3 ($\lambda_{10}$,  $\lambda_{30}$),
Class-4 ($\lambda_{11}$, $\lambda_{12}$, $\lambda_{13}$, $\lambda_{31}$ $\lambda_{32}$, $\lambda_{33}$),
Class-5 ($\lambda_{20}$), and Class-6 ($\lambda_{21}$,  $\lambda_{22}$,  $\lambda_{23}$), respectively.

Due to the complexity of real materials, parts interaction parameters may be deviated from the
same initial condition that is required in the Limit case.  To account for this, let us go beyond the
Limit case and consider a little more complicate case (i.e., Special case),
in which the interaction parameters within the same class still share the a starting value but instead
initial values of different classes can be independently tuned.

\begin{figure}[htbp]
    \centering
\subfigure[]{\includegraphics[scale = 0.25]{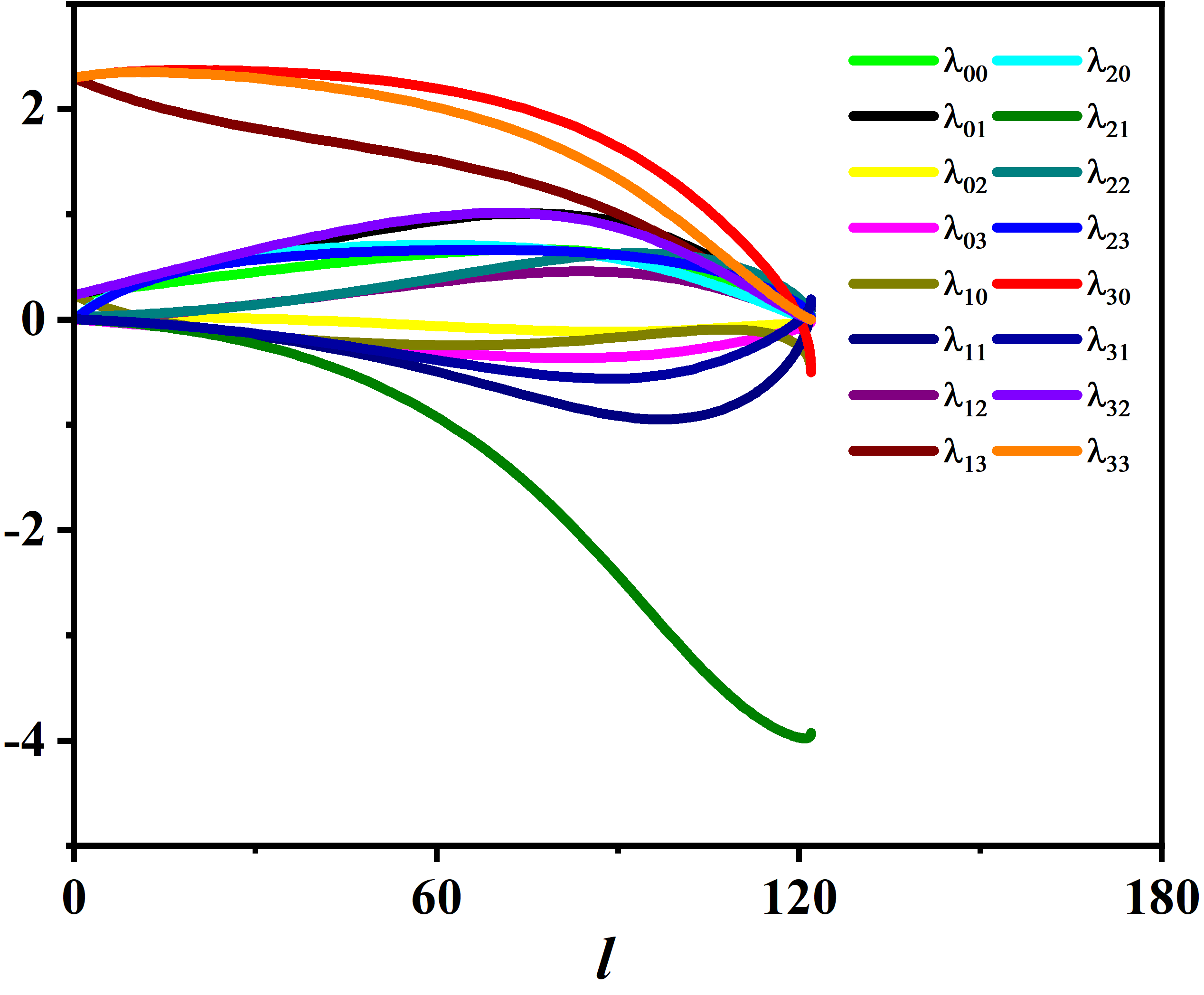}}\\ \vspace{0.13cm}
\subfigure[]{\includegraphics[scale = 0.25]{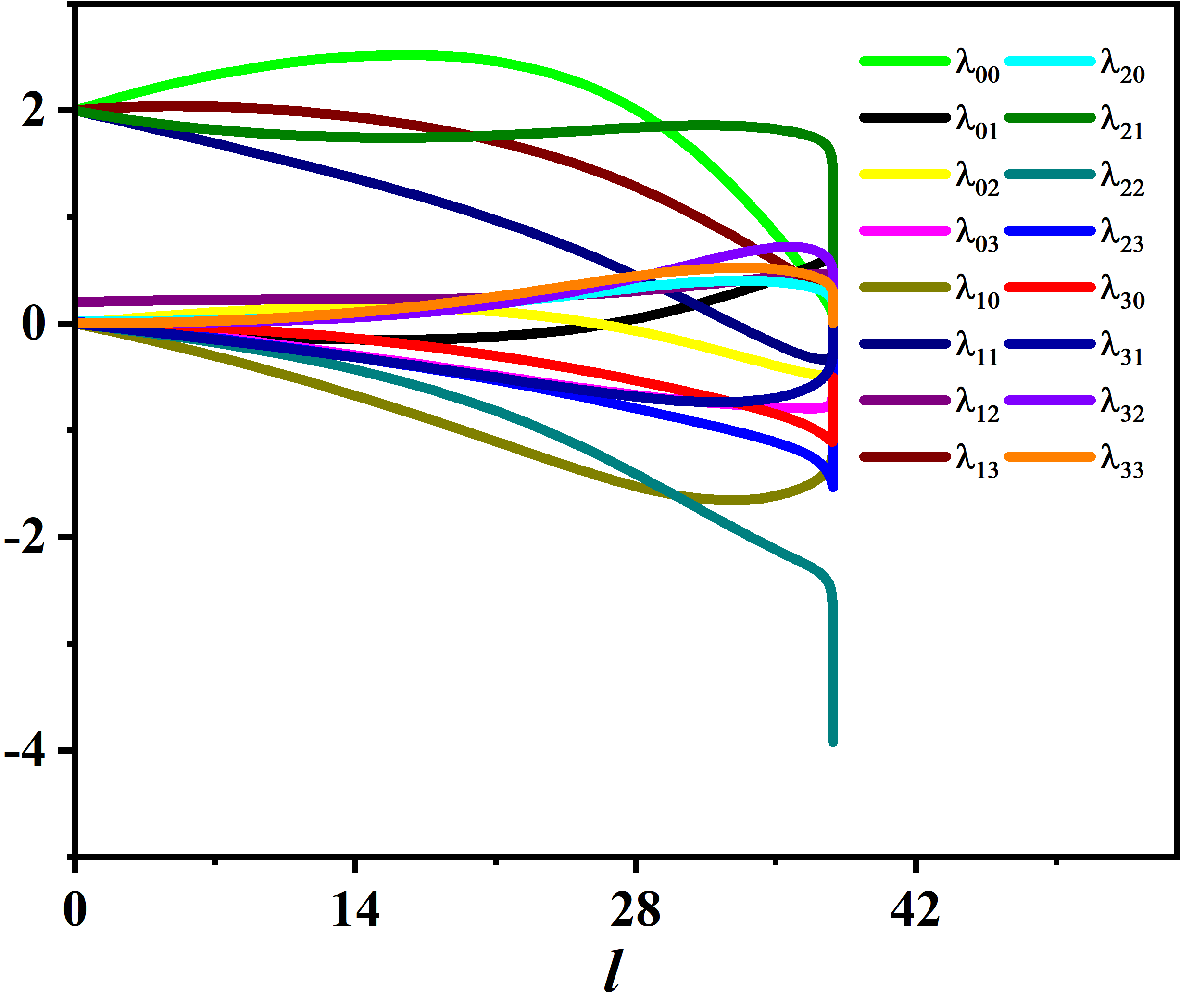}}\\ \vspace{0.13cm}
\subfigure[]{\includegraphics[scale = 0.25]{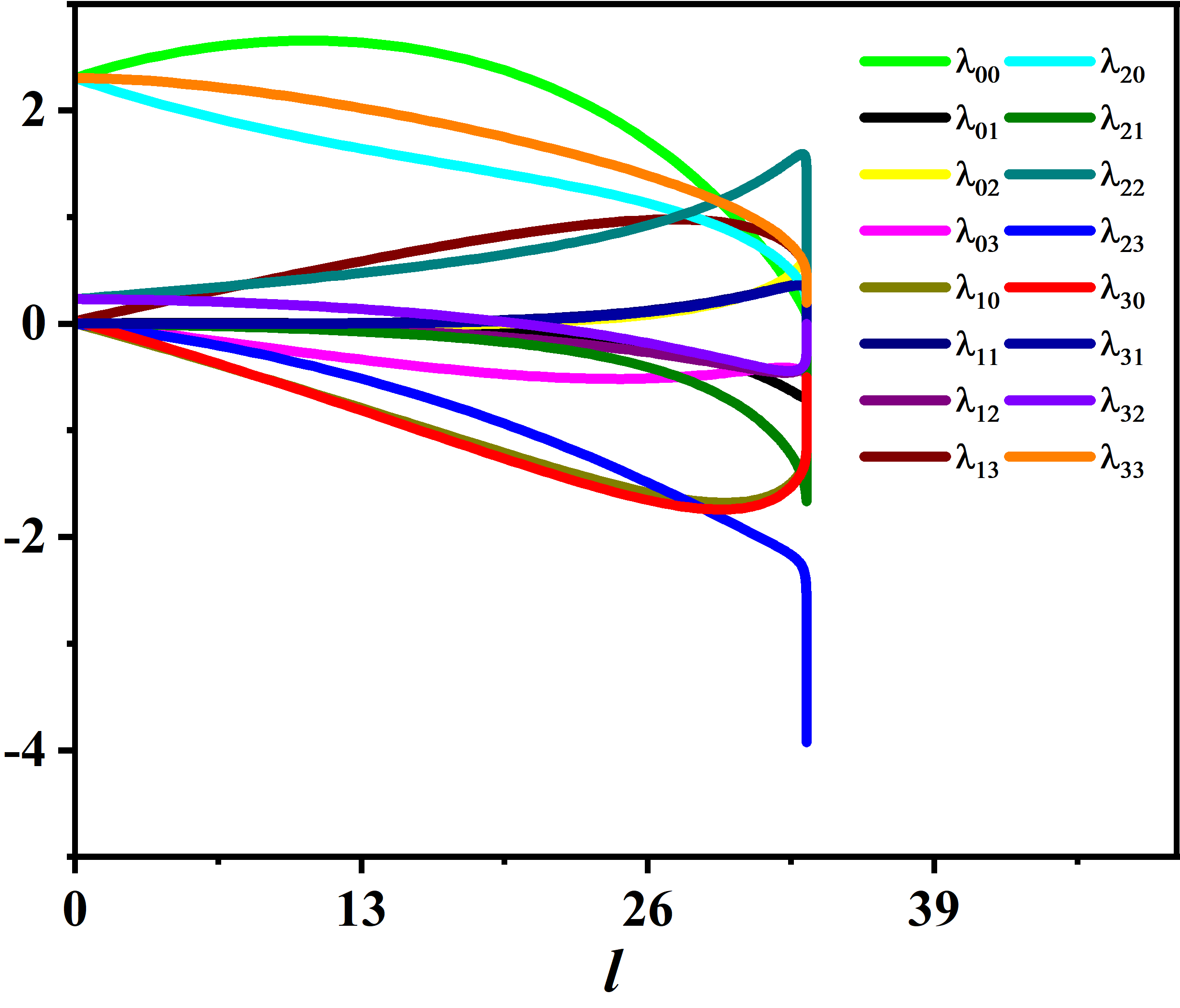}}\\
\vspace{-0.1cm}
\caption{(Color online) Energy-dependent flows of interaction parameters
in the General case at $t>0$ and fixed points towards (a) $\mathrm{FP}_{41}^+$,
(b) $\mathrm{FP}_{42}^+$, and (c) $\mathrm{FP}_{43}^+$
starting from the initial interaction parameters $(10^{-3},10^{-3},10^{-4},10^{-5},10^{-3},
10^{-5},10^{-6},10^{-2},10^{-6},10^{-5},\\
10^{-4},10^{-7},10^{-2},10^{-5},
10^{-3},10^{-2})$, and $(10^{-2},10^{-7},10^{-7},\\
10^{-4},10^{-5},10^{-2},10^{-3},10^{-2},10^{-4})$, and $(10^{-2},10^{-7},10^{-5},\\
10^{-5},10^{-5},10^{-6},10^{-7},10^{-4},10^{-2},10^{-6},10^{-3},10^{-7},\,\,10^{-4},\\
10^{-6},10^{-3},10^{-2})$, respectively.}
\label{fig6_FP_456}
\end{figure}

\begin{figure*}[htbp]
    \centering
\subfigure[]{\includegraphics[scale = 0.22]{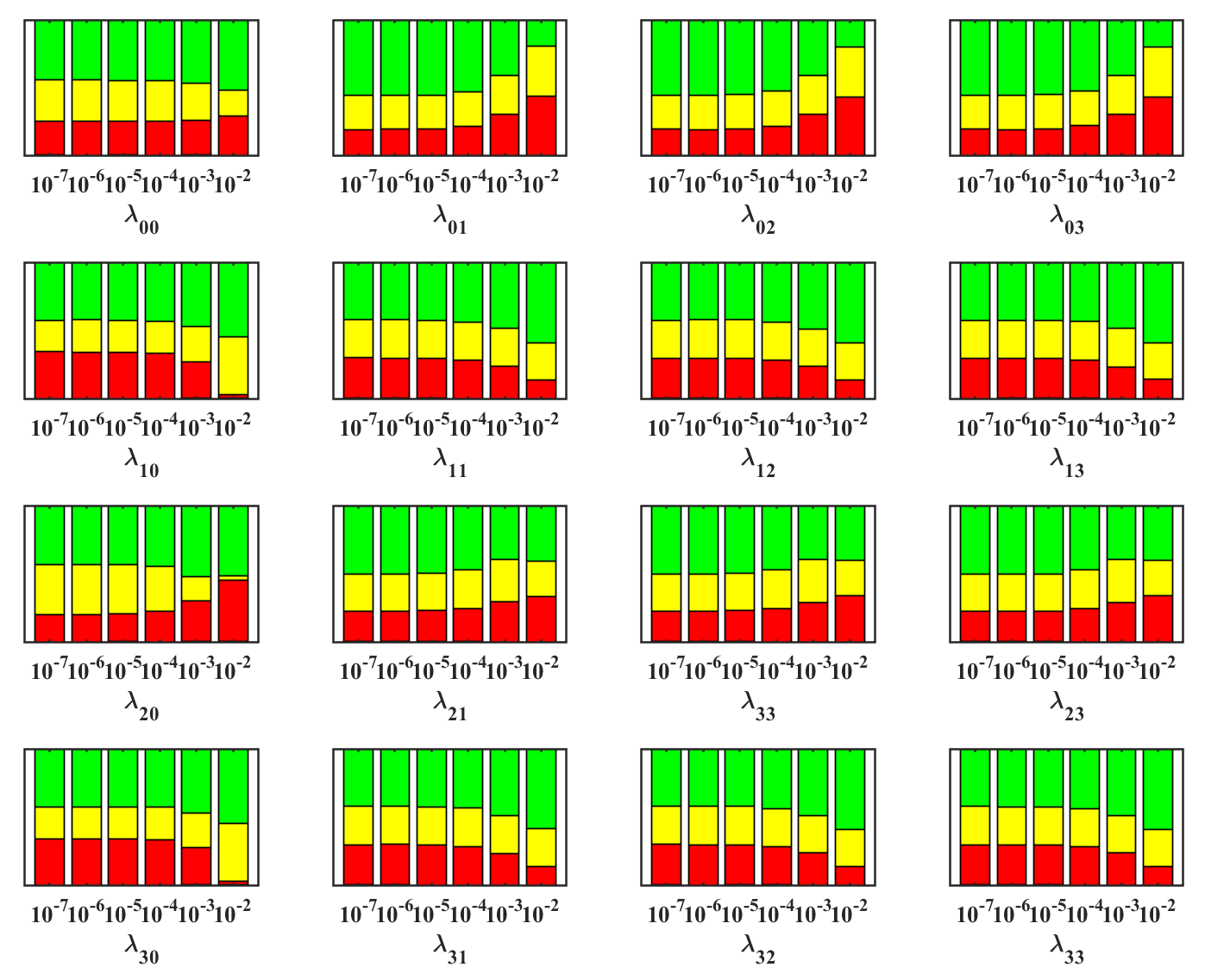}}\hspace{0.5cm}
\subfigure[]{\includegraphics[scale = 0.22]{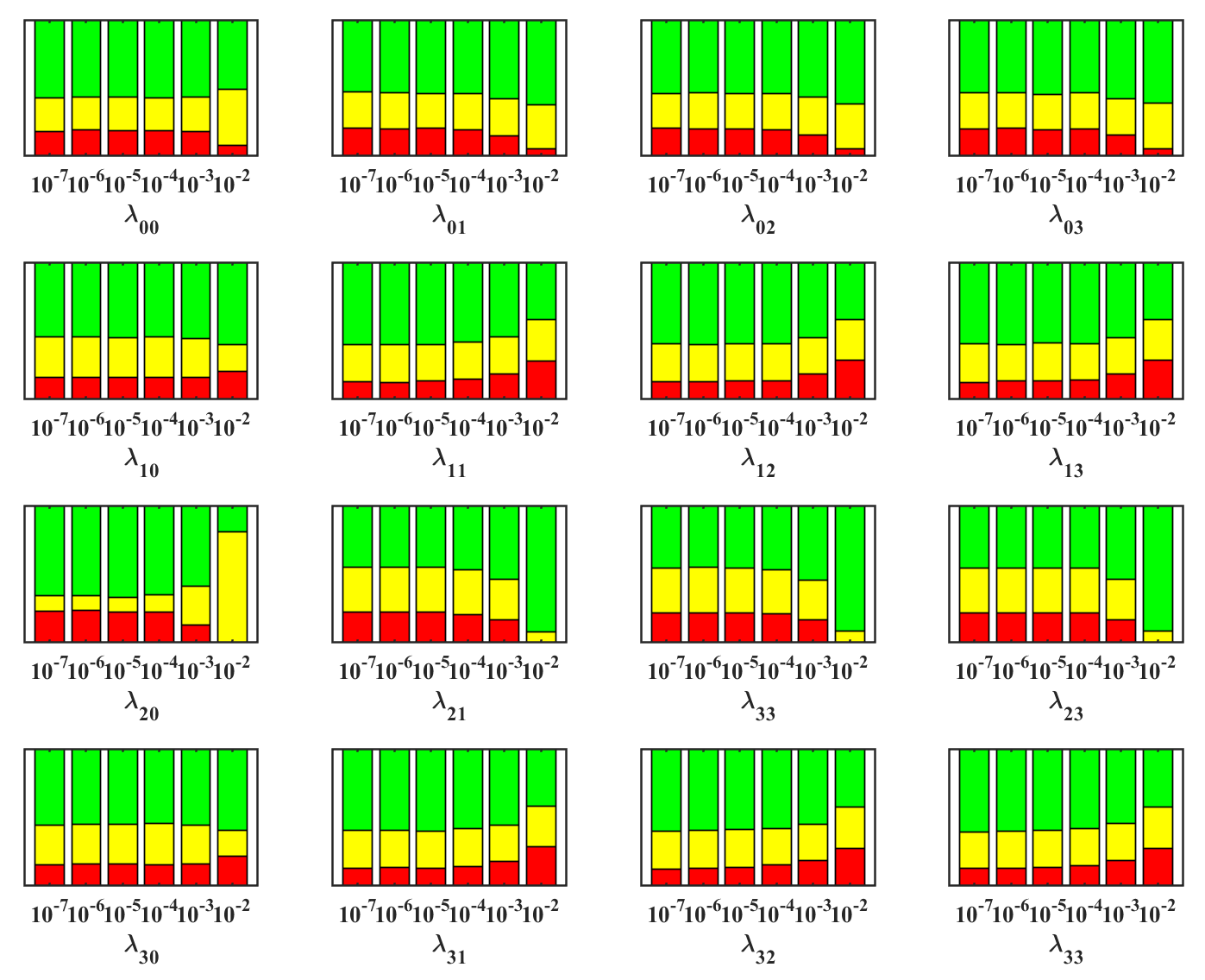}}\\
\vspace{0.1cm}
\caption{(Color online) Competition among different fixed points with varying the magnitude
of a single interaction parameter(horizontal axis) and fixing the others: (a) $t>0$ and (b) $t<0$,
where the vertical axis characterizes the possibility for fixed points, as well as
yellow, red, and green correspond to the  $\mathrm{FP}_1^\pm$, $\mathrm{FP}_3^\pm$,
and $\mathrm{FP}_4^\pm$ ($\mathrm{FP}_{41}^\pm$, $\mathrm{FP}_{42}^\pm$, or $\mathrm{FP}_{43}^\pm$), respectively.}
\label{fig7_para-influence-General-case}
\end{figure*}

After paralleling analogous numerical analysis in Limit case, we find that several new fixed points
can be generated. As to $t>0$, in addition to the $\mathrm{FP}_1^+$ obtained in the Limit case shown in Fig~\ref{fig2_Limit_case_FPs}(a),
fixed points $\mathrm{FP}_2^+$ and $\mathrm{FP}_3^+$ are induced as presented in Fig.~\ref{fig3_FP_123}, which are
appropriately expressed as
\begin{eqnarray}
\mathrm{FP}_2^+\approx
\begin{pmatrix}
-0.0078 & 0.4624 & 0.4624 & 0.4624\\
-0.3848 & 0.0680 & 0.0680 & 0.0680\\
0.0037 &  -2.2386 &  -2.2386 &  -2.2386\\
-0.3848 & 0.0680 & 0.0680 & 0.0680
\end{pmatrix},\nonumber \\
\mathrm{FP}_3^+\approx
\begin{pmatrix}
0.2189 & 0.5406 & 0.5406 & 0.5406\\
-1.3015 &  -0.9971 &  -0.9971 &  -0.9971\\
2.1702 & 0.5805 & 0.5805 & 0.5805\\
-1.3015 &  -0.9971 &  -0.9971 &  -0.9971
\end{pmatrix},\nonumber
\end{eqnarray}
with $i$ running from $1$ to $6$, which can also be compactly
expressed as
\begin{eqnarray}
&&\mathrm{FP}_2^+(\mathrm{Class}-i)\nonumber\\
&\approx&(-0.0078,\ 0.4624,\ -0.3848,\ 0.0680,\ 0.0037,\ -2.2386),\nonumber\\
&&\mathrm{FP}_3^+(\mathrm{Class}-i)\nonumber\\
&\approx&(0.2189,\ 0.5406,\ -1.3015,\ -0.9971,\ 2.1702,\ 0.5805).\nonumber
\end{eqnarray}
With respect to $t<0$, three more fixed points are found to be closely related to
their $t>0$ counterparts, including $\mathrm{FP}_1^-=-\mathrm{FP}_1^+$, $\mathrm{FP}_2^-
=-\mathrm{FP}_2^+$ and $\mathrm{FP}_3^-=-\mathrm{FP}_3^+$.

On the basis of above analysis, one can realize that the interplay among distinct types of interactions
coaxes the system to flow towards certain fixed points. As both the structure parameter $t$ and the interaction
couplings $\lambda_{ij}$ are involved in the coupled RG equations, their interplay and competition determine the structure
of fixed points. With respect to the Limit case shown in Fig.~\ref{fig2_Limit_case_FPs}, the sign of
parameter $t$ plays a more significant role in pinning down the fixed points than
the concrete initial values of interaction parameters.

\begin{figure}[htbp]
\centering
\subfigure[]{\includegraphics[width = 0.76\linewidth]{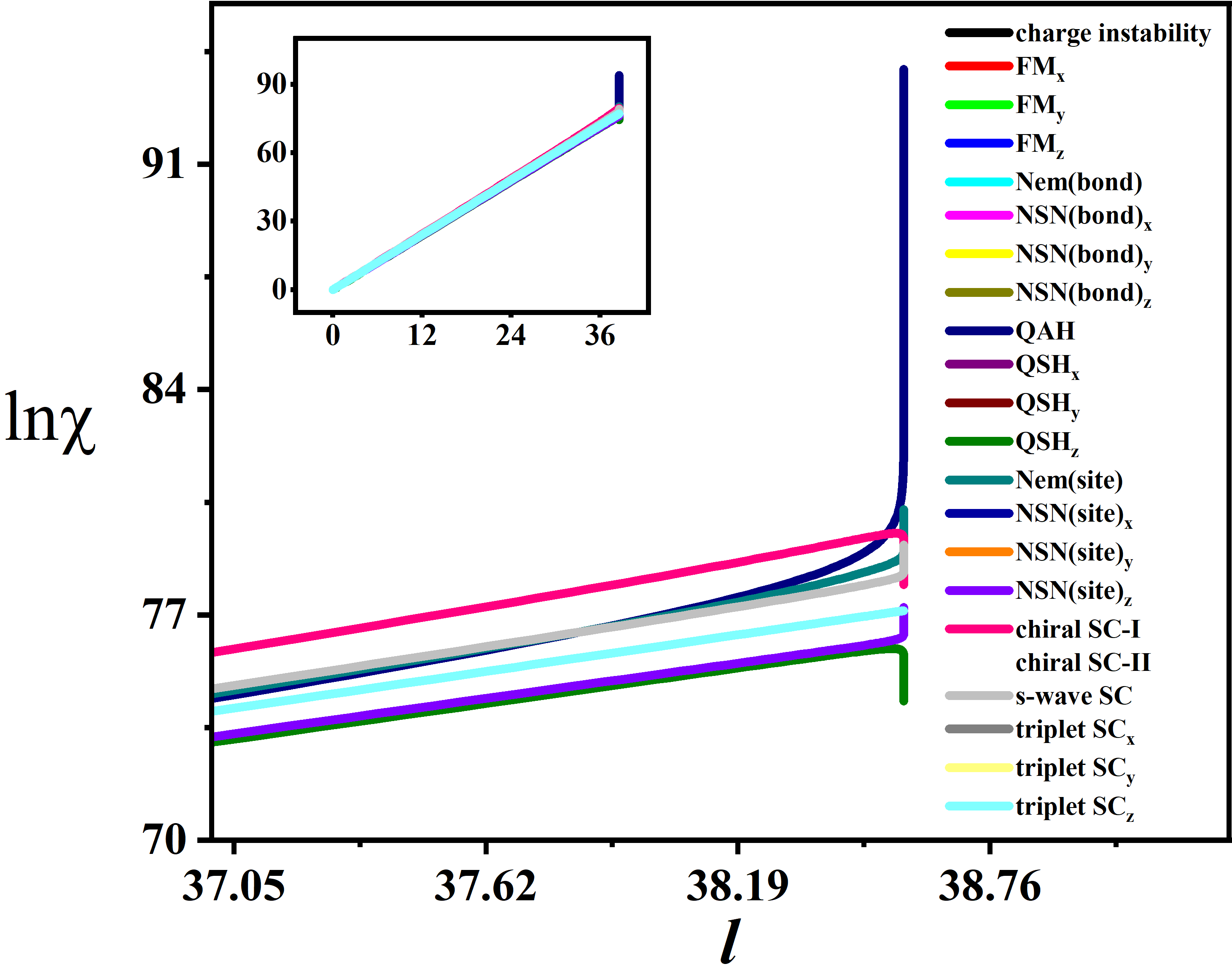}}\\ \vspace{0.05cm}
\subfigure[]{\includegraphics[width = 0.76\linewidth]{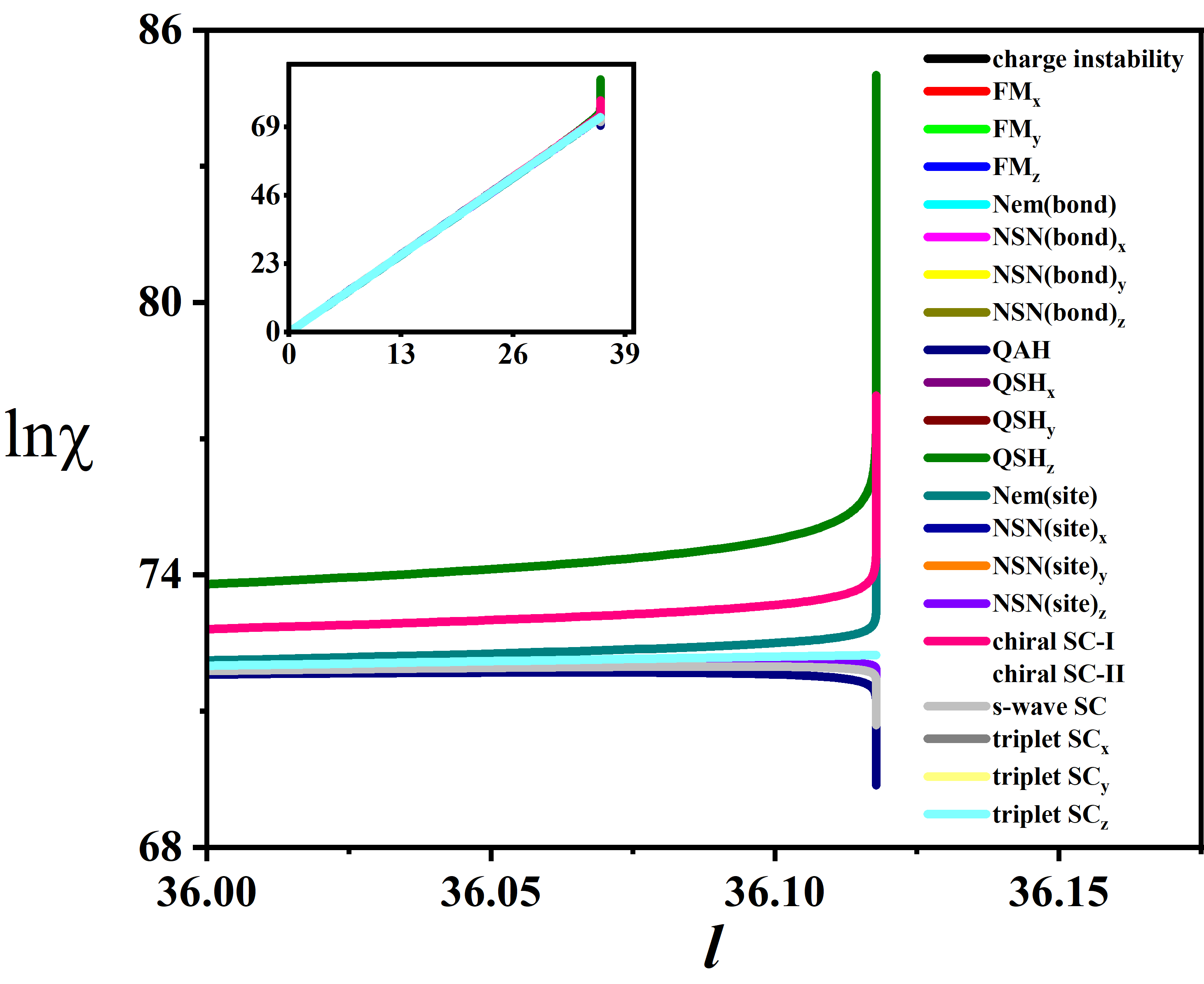}}
\vspace{0.05cm}
\caption{(Color online) Energy-dependent susceptibilities of all candidate instabilities presented in
Table~\ref{table:phase} as approaching (a) $\mathrm{FP}^+_1$ and (b) $\mathrm{FP}^-_2$, respectively.
The subscripts $(x,y,z)$ serve as the distinct components of corresponding states.} \label{fig:FP12-sus}
\end{figure}

In sharp contrast, regarding the Special case, six classes of interaction parameters can be adjusted independently at the
starting point, which in tandem with the sign of $t$ gives rise to more interesting consequences.

At first, we fix the initial values of five classes and vary the starting value of the
sixth class to study the influence on the fixed points as displayed in Fig.~\ref{fig4_single-para-influence-Specical-case}
where the proportion of each type of fixed point can exhibit the average distribution of fixed points. In the case of $t>0$ shown in Fig.~\ref{fig4_single-para-influence-Specical-case}(a),
it manifestly indicates that the increase of $\lambda_{00}, \lambda_{10}$ is profitable to flowing towards $\mathrm{FP}_1^+$, and particularly, $\lambda_{10}$ plays a critical role and also hinders the onset of $\mathrm{FP}_3^+$.
In comparison, the parameters $\lambda_{11}$, $\lambda_{20}$, and $\lambda_{21}$ with $t<0$ in
Fig.~\ref{fig4_single-para-influence-Specical-case}(b) dominate over other parameters.
It is unambiguous that the increase of $\lambda_{11}$ and $\lambda_{20}$ are very helpful to the
development of $\mathrm{FP}_3^-$ and $\mathrm{FP}_1^-$, respectively. Besides, tuning up the
$\lambda_{21}$ is of particular help to $\mathrm{FP}_2^-$.

In addition, we tune the starting values of three parameters simultaneously while keeping
the other three fixed to further examine the stabilities of
fixed points. For instance, with selecting $(\lambda_{00},\lambda_{01},\lambda_{10},\lambda_{11},\lambda_{20},\lambda_{21})$ from six classes, Fig.~\ref{fig5_three-para-influence-Specical-case} presents the competition among different fixed points
with variance of the sign of parameter $t$ and magnitudes of three parameters.

On one hand, one can notice that overall structures of fixed points for $t>0$ differ significantly from those for $t<0$.
In consequence, this implies that the sign of $t$ has an important contribution to the fixed points.
On the other hand, once the sign of $t$ is selected, it can also be clearly found that the initial
values of the parameters play a significant role in determining which fixed point the
system flows towards. As shown in Fig.~\ref{fig5_three-para-influence-Specical-case}(a) with $t>0$,
the increase of $\lambda_{11}$ and $\lambda_{21}$ is helpful to the onset of $\mathrm{FP}_3^+$ and
$\mathrm{FP}_1^+$, but instead $\mathrm{FP}_2^+$ once all three parameters are small enough.
The basic structure of Fig.~\ref{fig5_three-para-influence-Specical-case}(b)
is close in resemblance to that of Fig.~\ref{fig5_three-para-influence-Specical-case}(a).
Particularly, when the parameters are restricted to $10^{-4}-10^{-3}$,
there exists a ferocious competition among various fixed points and hence the
dominant FP, to a large extent, is sensitive to initial interaction strengths.
As to $t<0$, although it bears similarities
to Fig.~\ref{fig5_three-para-influence-Specical-case}(a),
Fig.~\ref{fig5_three-para-influence-Specical-case}(c) shows that
tuning up $\lambda_{20}$ and $\lambda_{21}$ are instructive to the generation of
$\mathrm{FP}_1^-$ and $\mathrm{FP}_3^-$, but rather decreasing them to $\mathrm{FP}_2^-$.
In sharp contrast to Fig.~\ref{fig5_three-para-influence-Specical-case}(b),
Fig.~\ref{fig5_three-para-influence-Specical-case}(d) exhibits that the system
is either attracted by $\mathrm{FP}^-_1$ or $\mathrm{FP}_2^-$, indicating the
sign of parameter $t$ plays a more crucial role.

\begin{figure}[htbp]
    \centering
    \subfigure[]{\includegraphics[width = \columnwidth]{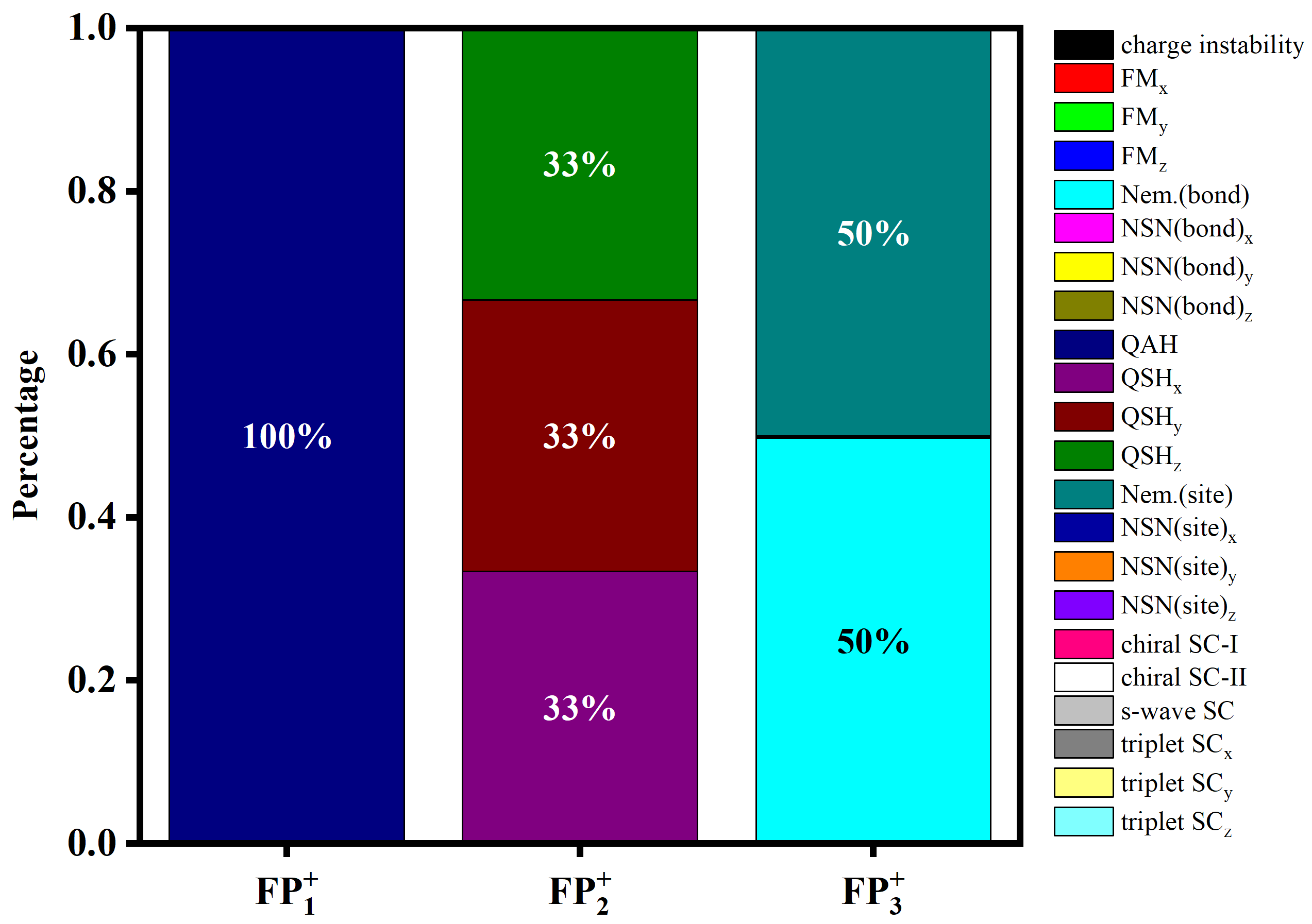}}\\ \vspace{0.05cm}
    \subfigure[]{\includegraphics[width = \columnwidth]{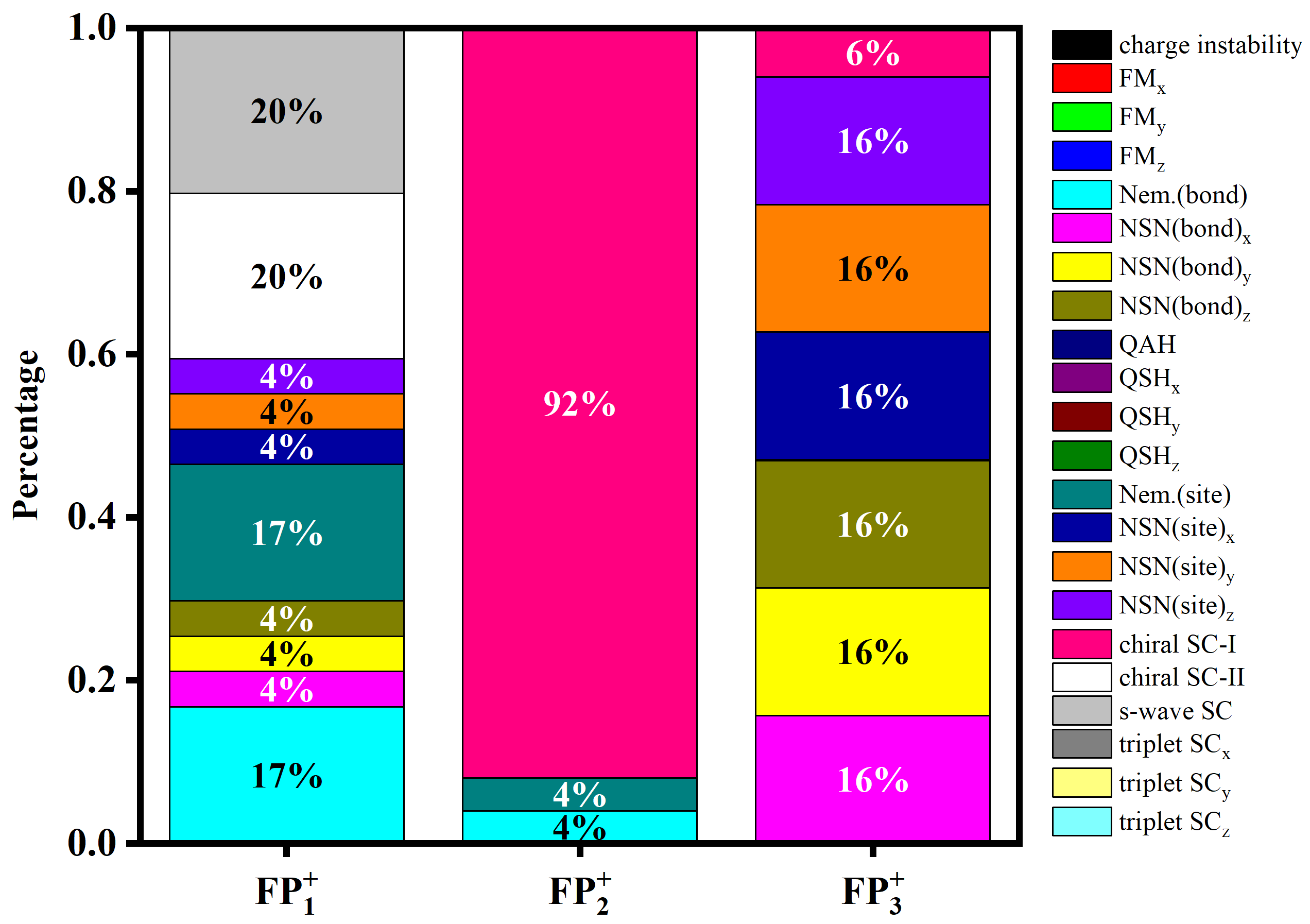}}\\ \vspace{0.05cm}
    \subfigure[]{\includegraphics[width = \columnwidth]{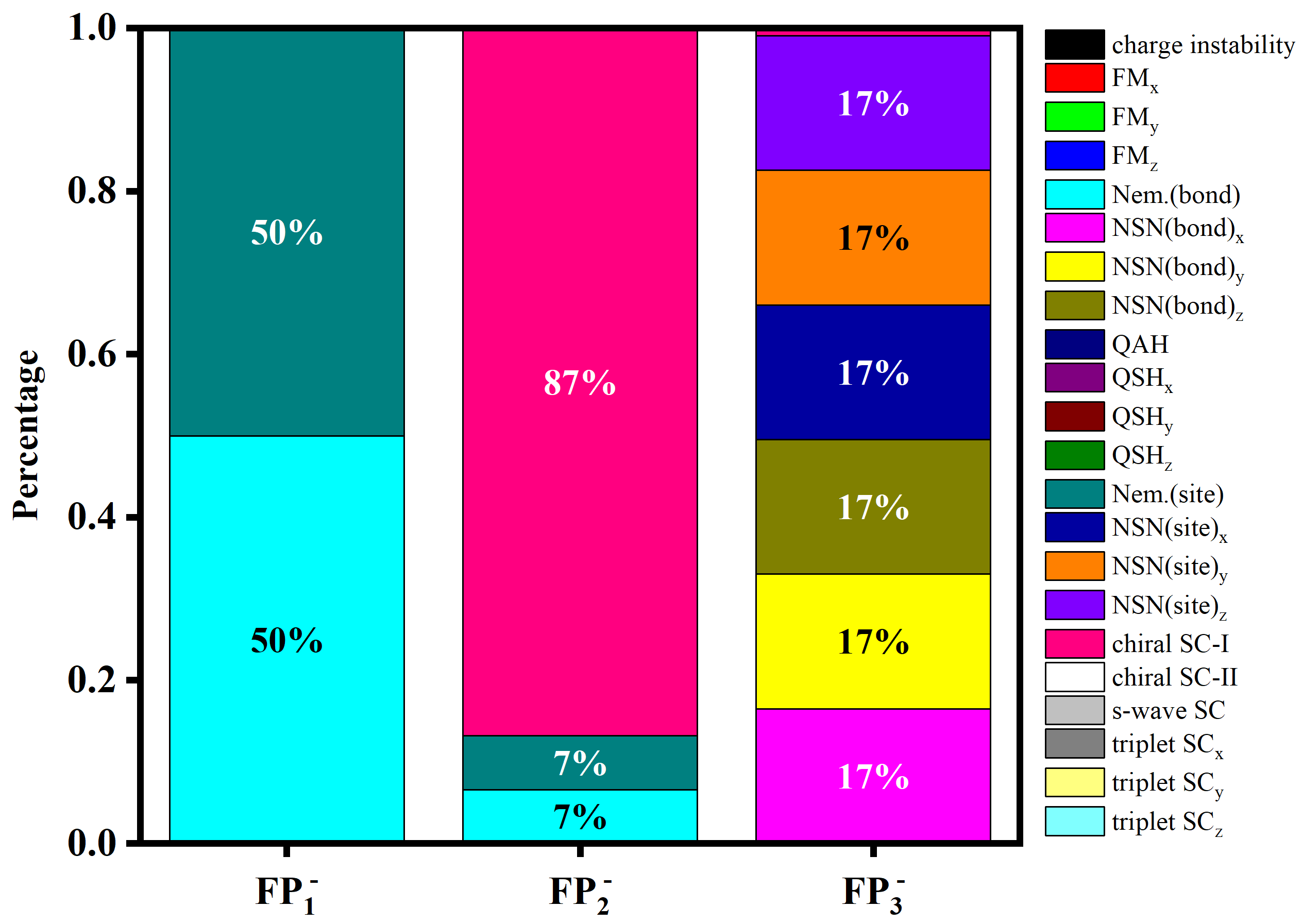}}\\
    \vspace{0.05cm}
    \caption{(Color online) Stabilities of (a) the leading phases at $t>0$, (b) the subleading phases at $t>0$, and
(c) the subleading phase at $t<0$ nearby the fixed points in the Special case
measured by the percentages with variation of initial conditions.}
    \label{fig:phase_FP_Specical_case}
\end{figure}

\subsection{General case}\label{Subsec_General-case-FP}

Furthermore, let us go beyond above two simplified cases and consider the General case in which
the initial values of all 16 interaction parameters can be independently assigned.
After carrying the similar analysis in Sec.~\ref{Subsection_FP_Limit}
and Sec.~\ref{Subsection_FP_special}, we find several interesting results in the low-energy regime.

At first, we realize that the intimate competition among all interactions melt the $\mathrm{FP}_2^\pm$
garnered in the Limit and Special cases. In other words, $\mathrm{FP}_2^\pm$ are only present in certain
special situations which demand the system to satisfy strict constrictions. As a result, the related
physics accompanied by such fixed point would be baldly sabotaged.
Additionally, it is of particular importance to highlight that another
three fixed points can be induced by the close interaction competition with the suitable initial conditions,
namely $\mathrm{FP}_{41}^\pm$, $\mathrm{FP}_{42}^\pm$, and $\mathrm{FP}_{43}^\pm$ with $\pm$ corresponding to
$t>0$ and $t<0$, respectively. Concretely, they appropriately take the form of
\begin{eqnarray}
\mathrm{FP}_{41}^+\approx
\begin{pmatrix}
            0.0000 & 0.0000 & 0.0000 & 0.0000 \\
            0.5015 & -0.1932 & 0.0000 & 0.0000 \\
            -0.0182 & 3.9271 & 0.0000 & 0.0000 \\
            0.5015 & -0.1932 & 0.0000 & 0.0000
\end{pmatrix}\\
\mathrm{FP}_{42}^+\approx
\begin{pmatrix}
            0.0000 &  0.0000 & 0.0000  & 0.0000 \\
            0.5015 &  0.0000 & -0.1932 & 0.0000 \\
            -0.0182 & 0.0000 & 3.9271  & 0.0000 \\
            0.5015 &  0.0000 & -0.1932 & 0.0000
\end{pmatrix}\\
\mathrm{FP}_{43}^+\approx
\begin{pmatrix}
            0.0000 &  0.0000 & 0.0000 & 0.0000 \\
            0.5015 &  0.0000 & 0.0000 & -0.1932  \\
            -0.0182 & 0.0000 & 0.0000 & 3.9271\\
            0.5015 &  0.0000 & 0.0000 & -0.1932
\end{pmatrix}
\end{eqnarray}
for $t>0$ as displayed in Fig.~\ref{fig6_FP_456}, and their $t<0$ counterparts share the same structures but
own the opposite values. What is more, the effects of initial parameters and sign of $t$ on the potential fixed
points of the system are examined and presented in Fig.~\ref{fig7_para-influence-General-case}.
One can clearly read that the $\mathrm{FP}^{\pm}_2$ vanishes due to strong interplay of interactions, and
the other kinds of fixed points compete strongly for both $t>0$ and $t<0$ as varying the
initial values of interaction parameters. The $t>0$ case displays a fierce competition between
distinct fixed points. In comparison, some interaction parameters play a more important role than others
in reshaping the fixed points for $t<0$. Particularly, $\lambda_{20}$ is helpful to
$\mathrm{FP}^-_{1}$, while $\lambda_{21}$, $\lambda_{22}$, and $\lambda_{23}$ prefer to drive the system
to $\mathrm{FP}^-_{4}$.

\begin{table*}[htbp]
\caption{Potential candidate instabilities and phases nearby the fixed points induced
by electron-electron interactions~\cite{Vafek2010PRB,Vafek2014PRB}.
Hereby, SC and FM serve as superconductivity and ferromagnetism,
QAH and QSH denote the quantum anomalous Hall state and quantum spin Hall,
as well as Nem and NSN correspond to the nematic and nematic-spin-nematic order, respectively.}\label{table:phase}
\vspace{0.3cm}
\resizebox{2\columnwidth}{!}{%
\setlength{\tabcolsep}{4mm}{
\begin{ruledtabular}
\begin{tabular}{@{}cccccc@{}}
\multicolumn{2}{c}{P-H charge channel} &
  \multicolumn{2}{c}{P-H spin channel} &
  \multicolumn{2}{c}{P-P channel} \\ \hline
$\tau_0 \otimes   \textbf{1}_{2\times 2}$ & \multicolumn{1}{c}{charge instability} & $\tau_0 \otimes \vec{\sigma}$ & FM & $\tau_0 \otimes \sigma_2$ & chiral SC-I \\
$\tau_1 \otimes   \textbf{1}_{2\times 2}$ & \multicolumn{1}{c}{Nem.(bond)} & $\tau_1 \otimes \vec{\sigma}$ & NSN(bond) & $\tau_1 \otimes \sigma_2$ & chiral SC-II \\
$\tau_2 \otimes   \textbf{1}_{2\times 2}$ & \multicolumn{1}{c}{QAH} & $\tau_2\otimes \vec{\sigma}$ & QSH & $\tau_3 \otimes \sigma_2$ & s-wave SC \\
$\tau_3 \otimes   \textbf{1}_{2\times 2}$ & \multicolumn{1}{c}{Nem.(site)} & $\tau_3 \otimes \vec{\sigma}$ & NSN(site) & $\tau_2 \otimes \sigma_{0,1,3}$ & triplet SC \\
\end{tabular}%
\end{ruledtabular}}}
\end{table*}

Before going further, we make brief comments on the underlying fixed points.
Compared to the spinless case~\cite{Vafek2014PRB,Wang2017PRB}, the close interplay of spinful interactions gives rise
to more systematical results including all potential fixed points and their complicated competition
in the low-energy regime. Besides, three new fixed points including $\mathrm{FP}_{41}^\pm$,
$\mathrm{FP}_{42}^\pm$, and $\mathrm{FP}_{43}^\pm$ can be developed by the intimate interplay of
spinful electron-electron interactions. Armed with these in hand, we can expect potential instabilities
around distinct kinds of fixed points, which we are going to deliver in the forthcoming
section~\ref{Sec_instab_PT}.

\begin{figure*}[htbp]
\centering
\subfigure[]{\includegraphics[width = \columnwidth]{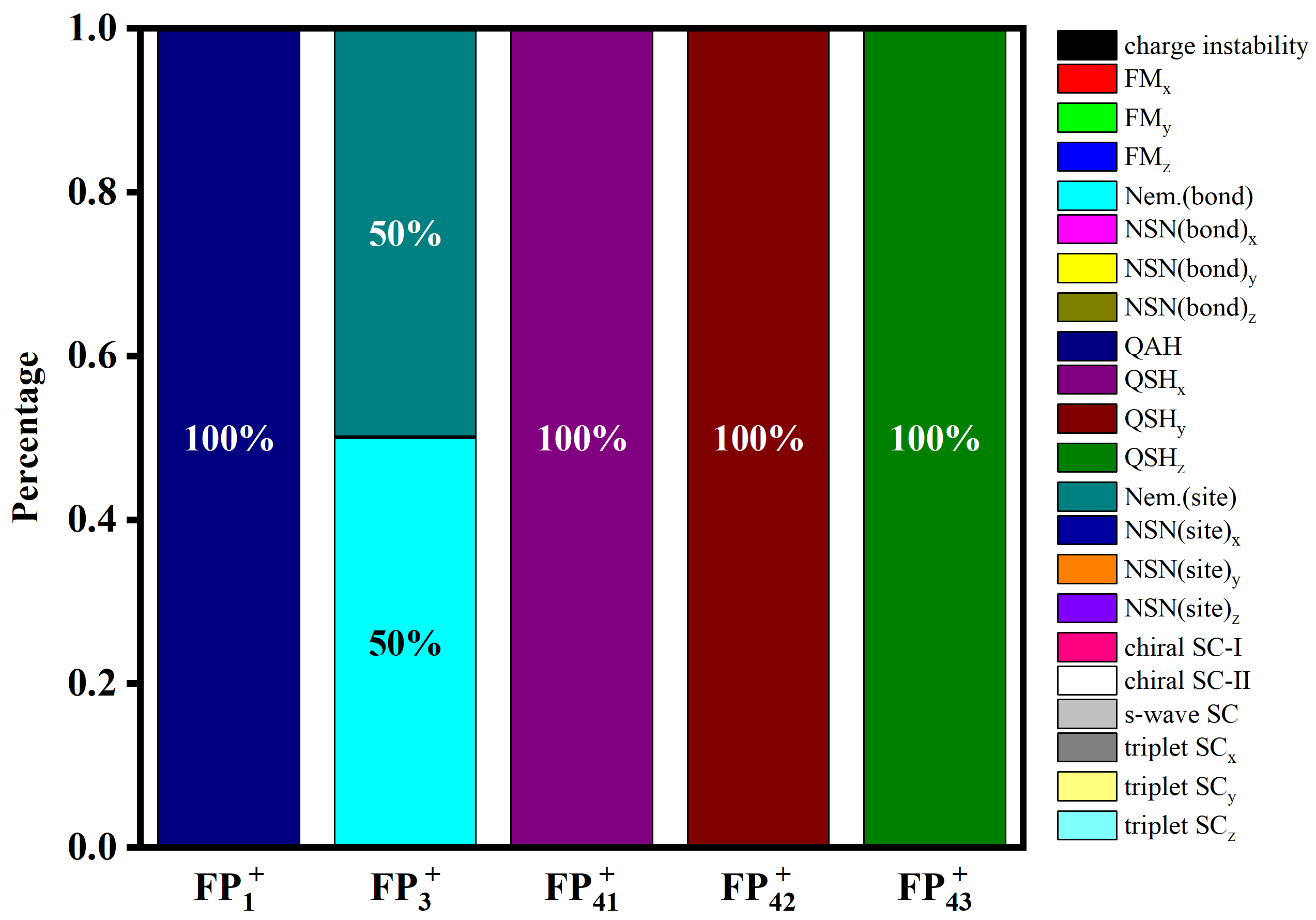}}\hspace{0.25cm}
\subfigure[]{\includegraphics[width = \columnwidth]{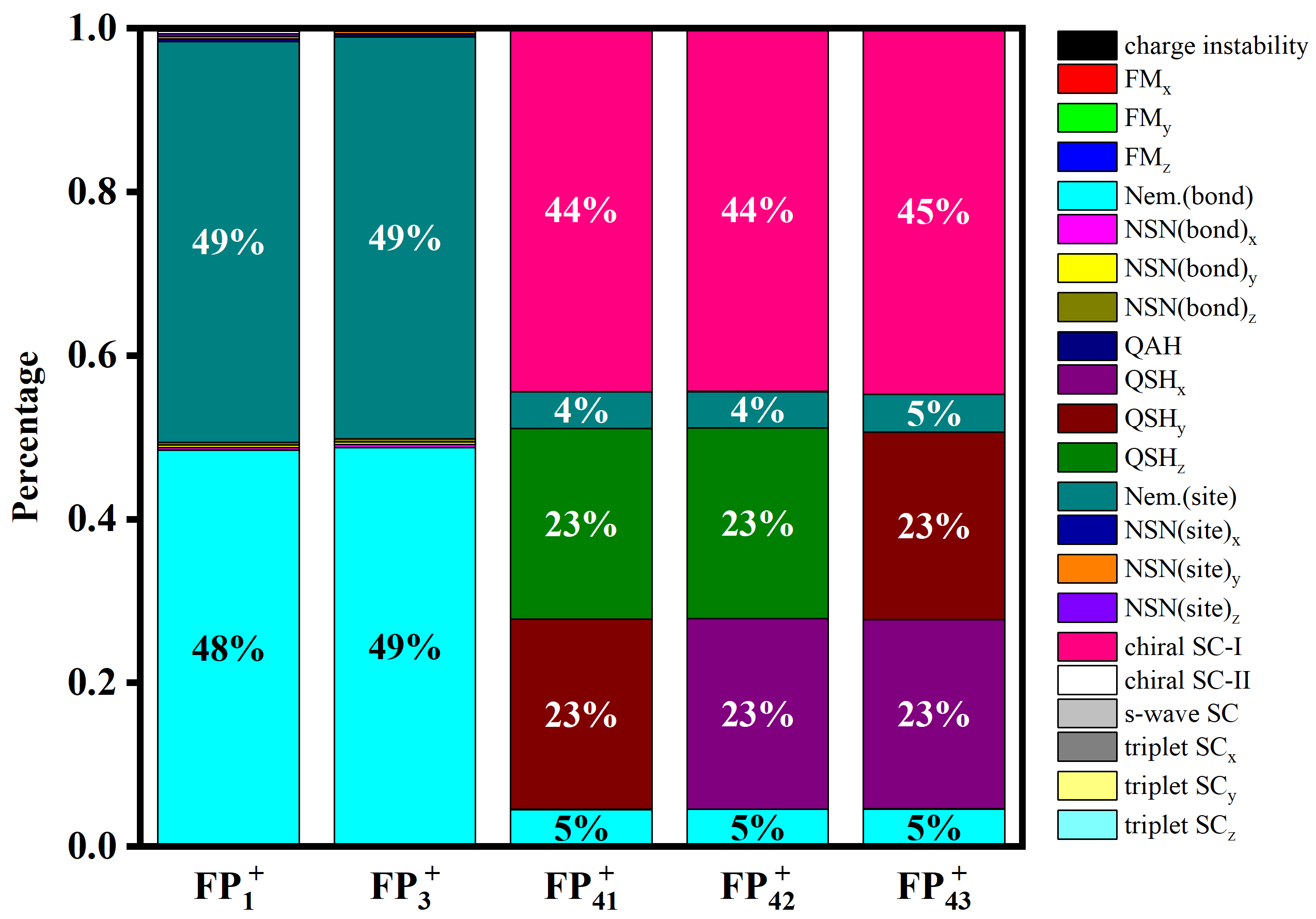}}\\ \vspace{0.1cm}
\subfigure[]{\includegraphics[width = \columnwidth]{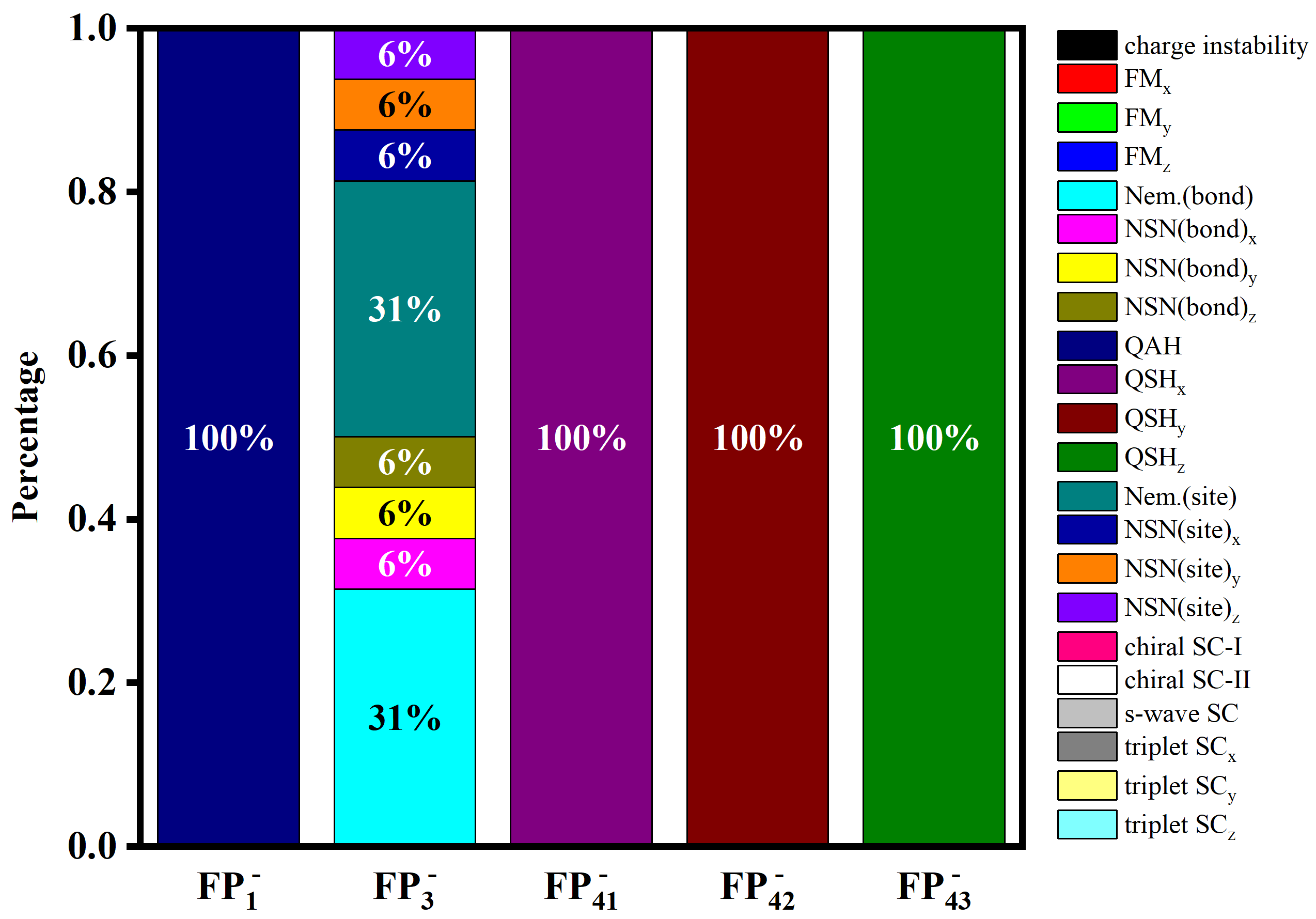}}\hspace{0.25cm}
\subfigure[]{\includegraphics[width = \columnwidth]{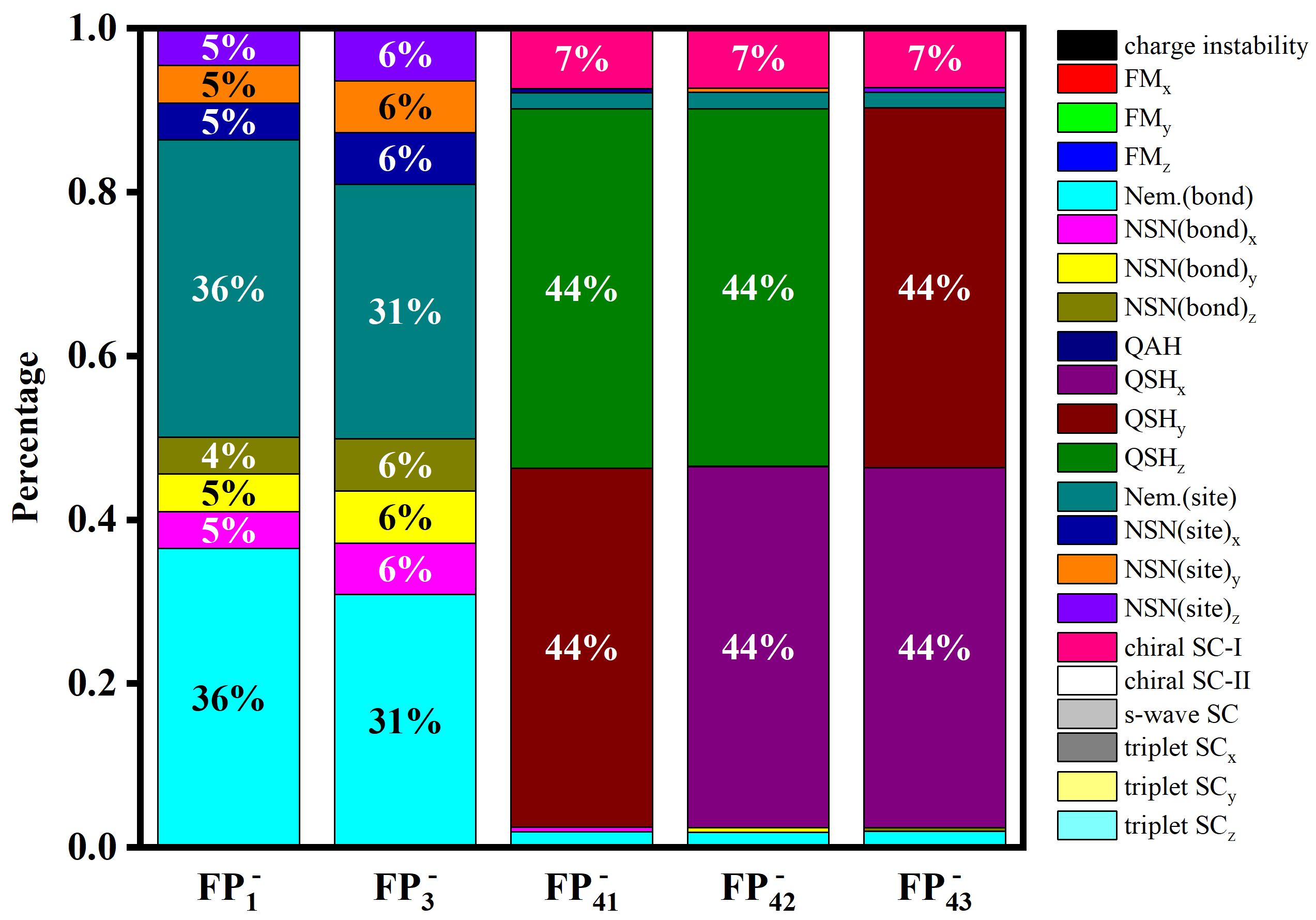}} \\
\vspace{0.05cm}
\caption{(Color online) Stabilities of (a) the leading phases at $t>0$, (b) the subleading phases at $t>0$,
(c) the leading phase at $t<0$, and (d) the subleading phase at $t<0$ nearby the fixed points in the General case
measured by the percentages with variation of initial conditions.}\label{fig:phase_FP_16_t-plus}
\end{figure*}

\section{Instabilities and phase transitions}\label{Sec_instab_PT}

Through a systematical analysis of the coupled RG equations~(\ref{Eq_RG-1})-(\ref{Eq_RG-2}) in
Sec.~\ref{Sec_Fixed_Points}, we present that the 2D QBCP system is attracted by
a series of fixed points (i.e., $\mathrm{FP}_{1,2,3,4}^\pm$) for all three distinct cases
due to the electron-electron interactions, which are primarily dependent upon the initial conditions.
Particularly, parts of the electron-electron couplings go towards divergence as approaching these
fixed points shown in Fig.~\ref{fig1_Limit_case_flows}.
In principle, such divergences are of close association with certain instabilities and
well-trodden signals for symmetry breakings~\cite{Cvetkovic2012PRB,Vafek2014PRB,Wang2017PRB, Maiti2010PRB, Altland2006Book,Vojta2003RPP,Halboth2000RPL,Halboth2000RPB, Eberlein2014PRB,Chubukov2012ARCMP,Nandkishore2012NP,Chubukov2016PRX,Roy2018RRX,Wang2020NPB}.
Accordingly, an important question naturally arises which instabilities and phase transitions
with certain symmetry breakings are dominant and preferable around these fixed points.
Clarifying this inquiry would be of particular help to improve our understandings on the low-energy
properties of 2D QBCP materials.

\subsection{Source terms and susceptibilities}

In order to examine the behaviors nearby the fixed points,
we adopt the following source terms consisting of fermionic bilinears to
characterize the potential candidates of
instabilities~\cite{Vafek2010PRB,Vafek2014PRB,Roy2018RRX,Roy2009.05055}  %for the charge and spin channels
\begin{eqnarray}
S_{\mathrm{sou}}
&=&\int d\tau \int d^2 \mathbf{x}
\left[
\sum_{\mu\nu}\Delta_{\mu\nu}^{\mathrm{PH}}\Psi^{\dagger}\mathcal{M}_{\mu\nu}\Psi \right.\nonumber\\
&&\left.+ \sum_{\mu\nu}\left(\Delta_{\mu\nu}^{\mathrm{PP}}\Psi^{\dagger}\mathcal{M}_{\mu\nu}\Psi^*+\mathrm{h.c.}\right)\right].\label{Eq_source-term}
\end{eqnarray}
Here, the matrix $\mathcal{M}_{\mu\nu}\equiv\tau_{\mu}\otimes\sigma_{\nu}$ with $\tau$ and $\sigma$ acting on space and spin
serve as the fermion bilinears for the candidates of symmetry breakings for our system as explicitly collected in
Table.~\ref{table:phase}~\cite{Vafek2010PRB,Vafek2014PRB}. In addition, $\Delta_{\mu\nu}^{\mathrm{PH/PP}}$
correspond to the strength of related fermion-source terms for the particle-hole and particle-particle channels,
respectively.

To proceed, the susceptibilities that are linked to the instabilities can be expressed as follows~\cite{Vafek2010PRB,Vafek2014PRB}
\begin{eqnarray}
\delta\chi_{\mu\nu}(l) = -\frac{\partial^{2}f}{\partial\Delta^{\mathrm{PH}/\mathrm{PP}}_{\mu\nu}(0)
\partial\Delta^{\ast\mathrm{PH}/\mathrm{PP}}_{\mu\nu}(0)},\label{Suscep}
\end{eqnarray}
where $f$ specifies the free energy density.
In order to identify the very dominant instabilities, we need to obtain the energy-dependent
susceptibilities as accessing the fixed points. To this end, we add the source terms~(\ref{Eq_source-term})
into our effective action~(\ref{Eq_S_eff}) and then derive the related
RG equations of $\Delta^{\mathrm{PH}/\mathrm{PP}}_{\mu\nu}$ by paralleling the analysis in Sec.~\ref{Sec_RGEqs},
which are provided in Appendix~\ref{Appendix_1L-Delta_i} for the details.

At current stage, as approaching certain fixed point, the energy-dependent
susceptibilities can be obtained via combining the RG evolutions of both fermionic couplings~(\ref{Eq_RG-1})-(\ref{Eq_RG-2}) and source
terms~(\ref{Eq_source-RG-1})-(\ref{Eq_source-RG-2}) and the relationship in Eq.~(\ref{Suscep}). To proceed, we are able to select the dominant instabilities from the candidates in Table.~\ref{table:phase} in that the ground state
can be characterized by the susceptibility with the strongest
divergence~\cite{Cvetkovic2012PRB,Vafek2014PRB,Wang2017PRB, Maiti2010PRB, Altland2006Book,
Vojta2003RPP,Halboth2000RPL,Halboth2000RPB, Eberlein2014PRB,Chubukov2012ARCMP,Nandkishore2012NP,
Chubukov2016PRX,Roy2018RRX,Nandkishore2008.05485}.
Before going further, it is of particular importance to emphasize that all the phases listed in
Table.~\ref{table:phase} are the potential candidates for an instability induced by some fixed point,
and accordingly, not all of them happen simultaneously, but instead only one of
them would win the competition and become the leading instability.
The corresponding results for three distinct cases will be addressed one by one in the following.

\subsection{Leading and subleading instabilities}

\subsubsection{Limit case}\label{Subsubsec_Limit_case}

At first, we consider the Limit case. As shown in Sec.~\ref{Subsection_FP_Limit}, there exist two types of
fixed points, namely $\mathrm{FP}^+_1$ for $t>0$ and $\mathrm{FP}^-_2$ for $t<0$, respectively.

Fig.~\ref{fig:FP12-sus} presents the energy-dependent susceptibilities as the system approaches such two fixed points.
We can clearly read from Fig.~\ref{fig:FP12-sus} that the leading instability corresponds to the QAH phase around
$\mathrm{FP}^+_1$ but instead the isotropic QSH phase (with the contributions
from $x,y,z$ directions being degenerate) in the vicinity of
$\mathrm{FP}^-_2$. Besides, it is also of particular importance to comment on
the subleading instabilities, which are currently subordinate to the leading ones but may compete
with the leading ones and dominate over them under certain adjusted conditions. Clearly,
the subleading phases for Limit case are the Nem.site(bond) and chiral SC-I for accessing
$\mathrm{FP}^+_1$ and $\mathrm{FP}^-_2$, respectively.

\subsubsection{Special case}\label{Subsubsec_Special_case}

Subsequently, we move to the Special case which owns three distinct sorts of fixed points including $\mathrm{FP}^+_1$,
$\mathrm{FP}^+_2$ and $\mathrm{FP}^+_3$.

With respect to $t>0$, paralleling the analysis in Sec.~\ref{Subsubsec_Limit_case},
we notice that the leading instabilities around $\mathrm{FP}_1^+$, $\mathrm{FP}^+_2$ and $\mathrm{FP}^+_3$, are occupied by
the QAH, the isotropic $\mathrm{QSH}$ ($\mathrm{QSH}_{x}$, $\mathrm{QSH}_{y}$, and $\mathrm{QSH}_{z}$ are equivalent),
and Nem.(site)/Nem.(bond) (these two phases are degenerate), respectively.

In addition, as mentioned in Sec.~\ref{Sec_Fixed_Points}, each fixed point governs a regime
in the interaction-parameter space. In this sense, we need to examine the stability of leading phase
for certain fixed point with variation of the initial conditions.  Fig.~\ref{fig:phase_FP_Specical_case}(a)
displays the proportion of leading phases around three fixed points with tuning the
initial values of interaction parameters ranging from $10^{-2}$ to $10^{-7}$. This implies that the leading
phases of such three fixed points are adequately stable.

In comparison, the subleading phases around these three fixed points displayed
in Fig.~\ref{fig:phase_FP_Specical_case}(b) are much more susceptible to the starting conditions.
We can find that there are multiple candidates for subleading phases nearby $\mathrm{FP}_1^+$ and
chiral SC-II/s-wave SC and Nem.(site)/Nem.(bond) take a slight advantage. As to
$\mathrm{FP}_3^+$, it is similar to $\mathrm{FP}_1^+$, but for $\mathrm{FP}_2^+$ the chiral SC-I dominates
the subleading phase.

For completeness, we provide several comments on the $t<0$ situation.
In analogous to their $t>0$ counterparts, the leading phases are robust enough.
Fig.~\ref{fig:phase_FP_Specical_case}(c) suggests that the basic results for $\mathrm{FP}_2^-$/$\mathrm{FP}_3^-$ are
similar to those of $\mathrm{FP}_2^+$/$\mathrm{FP}_3^+$, while the Nem.site(bond) state around $\mathrm{FP}_1^{-}$
wins the competition among other phases in $t>0$ case and become the manifestly subleading phases.

\begin{table*}[]
\caption{Collections of the leading (blue) and subleading (red) phases as approaching the corresponding fixed points
for both $t>0$ and $t<0$ situations. Hereby, L, S, and G cases are abbreviations for the Limit, Special,
and General cases, respectively.}\label{table-phase-summary}
\vspace{0.3cm}
\resizebox{\linewidth}{!}{%
\begin{tabular}{@{}ccccccc@{}}
\hline
\hline
\renewcommand{\arraystretch}{1}
\rule{0pt}{13pt}&
  $\mathrm{FP_{1}^+}$ \vspace{4pt}&
  $\mathrm{FP_{2}^+}$ &
  $\mathrm{FP_{3}^+}$ &
  $\mathrm{FP_{41}^+}$ &
  $\mathrm{FP_{42}^+}$ &
  $\mathrm{FP_{43}^+}$  \\ \hline
\rule{0pt}{25pt}L case &
  \begin{tabular}[c]{@{}c@{}}\blue{QAH}\\\rule{0pt}{13pt}     \red{Nem.(site)/(bond)}\end{tabular} &
  --- &
  --- &
  --- &
  --- &
  --- \\\rule{0pt}{1pt} \\\hline
\rule{0pt}{30pt}S case &
  \begin{tabular}[c]{@{}c@{}}\blue{QAH}\\\rule{0pt}{13pt}    \red{Nem.(site)/(bond)} \\ \red{Chiral SC II/s-wave SC}\end{tabular} &
  \begin{tabular}[c]{@{}c@{}}\blue{$\mathrm{QSH}_{xyz}$}\\\rule{0pt}{13pt}         \red{Chiral SC I}\end{tabular} &
  \begin{tabular}[c]{@{}c@{}}\blue{Nem.(site)/(bond)}\\\rule{0pt}{13pt}         \red{NSN.(site)/(bond)}\end{tabular} &
  --- &
  --- &
  --- \\\rule{0pt}{1pt} \\\hline
\rule{0pt}{27pt}G case &
  \begin{tabular}[c]{@{}c@{}}\blue{QAH}\\\rule{0pt}{13pt}         \red{Nem.(site)/(bond)}\end{tabular} &
  --- &
  \begin{tabular}[c]{@{}c@{}}\blue{Nem.(site)/(bond)}\\\rule{0pt}{13pt}         \red{Nem.(site)/(bond)}\end{tabular} &
  \begin{tabular}[c]{@{}c@{}}\blue{$\mathrm{QSH}_{x}$}\\\rule{0pt}{13pt}         \red{$\mathrm{QSH}_{y,z}$/Chiral SC I}\end{tabular} &
  \begin{tabular}[c]{@{}c@{}}\blue{$\mathrm{QSH}_{y}$}\\\rule{0pt}{13pt}         \red{$\mathrm{QSH}_{z,x}$/Chiral SC I}\end{tabular} &
  \begin{tabular}[c]{@{}c@{}}\blue{$\mathrm{QSH}_{z}$}\\\rule{0pt}{13pt}
  \red{$\mathrm{QSH}_{x,y}$/Chiral SC I}\end{tabular} \\\rule{0pt}{1pt} \\\hline
%\hline
\hline
\rule{0pt}{13pt}&
  $\mathrm{FP_{1}^-}$ \vspace{4pt}&
  $\mathrm{FP_{2}^-}$ &
  $\mathrm{FP_{3}^-}$ &
  $\mathrm{FP_{41}^-}$ &
  $\mathrm{FP_{42}^-}$ &
  $\mathrm{FP_{43}^-}$ \\ \hline
\rule{0pt}{25pt}L case &
  --- &
  \begin{tabular}[c]{@{}c@{}}\blue{$\mathrm{QSH}_{xyz}$}\\\rule{0pt}{13pt}     \red{Chiral SC I}\end{tabular} &
  --- &
  --- &
  --- &
  --- \\\rule{0pt}{1pt} \\\hline
\rule{0pt}{30pt}S case &
  \begin{tabular}[c]{@{}c@{}}\blue{QAH}\\\rule{0pt}{13pt}    \red{Nem.(site)/(bond)} \end{tabular} &
  \begin{tabular}[c]{@{}c@{}}\blue{$\mathrm{QSH}_{xyz}$}\\\rule{0pt}{13pt}         \red{Chiral SC I}\end{tabular} &
  \begin{tabular}[c]{@{}c@{}}\blue{Nem.(site)/(bond)}\\\rule{0pt}{13pt}         \red{NSN.(site)/(bond)}\end{tabular} &
  --- &
  --- &
  --- \\\rule{0pt}{1pt} \\\hline
\rule{0pt}{30pt}G case &
  \begin{tabular}[c]{@{}c@{}}\blue{QAH}\\        \red{Nem.(site)/(bond)} \\\red{NSN.(site)/(bond)} \end{tabular} &
  --- &
  \begin{tabular}[c]{@{}c@{}}\blue{Nem.(site)/(bond)}\\\blue{NSN.(site)/(bond)}\\        \red{Nem.(site)/(bond)}\\\red{NSN.(site)/(bond)}\end{tabular} &
  \begin{tabular}[c]{@{}c@{}}\blue{$\mathrm{QSH}_{x}$}\\\rule{0pt}{13pt}         \red{$\mathrm{QSH}_{y,z}$}\end{tabular} &
  \begin{tabular}[c]{@{}c@{}}\blue{$\mathrm{QSH}_{y}$}\\\rule{0pt}{13pt}         \red{$\mathrm{QSH}_{z,x}$}\end{tabular} &
  \begin{tabular}[c]{@{}c@{}}\blue{$\mathrm{QSH}_{z}$}\\\rule{0pt}{13pt}         \red{$\mathrm{QSH}_{x,y}$}\end{tabular} \\\rule{0pt}{1pt} \\\hline\hline

\end{tabular}%
}
\end{table*}

\subsubsection{General case}

At last, let us put our focus on the General case. In this circumstance, it shows in Sec.~\ref{Subsec_General-case-FP}
that both $\mathrm{FP}^\pm_1$ and $\mathrm{FP}^\pm_3$ can be reached as well, but $\mathrm{FP}^\pm_2$ are
replaced by three new fixed points including $\mathrm{FP}^\pm_{41}$, $\mathrm{FP}^\pm_{42}$ and $\mathrm{FP}^\pm_{43}$.

Considering $t>0$, we carry out the analogous analysis in Sec.~\ref{Subsubsec_Special_case} and then
figure out that the most preferable states that the system flows towards around $\mathrm{FP}^+_1$ and $\mathrm{FP}^+_3$
are still the QAH and Nem.(site)/Nem.(bond), respectively.
However, in sharp contrast, the leading instability accompanied by $\mathrm{FP}^+_{41}$
corresponds to the $\mathrm{QSH}_x$, in which the $\mathrm{QSH}$ susceptibility
becomes anisotropic and the $x-$direction component dominates over the other two directions.
Similarly, $\mathrm{QSH}_y$ and $\mathrm{QSH}_z$ occupy the most favorable phases in the vicinity of
$\mathrm{FP}^+_{42}$ and $\mathrm{FP}^+_{43}$, respectively.
It is therefore of remark significance to point out that the rotation symmetry of spin space is broken by the
spinful electron-electron interactions. As a result, the $\mathrm{QSH}_{x,y,z}$
are no longer degenerate but instead split and become anisotropic.
Again, we investigate the stabilities of leading phases and present Fig.~\ref{fig:phase_FP_16_t-plus}(a)
to show that these leading phases are stable under the variation of initial conditions.

In addition, we briefly give several comments on the subleading phases around these fixed points.
Comparing with the Special case where several subleading instabilities are observed around $\mathrm{FP}^+_1$
and $\mathrm{FP}^+_3$, Fig.~\ref{fig:phase_FP_16_t-plus}(b) indicates that only Nem.(site) and Nem.(bond)
compete for the subleading phases in General case.
However, a number of phases including the other two components of $\mathrm{QSH}$ as well as chiral SC-I
have an opportunity to be the subleading instabilities nearby $\mathrm{FP}^+_{41}$, $\mathrm{FP}^+_{42}$,
and $\mathrm{FP}^+_{43}$.

As to the $t<0$ situation, Fig.~\ref{fig:phase_FP_16_t-plus}(c) shows that the leading phases for
$\mathrm{FP}^-_1$, $\mathrm{FP}^-_{41}$, $\mathrm{FP}^-_{42}$,
and $\mathrm{FP}^-_{43}$ are analogous to their $t>0$ case. But rather for $\mathrm{FP}^-_3$,
there are additional candidates including Nem.site(bond) and NSN.site(bond) that
compete for the leading phases. Besides, we notice from Fig.~\ref{fig:phase_FP_16_t-plus}(d) that
the other two QSH components dominate over the chiral SC-I and have a bigger chance to be the subleading phases
around $\mathrm{FP}^-_{41}$, $\mathrm{FP}^-_{42}$,
and $\mathrm{FP}^-_{43}$. Different from the $t>0$ case, there exist more phases can be the candidates for
the subleading states for $\mathrm{FP}^-_1$ and $\mathrm{FP}^-_3$.

To recapitulate, Table~\ref{table-phase-summary} summarizes our basic conclusions for the leading and subleading instabilities
around all the potential fixed points induced by spinful electron-electron interactions.

\subsection{Brief discussions}

Before closing this section, we would like to address several comments on the basic results.
On one hand, the inclusion of spinful electron-electron interactions, as compared to the
spinless case~\cite{Vafek2014PRB,Wang2017PRB}, can be capable of generating more fixed points including
$\mathrm{FP}^\pm_{1,2,3}$ and $\mathrm{FP}^\pm_{41,42,43}$ as presented in Sec.~\ref{Sec_Fixed_Points},
which dictate the low-energy fate of the 2D QBCP system.
On the other hand, as approaching these fixed points, we find that a series of instabilities can be induced by
the spinful electron-electron interactions as catalogued in Table~\ref{table-phase-summary}.
As to the leading phases, in addition to the QAH and isotropic QSH~\cite{Vafek2014PRB},
the 2D QBCP system can undergo a phase transition to either an anisotropic QSH or a Nem.site(bond) state.
Besides, a plethora of candidate instabilities exhibited in Table~\ref{table-phase-summary}
endeavor to run for the subleading phases, which can compete with the leading ones and may become
dominant instabilities under certain modified conditions.
To wrap up, the spinful electron-electron interactions play an essential role in
inducing the underlying instabilities and
reshaping the low-energy behavior of 2D QBCP materials.

Subsequently, let us address several underlying explanations for these new behavior.
Fixing a certain model, taking into account more or less physical ingredients is of particular importance
to reveal the low-energy behavior. In Ref.~\cite{Fradkin2009PRL}, the authors considered the spin effects but worked
at the mean-field level without including the quantum fluctuations, which typically provide basic contributions. Although Ref.~\cite{Vafek2014PRB} considered spin effects, the authors used a $2 \times 2$ spinor to describe the quasiparticle, implying that the contributions from spin-up and spin-down are treated equally in low-energy properties. Consequently, the spinful effects and their interplay with electron-electron interactions cannot be fully included.
Working in the $2\times 2$ space implies that the spinful effects may only be partially taken into account.
In sharp contrast, we explicitly employ a $4$-component spinor to characterize the low-energy
excitations and work in the $4 \times 4$ space.  This approach necessitates dealing with 16 components of interaction couplings compared to 4 couplings in previous works~\cite{Fradkin2009PRL, Vafek2014PRB}. Accordingly, our renormalization group (RG) equations incorporate one-loop corrections beyond the mean-field level, fully capturing the spinful ingredients to provide more accurate physical information.

\section{Summary}\label{Sec_summary}

In summary, our study presents a systematical investigation of the interplay of sixteen types of marginal
spinful electron-electron interactions and the low-energy instabilities of 2D QBCP semimetals
by virtue of the RG approach~\cite{Shankar1994RMP,Wilson1975RMP,Polchinski9210046}.
After considering all one-loop corrections, we establish the energy-dependent RG evolutions of all
interaction parameters, which are closely coupled and dictate the
low-energy physics of 2D QBCP system. A detailed numerical analysis addresses a series of
interesting behaviors induced by these interactions that exhibit significant differences
compared to those of the spinless situation.

To begin with, we find that the 2D QBCP systems are attracted by several distinct kinds of fixed points
in the interaction-parameter space. In particular, they are heavily dependent upon the initial conditions, including the
value of interaction parameters and structure parameter $t$. These fall into three categories consisting of Limit case,
Special case, and General case as demonstrated in Sec.~\ref{Sec_Fixed_Points}. Specifically, there exist the fixed
points $\mathrm{FP}_1^{+}$ and $\mathrm{FP}_2^{-}$ in the Limit case,
but instead $\mathrm{FP}_1^{\pm}$, $\mathrm{FP}_2^{\pm}$, and $\mathrm{FP}_3^{\pm}$ in the Special case.
In contrast, the General case gives rise to $\mathrm{FP}_1^{\pm}$, $\mathrm{FP}_3^{\pm}$, and
$\mathrm{FP}_{41,42,43}^{\pm}$. Besides, the stabilities of fixed points are also provided in
Figs.~\ref{fig4_single-para-influence-Specical-case}-\ref{fig6_FP_456}
with the variation of parts of interaction parameters. In principle, certain instabilities with certain symmetry breakings
that are accompanied by phase transitions can be expected as approaching these fixed points.
Subsequently, we bring out the source terms composed of the fermionic bilinears
to capture the potential instabilities~\cite{Vafek2010PRB,Vafek2014PRB,Roy2018RRX,Roy2009.05055}.
After evaluating the susceptibilities of all
candidate states by combining the source terms and RG equations of interaction parameters,
we find that the spinful fermion-fermion interactions can induce sorts of
favorable instabilities in the vicinity of these fixed points as summarized in Table~\ref{table-phase-summary}.
In the vicinity of $\mathrm{FP}_{1}^{\pm}$, $\mathrm{FP}_{2}^{\pm}$, and $\mathrm{FP}_{3}^{\pm}$,
it clearly indicates that the QAH, QSH, and Nem.site(bond) states are dominant,
and correspondingly, Nem.site(bond), Chiral SC-I, and NSN.site(bond) are
the most probable candidates to run for the subleading phases, respectively.
In comparison, QSH becomes anisotropic nearby the $\mathrm{FP}_{41,42,43}^{\pm}$, around which only one
component of QSH plays a leading role but the other two components only own
the chance to compete for the subleading instabilities with Chiral SC-I.
To be brief, the spinful fermion-fermion interactions are of particular importance to pinpoint
the low-energy behavior of 2D QBCP materials. Compared to the spinless
case~\cite{Vafek2014PRB,Wang2017PRB}, the spinful fermion-fermion interactions and their intimate competitions bring a series of
new critical behavior in the low-energy regime, including more fixed points and more favorable phase transitions which are collected in
Table~\ref{table-phase-summary}. We wish these findings would be instructive to
improve our understandings of 2D QBCP semimetals and helpful to study the analogous materials.

\section*{ACKNOWLEDGEMENTS}

We thank Wen-Hao Bian for the helpful discussions. J.W. was partially supported by the National Natural
Science Foundation of China under Grant No. 11504360.

\appendix

\section{One-loop corrections to the electron-electron couplings}\label{Appendix-one-loop-corrections}

\begin{figure}[htbp]
\centering
\includegraphics[scale=0.09]{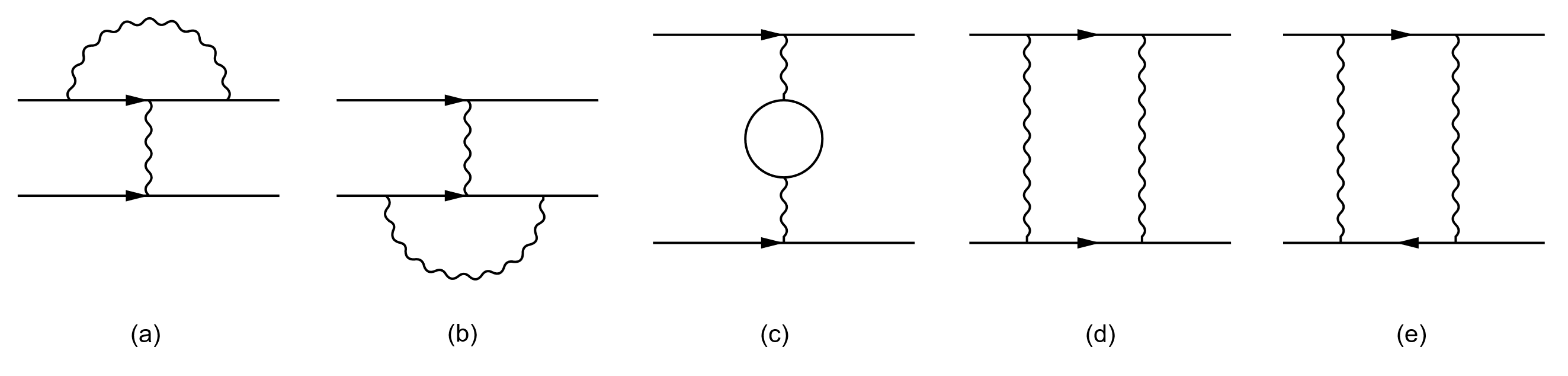}\\ \vspace{-0.2cm}
\caption{One-loop corrections to the electron-electron interaction couplings (a)-(e)
due to the electron-electron interactions. The solid and wavy lines denote the electronic
propagator and electron-electron interaction, respectively.}\label{Fig_Feynman Diagram}
\end{figure}

On the basis of our effective field action~(\ref{Eq_S_eff}), the one-loop diagrams that contribute
to the electron-electron couplings are exhibited in Fig.~\ref{Fig_Feynman Diagram}.
After performing the long but standard calculations~\cite{Cvetkovic2012PRB,Vafek2014PRB,Wang2017PRB},
we are left with the following one-loop corrections to electron-electron interaction parameters
\begin{widetext}
\begin{eqnarray}
\delta S_{\lambda_{00}}
&=&\frac{-l}{8\pi \left|t\right|}(\lambda_{00} \lambda_{10} +\lambda_{01} \lambda_{11} +\lambda_{02} \lambda_{12} +\lambda_{03} \lambda_{13} +\lambda_{00} \lambda_{30} +\lambda_{01} \lambda_{31} +\lambda_{02} \lambda_{32} +\lambda_{03} \lambda_{33} )\mathcal{A}_{00},\label{Eq_A1}\\
\delta S_{\lambda_{01}}&=&\frac{-l}{8 \pi  \left|t\right|} ( \lambda_{00} \lambda_{11} -2\lambda_{02} \lambda_{03} +\lambda_{01} \lambda_{10} -2\lambda_{12} \lambda_{13} +\lambda_{00} \lambda_{31} +\lambda_{01} \lambda_{30} +\lambda_{12} \lambda_{23} +\lambda_{13} \lambda_{22}-2\lambda_{22} \lambda_{23}\nonumber\\&&+\lambda_{22} \lambda_{33} +\lambda_{23} \lambda_{32} -2\lambda_{32} \lambda_{33})\mathcal{A}_{01},\\
\delta S_{\lambda_{02}}&=&\frac{-l}{8 \pi  \left|t\right|} (\lambda_{00} \lambda_{12} -2\lambda_{01} \lambda_{03} +\lambda_{02} \lambda_{10} -2\lambda_{11} \lambda_{13} +\lambda_{00} \lambda_{32} +\lambda_{02} \lambda_{30} +\lambda_{11} \lambda_{23} +\lambda_{13} \lambda_{21} -2\lambda_{21} \lambda_{23}\nonumber\\&& +\lambda_{21} \lambda_{33} +\lambda_{23} \lambda_{31} -2\lambda_{31} \lambda_{33} )\mathcal{A}_{02},\\
\delta S_{\lambda_{03}}&=&\frac{-l}{8 \pi  \left|t\right|} (\lambda_{00} \lambda_{13} -2\lambda_{01} \lambda_{02} +\lambda_{03} \lambda_{10} -2\lambda_{11} \lambda_{12} +\lambda_{00} \lambda_{33} +\lambda_{03} \lambda_{30} +\lambda_{11} \lambda_{22} +\lambda_{12} \lambda_{21} -2\lambda_{21} \lambda_{22}\nonumber\\&& +\lambda_{21} \lambda_{32} +\lambda_{22} \lambda_{31} -2\lambda_{31} \lambda_{32} )\mathcal{A}_{03},\\
\delta S_{\lambda_{10}}&=&\frac{-l}{16\pi \left|t\right|}(\lambda_{00} \lambda_{00} -2\lambda_{00} \lambda_{10} +2\lambda_{00} \lambda_{20} +\lambda_{01} \lambda_{01} -2\lambda_{01} \lambda_{10} +2\lambda_{01} \lambda_{21} +\lambda_{02} \lambda_{02} -2\lambda_{02} \lambda_{10}
+2\lambda_{02} \lambda_{22}\nonumber\\&&
+\lambda_{03} \lambda_{03} -2\lambda_{03} \lambda_{10} +2\lambda_{03} \lambda_{23} +7\lambda_{10} \lambda_{10} -2\lambda_{10} \lambda_{11} -2\lambda_{10} \lambda_{12} -2\lambda_{10} \lambda_{13} + 2\lambda_{10} \lambda_{20}+2\lambda_{10} \lambda_{21} \nonumber\\&&
+2\lambda_{10} \lambda_{22} + 2\lambda_{10} \lambda_{23} +2\lambda_{10} \lambda_{30}
+ 2\lambda_{10} \lambda_{31} +2\lambda_{10} \lambda_{32} +2\lambda_{10} \lambda_{33} +\lambda_{11} \lambda_{11} +\lambda_{12} \lambda_{12} +\lambda_{13} \lambda_{13} +\lambda_{20} \lambda_{20}\nonumber\\&&-4\lambda_{20} \lambda_{30} +\lambda_{21} \lambda_{21} -4\lambda_{21} \lambda_{31} +\lambda_{22} \lambda_{22} -4\lambda_{22} \lambda_{32} +\lambda_{23} \lambda_{23} -4\lambda_{23} \lambda_{33}+\lambda_{30} \lambda_{30}+\lambda_{31} \lambda_{31} +\lambda_{32} \lambda_{32}\nonumber\\&& +\lambda_{33} \lambda_{33} )\mathcal{A}_{10},\\
\delta S_{\lambda_{11}}&=&\frac{-l}{8 \pi  \left|t\right|} (\lambda_{00} \lambda_{01} -\lambda_{00} \lambda_{11} -\lambda_{01} \lambda_{11} +\lambda_{02} \lambda_{11} +\lambda_{03} \lambda_{11} -2\lambda_{02} \lambda_{13} -2\lambda_{03} \lambda_{12} +\lambda_{00} \lambda_{21} +\lambda_{01} \lambda_{20} +\lambda_{11} \lambda_{12}\nonumber\\&& +\lambda_{11} \lambda_{13} +\lambda_{11} \lambda_{20} +\lambda_{11} \lambda_{21} -\lambda_{11} \lambda_{22} -\lambda_{11} \lambda_{23} +\lambda_{11} \lambda_{30} +\lambda_{20} \lambda_{21} +\lambda_{11} \lambda_{31} -\lambda_{11} \lambda_{32} -\lambda_{11} \lambda_{33}\nonumber\\&& +\lambda_{12} \lambda_{33} +\lambda_{13} \lambda_{32} -2\lambda_{20} \lambda_{31} -2\lambda_{21} \lambda_{30} +\lambda_{30} \lambda_{31} +3\lambda_{11}\lambda_{11} )\mathcal{A}_{11},\\
\delta S_{\lambda_{12}}&=&\frac{-l}{8 \pi  \left|t\right|} (\lambda_{00} \lambda_{02} -\lambda_{00} \lambda_{12} +\lambda_{01} \lambda_{12} -2\lambda_{01} \lambda_{13} -\lambda_{02} \lambda_{12} -2\lambda_{03} \lambda_{11} +\lambda_{03} \lambda_{12} +\lambda_{00} \lambda_{22} +\lambda_{02} \lambda_{20}\nonumber\\&& +\lambda_{11} \lambda_{12} +\lambda_{12} \lambda_{13} +\lambda_{12} \lambda_{20} -\lambda_{12} \lambda_{21} +\lambda_{12} \lambda_{22} -\lambda_{12} \lambda_{23} +\lambda_{12} \lambda_{30} +\lambda_{20} \lambda_{22} -\lambda_{12} \lambda_{31} +\lambda_{11} \lambda_{33}\nonumber\\&& +\lambda_{12} \lambda_{32} +\lambda_{13} \lambda_{31} -\lambda_{12} \lambda_{33} -2\lambda_{20} \lambda_{32} -2\lambda_{22} \lambda_{30} +\lambda_{30} \lambda_{32} +3\lambda_{12}\lambda_{12} )\mathcal{A}_{12},\\
\delta S_{\lambda_{13}}&=&\frac{-l}{8 \pi  \left|t\right|} (\lambda_{00} \lambda_{03} -\lambda_{00} \lambda_{13} -2\lambda_{01} \lambda_{12} -2\lambda_{02} \lambda_{11} +\lambda_{01} \lambda_{13} +\lambda_{02} \lambda_{13} -\lambda_{03} \lambda_{13} +\lambda_{00} \lambda_{23} +\lambda_{03} \lambda_{20}\nonumber\\&& +\lambda_{11} \lambda_{13} +\lambda_{12} \lambda_{13} +\lambda_{13} \lambda_{20} -\lambda_{13} \lambda_{21} -\lambda_{13} \lambda_{22} +\lambda_{13} \lambda_{23} +\lambda_{11} \lambda_{32} +\lambda_{12} \lambda_{31} +\lambda_{13} \lambda_{30}  +\lambda_{20} \lambda_{23}\nonumber\\&& -\lambda_{13} \lambda_{31}-\lambda_{13} \lambda_{32} +\lambda_{13} \lambda_{33} -2\lambda_{20} \lambda_{33} -2\lambda_{23} \lambda_{30} +\lambda_{30} \lambda_{33} +3\lambda_{13}\lambda_{13} )\mathcal{A}_{13},\\
\delta S_{\lambda_{20}}&=&\frac{-l}{8 \pi  \left|t\right|} (\lambda_{00} \lambda_{10} +\lambda_{01} \lambda_{11} +\lambda_{02} \lambda_{12} +\lambda_{03} \lambda_{13} -2\lambda_{00} \lambda_{20} -2\lambda_{01} \lambda_{20} -2\lambda_{02} \lambda_{20} -2\lambda_{03} \lambda_{20} +\lambda_{00} \lambda_{30}\nonumber\\&& +2\lambda_{10} \lambda_{20} +2\lambda_{11} \lambda_{20} +\lambda_{01} \lambda_{31} +2\lambda_{12} \lambda_{20} +2\lambda_{13} \lambda_{20} +\lambda_{02} \lambda_{32} +\lambda_{03} \lambda_{33} -2\lambda_{10} \lambda_{30} -2\lambda_{20} \lambda_{21}\nonumber\\&& -2\lambda_{11} \lambda_{31} -2\lambda_{20} \lambda_{22} -2\lambda_{20} \lambda_{23} -2\lambda_{12} \lambda_{32} -2\lambda_{13} \lambda_{33} +2\lambda_{20} \lambda_{30} +2\lambda_{20} \lambda_{31} +2\lambda_{20} \lambda_{32} +2\lambda_{20} \lambda_{33}\nonumber\\&& +6\lambda_{20}\lambda_{20} )\mathcal{A}_{20},\\
\delta S_{\lambda_{21}}&=&\frac{-l}{8 \pi  \left|t\right|} (\lambda_{00} \lambda_{11} +\lambda_{01} \lambda_{10} -2\lambda_{00} \lambda_{21} -2\lambda_{01} \lambda_{21} +2\lambda_{02} \lambda_{21} +2\lambda_{03} \lambda_{21} -2\lambda_{02} \lambda_{23} -2\lambda_{03} \lambda_{22}\nonumber\\&& +\lambda_{00} \lambda_{31} +\lambda_{01} \lambda_{30} +2\lambda_{10} \lambda_{21} +2\lambda_{11} \lambda_{21} -2\lambda_{12} \lambda_{21} -2\lambda_{13} \lambda_{21} +\lambda_{12} \lambda_{23} +\lambda_{13} \lambda_{22} -2\lambda_{10} \lambda_{31}\nonumber\\&& -2\lambda_{11} \lambda_{30} -2\lambda_{20} \lambda_{21} +2\lambda_{21} \lambda_{22} +2\lambda_{21} \lambda_{23} +2\lambda_{21} \lambda_{30} +2\lambda_{21} \lambda_{31} -2\lambda_{21} \lambda_{32} -2\lambda_{21} \lambda_{33}\nonumber\\&& +\lambda_{22} \lambda_{33} +\lambda_{23} \lambda_{32} +6\lambda_{21}\lambda_{21} )\mathcal{A}_{21},\\
\delta S_{\lambda_{22}}&=&\frac{-l}{8 \pi  \left|t\right|} (\lambda_{00} \lambda_{12} +\lambda_{02} \lambda_{10} -2\lambda_{00} \lambda_{22} +2\lambda_{01} \lambda_{22} -2\lambda_{01} \lambda_{23} -2\lambda_{02} \lambda_{22} -2\lambda_{03} \lambda_{21} +2\lambda_{03} \lambda_{22}\nonumber\\&& +\lambda_{00} \lambda_{32} +\lambda_{02} \lambda_{30} +2\lambda_{10} \lambda_{22} -2\lambda_{11} \lambda_{22} +\lambda_{11} \lambda_{23} +2\lambda_{12} \lambda_{22} +\lambda_{13} \lambda_{21} -2\lambda_{13} \lambda_{22} -2\lambda_{10} \lambda_{32}\nonumber\\&& -2\lambda_{12} \lambda_{30} -2\lambda_{20} \lambda_{22} +2\lambda_{21} \lambda_{22} +2\lambda_{22} \lambda_{23} +2\lambda_{22} \lambda_{30} -2\lambda_{22} \lambda_{31} +\lambda_{21} \lambda_{33} +2\lambda_{22} \lambda_{32}\nonumber\\&& +\lambda_{23} \lambda_{31} -2\lambda_{22} \lambda_{33} +6\lambda_{22}\lambda_{22} )\mathcal{A}_{22},\\
\delta S_{\lambda_{23}}&=&\frac{-l}{8 \pi  \left|t\right|} (\lambda_{00} \lambda_{13} +\lambda_{03} \lambda_{10} -2\lambda_{00} \lambda_{23} -2\lambda_{01} \lambda_{22} -2\lambda_{02} \lambda_{21} +2\lambda_{01} \lambda_{23} +2\lambda_{02} \lambda_{23} -2\lambda_{03} \lambda_{23}\nonumber\\&& +\lambda_{00} \lambda_{33} +\lambda_{03} \lambda_{30} +2\lambda_{10} \lambda_{23} +\lambda_{11} \lambda_{22} +\lambda_{12} \lambda_{21} -2\lambda_{11} \lambda_{23} -2\lambda_{12} \lambda_{23} +2\lambda_{13} \lambda_{23} -2\lambda_{10} \lambda_{33}\nonumber\\&& -2\lambda_{13} \lambda_{30} -2\lambda_{20} \lambda_{23} +2\lambda_{21} \lambda_{23} +2\lambda_{22} \lambda_{23} +\lambda_{21} \lambda_{32} +\lambda_{22} \lambda_{31} +2\lambda_{23} \lambda_{30} -2\lambda_{23} \lambda_{31}\nonumber\\&& -2\lambda_{23} \lambda_{32} +2\lambda_{23} \lambda_{33} +6\lambda_{23}\lambda_{23} )\mathcal{A}_{23},\\
\delta S_{\lambda_{30}}&=&\frac{-l}{16 \pi  \left|t\right|} (\lambda_{00}\lambda_{00} +2\lambda_{00} \lambda_{20} -2\lambda_{00} \lambda_{30} +\lambda_{01}\lambda_{01} +2\lambda_{01} \lambda_{21} -2\lambda_{01} \lambda_{30} +\lambda_{02}\lambda_{02} +2\lambda_{02} \lambda_{22}\nonumber\\&& -2\lambda_{02} \lambda_{30} +\lambda_{03}\lambda_{03} +2\lambda_{03} \lambda_{23} -2\lambda_{03} \lambda_{30} +\lambda_{10}\lambda_{10} -4\lambda_{10} \lambda_{20} +2\lambda_{10} \lambda_{30} +\lambda_{11}\lambda_{11} -4\lambda_{11} \lambda_{21}\nonumber\\&& +2\lambda_{11} \lambda_{30} +\lambda_{12}\lambda_{12} -4\lambda_{12} \lambda_{22} +2\lambda_{12} \lambda_{30} +\lambda_{13}\lambda_{13} -4\lambda_{13} \lambda_{23} +2\lambda_{13} \lambda_{30} +\lambda_{20}\lambda_{20} +2\lambda_{20} \lambda_{30}\nonumber\\&& +\lambda_{21}\lambda_{21} +2\lambda_{21} \lambda_{30} +\lambda_{22}\lambda_{22} +2\lambda_{22} \lambda_{30} +\lambda_{23}\lambda_{23} +2\lambda_{23} \lambda_{30} +7\lambda_{30}\lambda_{30} -2\lambda_{30} \lambda_{31} -2\lambda_{30} \lambda_{32}\nonumber\\&& -2\lambda_{30} \lambda_{33} +\lambda_{31}\lambda_{31} +\lambda_{32}\lambda_{32} +\lambda_{33}\lambda_{33} )\mathcal{A}_{30},\\
\delta S_{\lambda_{31}}&=&\frac{-l}{8 \pi  \left|t\right|} (\lambda_{00} \lambda_{01} +\lambda_{00} \lambda_{21} +\lambda_{01} \lambda_{20} +\lambda_{10} \lambda_{11} -\lambda_{00} \lambda_{31} -2\lambda_{10} \lambda_{21} -2\lambda_{11} \lambda_{20} -\lambda_{01} \lambda_{31} +\lambda_{02} \lambda_{31}\nonumber\\&& +\lambda_{03} \lambda_{31} -2\lambda_{02} \lambda_{33} -2\lambda_{03} \lambda_{32} +\lambda_{10} \lambda_{31} +\lambda_{20} \lambda_{21} +\lambda_{11} \lambda_{31} -\lambda_{12} \lambda_{31} -\lambda_{13} \lambda_{31} +\lambda_{12} \lambda_{33}  +\lambda_{13} \lambda_{32}\nonumber\\&&+\lambda_{20} \lambda_{31} +\lambda_{21} \lambda_{31} -\lambda_{22} \lambda_{31} -\lambda_{23} \lambda_{31} +\lambda_{31} \lambda_{32} +\lambda_{31} \lambda_{33} +3\lambda_{31}\lambda_{31} )\mathcal{A}_{31},\\
\delta S_{\lambda_{32}}&=&\frac{-l}{8 \pi  \left|t\right|} (\lambda_{00} \lambda_{02} +\lambda_{00} \lambda_{22} +\lambda_{02} \lambda_{20} +\lambda_{10} \lambda_{12} -\lambda_{00} \lambda_{32} -2\lambda_{10} \lambda_{22} -2\lambda_{12} \lambda_{20} +\lambda_{01} \lambda_{32}  -2\lambda_{01} \lambda_{33}\nonumber\\&& -\lambda_{02} \lambda_{32} -2\lambda_{03} \lambda_{31} +\lambda_{03} \lambda_{32} +\lambda_{10} \lambda_{32} +\lambda_{20} \lambda_{22} -\lambda_{11} \lambda_{32} +\lambda_{11} \lambda_{33} +\lambda_{12} \lambda_{32} +\lambda_{13} \lambda_{31} -\lambda_{13} \lambda_{32}\nonumber\\&& +\lambda_{20} \lambda_{32} -\lambda_{21} \lambda_{32} +\lambda_{22} \lambda_{32} -\lambda_{23} \lambda_{32} +\lambda_{31} \lambda_{32} +\lambda_{32} \lambda_{33} +3\lambda_{32}\lambda_{32} )\mathcal{A}_{32},\\
\delta S_{\lambda_{33}}&=&\frac{-l}{8 \pi  \left|t\right|} (\lambda_{00} \lambda_{03} +\lambda_{00} \lambda_{23} +\lambda_{03} \lambda_{20} +\lambda_{10} \lambda_{13} -\lambda_{00} \lambda_{33} -2\lambda_{01} \lambda_{32} -2\lambda_{02} \lambda_{31} -2\lambda_{10} \lambda_{23} -2\lambda_{13} \lambda_{20}\nonumber\\&& +\lambda_{01} \lambda_{33} +\lambda_{02} \lambda_{33} -\lambda_{03} \lambda_{33} +\lambda_{10} \lambda_{33} +\lambda_{11} \lambda_{32} +\lambda_{12} \lambda_{31} +\lambda_{20} \lambda_{23} -\lambda_{11} \lambda_{33} -\lambda_{12} \lambda_{33} +\lambda_{13} \lambda_{33}\nonumber\\&& +\lambda_{20} \lambda_{33} -\lambda_{21} \lambda_{33} -\lambda_{22} \lambda_{33} +\lambda_{23} \lambda_{33} +\lambda_{31} \lambda_{33} +\lambda_{32} \lambda_{33} +3\lambda_{33}\lambda_{33} )\mathcal{A}_{33}.\label{Eq_A16}
\end{eqnarray}
with
where $\mathcal{A}_{\mu\nu}$ are defined as
\begin{eqnarray}
\mathcal{A}_{\mu\nu}
&\equiv& \int_{-\infty}^{\infty}\frac{d\omega_{1}d\omega_{2}d\omega_{3}}{(2\pi)^3}
\int^{b\Lambda}\frac{d^2\mathbf{k}_{1}d^2\mathbf{k}_{2}d^2\mathbf{k}_{3}}{(2\pi)^6}
\Psi^{\dagger}(\omega_1,\mathbf{k}_1)\Sigma_{\mu\nu}\Psi(\omega_2,\mathbf{k}_2)\nonumber
\Psi^{\dagger}(\omega_3,\mathbf{k}_3)\nonumber\\
&&\times\Sigma_{\mu\nu}\Psi(\omega_1+\omega_2-\omega_3,\mathbf{k}_1
+\mathbf{k}_2-\mathbf{k}_3).\label{Eq_A_mu_nu}
\end{eqnarray}

\end{widetext}

\begin{figure}[htbp]
\hspace{-1.2cm}
\includegraphics[scale=0.1]{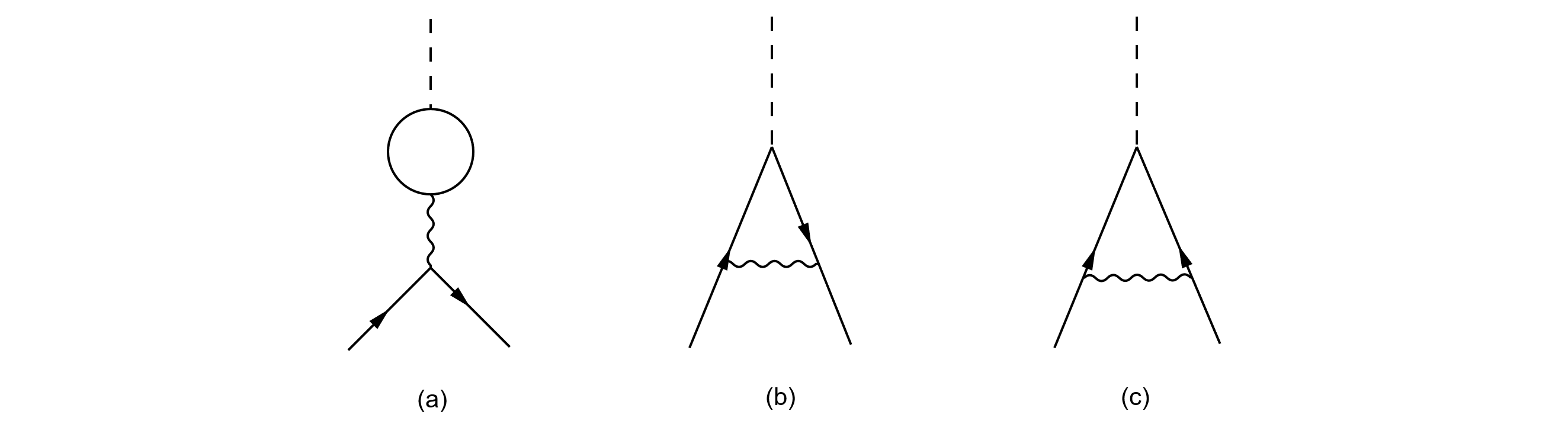} \\ \vspace{-0.1cm}
\caption{One-loop corrections to the bilinear fermionic source
terms for the particle-hole channel (a)-(b) and the particle-particle channel (c).
The solid, wave, and dash lines correspond to the electronic,
electron-electron interaction and source term,
respectively.}\label{Fig_Feynman Diagram_2}
\end{figure}

\section{One-loop flows of source terms}\label{Appendix_1L-Delta_i}

According to the effective field action~(\ref{Eq_S_eff}) and the source terms~(\ref{Eq_source-term}),
the electron-electron interactions can contribute to the source terms as illustrated in Fig.~\ref{Fig_Feynman Diagram_2}
for the one-loop level~\cite{Cvetkovic2012PRB,Vafek2014PRB,Wang2017PRB}. After carrying out the analogous
calculations in Appendix~\ref{Appendix-one-loop-corrections} in tandem with
the RG scalings in Sec.~\ref{Sec_RGEqs}, we obtain the energy-dependent evolutions of source terms as
follows (to be convenient and consistent with notations in Table~\ref{table:phase}, we hereby add the
scripts PH and PP to denote the particle-hole and particle-particle channels with $c$ and $s$ corresponding
to the charge and spin cases, respectively)
\begin{eqnarray}
\frac{d\Delta^{\mathrm{PH}}_{c1}}{dl}
&=&2\Delta^{\mathrm{PH}}_{c1},\label{Eq_source-RG-1}\\
\frac{d\Delta^{\mathrm{PH}}_{s1-1}}{dl}
&=&2\Delta^{\mathrm{PH}}_{s1-1},\\
\frac{d\Delta^{\mathrm{PH}}_{s1-2}}{dl}
&=&2\Delta^{\mathrm{PH}}_{s1-2},\\
\frac{d\Delta^{\mathrm{PH}}_{s1-3}}{dl}
&=&2\Delta^{\mathrm{PH}}_{s1-3},\\
\frac{d\Delta^{\mathrm{PP}}_{4-1}}{dl}
&=&2\Delta^{\mathrm{PP}}_{4-1},\\\nonumber\\
\frac{d\Delta^{\mathrm{PP}}_{4-2}}{dl}
&=&2\Delta^{\mathrm{PP}}_{4-2},\\\nonumber\\
\frac{d\Delta^{\mathrm{PP}}_{4-3}}{dl}
&=&2\Delta^{\mathrm{PP}}_{4-3},
\end{eqnarray}
and
\begin{widetext}
\begin{eqnarray}
\frac{d\Delta^{\mathrm{PH}}_{c2}}{dl}
&=&\Bigl[2-\frac{t}{4|t|}(7\lambda_{10} -\lambda_{01} -\lambda_{02} -\lambda_{03} -\lambda_{00} -\lambda_{11} -\lambda_{12} -\lambda_{13} +\lambda_{20} +\lambda_{21}  +\lambda_{22} +\lambda_{23} +\lambda_{30}\nonumber\\&& +\lambda_{31} +\lambda_{32} +\lambda_{33} )\Bigr]\Delta^{\mathrm{PH}}_{c2},\\
\frac{d\Delta^{\mathrm{PH}}_{c3}}{dl}
&=&\Bigl[2-\frac{t}{2|t|}(\lambda_{10} -\lambda_{01} -\lambda_{02} -\lambda_{03} -\lambda_{00} +\lambda_{11} +\lambda_{12} +\lambda_{13} +7\lambda_{20} -\lambda_{21} -\lambda_{22} -\lambda_{23} +\lambda_{30}\nonumber\\&& +\lambda_{31} +\lambda_{32} +\lambda_{33} )\Bigr]\Delta^{\mathrm{PH}}_{c3},\\\nonumber\\
\frac{d\Delta^{\mathrm{PH}}_{c4}}{dl}
&=&\Bigl[2-\frac{t}{4|t|}(\lambda_{10} -\lambda_{01} -\lambda_{02} -\lambda_{03} -\lambda_{00} +\lambda_{11} +\lambda_{12} +\lambda_{13} +\lambda_{20} +\lambda_{21} +\lambda_{22} +\lambda_{23} +7\lambda_{30}\nonumber\\&& -\lambda_{31} -\lambda_{32} -\lambda_{33} )\Bigr]\Delta^{\mathrm{PH}}_{c4},\\\nonumber\\
\frac{d\Delta^{\mathrm{PH}}_{s2-1}}{dl}
&=&\Bigl[2-\frac{t}{4|t|}(\lambda_{02} -\lambda_{01} -\lambda_{00} +\lambda_{03} -\lambda_{10} +7\lambda_{11} +\lambda_{12} +\lambda_{13} +\lambda_{20} +\lambda_{21} -\lambda_{22} -\lambda_{23} +\lambda_{30}\nonumber\\&& +\lambda_{31} -\lambda_{32} -\lambda_{33} )\Bigr]\Delta^{\mathrm{PH}}_{s2-1},\\\nonumber\\
\frac{d\Delta^{\mathrm{PH}}_{s2-2}}{dl}
&=&\Bigl[2-\frac{t}{4|t|}(\lambda_{01} -\lambda_{00} -\lambda_{02} +\lambda_{03} -\lambda_{10} +\lambda_{11} +7\lambda_{12} +\lambda_{13} +\lambda_{20} -\lambda_{21} +\lambda_{22} -\lambda_{23} +\lambda_{30}\nonumber\\&& -\lambda_{31} +\lambda_{32} -\lambda_{33} )\Bigr]\Delta^{\mathrm{PH}}_{s2-2}, \\\nonumber\\
\frac{d\Delta^{\mathrm{PH}}_{s2-3}}{dl}
&=&\Bigl[2-\frac{t}{4|t|}(\lambda_{01} -\lambda_{00} +\lambda_{02} -\lambda_{03} -\lambda_{10} +\lambda_{11} +\lambda_{12} +7\lambda_{13} +\lambda_{20} -\lambda_{21} -\lambda_{22} +\lambda_{23} +\lambda_{30}\nonumber\\&& -\lambda_{31} -\lambda_{32} +\lambda_{33} )\Bigr]\Delta^{\mathrm{PH}}_{s2-3},\\
\frac{d\Delta^{\mathrm{PH}}_{s3-1}}{dl}
&=&\Bigl[2-\frac{t}{2|t|}(\lambda_{02} -\lambda_{01} -\lambda_{00} +\lambda_{03} +\lambda_{10} +\lambda_{11} -\lambda_{12} -\lambda_{13} -\lambda_{20} +7\lambda_{21} +\lambda_{22} +\lambda_{23} +\lambda_{30}\nonumber\\&& +\lambda_{31} -\lambda_{32} -\lambda_{33} )\Bigr]\Delta^{\mathrm{PH}}_{s3-1},\\\nonumber\\
\frac{d\Delta^{\mathrm{PH}}_{s3-2}}{dl}
&=&\Bigl[2-\frac{t}{2|t|}(\lambda_{01} -\lambda_{00} -\lambda_{02} +\lambda_{03} +\lambda_{10} -\lambda_{11} +\lambda_{12} -\lambda_{13} -\lambda_{20} +\lambda_{21} +7\lambda_{22} +\lambda_{23} +\lambda_{30}\nonumber\\&& -\lambda_{31} +\lambda_{32} -\lambda_{33} )\Bigr]\Delta^{\mathrm{PH}}_{s3-2},\\\nonumber\\
\frac{d\Delta^{\mathrm{PH}}_{s3-3}}{dl}
&=&\Bigl[2-\frac{t}{2|t|}(\lambda_{01} -\lambda_{00} +\lambda_{02} -\lambda_{03} +\lambda_{10} -\lambda_{11} -\lambda_{12} +\lambda_{13} -\lambda_{20} +\lambda_{21} +\lambda_{22} +7\lambda_{23} +\lambda_{30}\nonumber\\&& -\lambda_{31} -\lambda_{32} +\lambda_{33} )\Bigr]\Delta^{\mathrm{PH}}_{s3-3},\\\nonumber\\
\frac{d\Delta^{\mathrm{PH}}_{s4-1}}{dl}
&=&\Bigl[2-\frac{t}{4|t|}(\lambda_{02} -\lambda_{01} -\lambda_{00} +\lambda_{03} +\lambda_{10} +\lambda_{11} -\lambda_{12} -\lambda_{13} +\lambda_{20} +\lambda_{21} -\lambda_{22} -\lambda_{23} -\lambda_{30}\nonumber\\&& +7\lambda_{31} +\lambda_{32} +\lambda_{33} )\Bigr]\Delta^{\mathrm{PH}}_{s4-1},\\\nonumber\\
\frac{d\Delta^{\mathrm{PH}}_{s4-2}}{dl}
&=&\Bigl[2-\frac{t}{4|t|}(\lambda_{01} -\lambda_{00} -\lambda_{02} +\lambda_{03} +\lambda_{10} -\lambda_{11} +\lambda_{12} -\lambda_{13} +\lambda_{20} -\lambda_{21} +\lambda_{22} -\lambda_{23} -\lambda_{30}\nonumber\\&& +\lambda_{31} +7\lambda_{32} +\lambda_{33} )\Bigr]\Delta^{\mathrm{PH}}_{s4-2},\\\nonumber\\
\frac{d\Delta^{\mathrm{PH}}_{s4-3}}{dl}
&=&\Bigl[2-\frac{t}{4|t|}(\lambda_{01} -\lambda_{00} +\lambda_{02} -\lambda_{03} +\lambda_{10} -\lambda_{11} -\lambda_{12} +\lambda_{13} +\lambda_{20} -\lambda_{21} -\lambda_{22} +\lambda_{23} -\lambda_{30}\nonumber\\&& +\lambda_{31} +\lambda_{32} +7\lambda_{33} )\Bigr]\Delta^{\mathrm{PH}}_{s4-3},\\\nonumber\\
\frac{d\Delta^{\mathrm{PP}}_{1}}{dl}
&=&\Bigl[2+\frac{t}{2|t|}(\lambda_{01} -\lambda_{00} +\lambda_{02} +\lambda_{03} -\lambda_{10} +\lambda_{11} +\lambda_{12} +\lambda_{13} +\lambda_{20} -\lambda_{21} -\lambda_{22} -\lambda_{23} -\lambda_{30}\nonumber\\&& +\lambda_{31} +\lambda_{32} +\lambda_{33} )\Bigr]\Delta^{\mathrm{PP}}_{1},\\\nonumber\\
\frac{d\Delta^{\mathrm{PP}}_{2}}{dl}
&=&\Bigl[2+\frac{t}{4|t|}(\lambda_{01} -\lambda_{00} +\lambda_{02} +\lambda_{03} -\lambda_{10} +\lambda_{11} +\lambda_{12} +\lambda_{13} -\lambda_{20} +\lambda_{21} +\lambda_{22} +\lambda_{23} +\lambda_{30}\nonumber\\&& -\lambda_{31} -\lambda_{32} -\lambda_{33} )\Bigl]\Delta^{\mathrm{PP}}_{2},\\\nonumber\\
\frac{d\Delta^{\mathrm{PP}}_{3}}{dl}
&=&\Bigl[2+\frac{t}{4|t|}(\lambda_{01} -\lambda_{00} +\lambda_{02} +\lambda_{03} +\lambda_{10} -\lambda_{11} -\lambda_{12} -\lambda_{13} -\lambda_{20} +\lambda_{21} +\lambda_{22} +\lambda_{23} -\lambda_{30}\nonumber\\&& +\lambda_{31} +\lambda_{32} +\lambda_{33} )\Bigr]\Delta^{\mathrm{PP}}_{3}.\label{Eq_source-RG-2}
\end{eqnarray}
\end{widetext}

%\section*{Data Availability Statement}
%
%Data Availability Statement: No Data associated in
%the manuscript.

%%%%%%%%%%%%%%%%%%%%%%%%%%%%%%%%%%%%%%%%%%%%%%%%%%%%%%%%%%%%%%%%%%%%%%%%%%%%%%%%%%%%%%%%%
%%%%%%%%%%%%%%%%%%%%%%%%%%%%%%%%%%%%%%%%%%%%%%%%%%%%%%%%%%%%%%%%%%%%%%%%%%%%%%%%%%%%%%%%%

%\end{CJK*}

\end{document}